\documentclass[12pt]{article}

\usepackage{latexsym}
\usepackage{verbatim}
\usepackage{amssymb}
\usepackage{multirow}
\usepackage{XCharter}
\usepackage[dvipsnames]{xcolor}
\definecolor{ColorCite}{named}{BrickRed}
\definecolor{ColorLink}{named}{NavyBlue}
\definecolor{ColorURL}{named}{RoyalBlue}
\definecolor{arXiv}{named}{Maroon}
\usepackage{mathdots}
\usepackage{tabularray}
\usepackage{tikz-cd}
\usepackage{slashed}
\usepackage{nicefrac}
\usepackage{enumitem}
\usepackage{physics}
\usepackage{mathrsfs}
\usepackage{fancyhdr}
\usepackage{hyperref}
\hypersetup{colorlinks=true,
            linkcolor=ColorLink,
            citecolor=ColorCite,
            urlcolor=ColorURL}
\usepackage{cite}
\usepackage{lastpage}
\usepackage[
    top     = 2.75cm,
    bottom  = 2.75cm,
    left    = 20mm,
    right   = 20mm]{geometry}

\usepackage{genyoungtabtikz}
\Yboxdim{8pt}
\Ylinethick{1pt}
\newcommand{\sYng}[1]{{\Yboxdim{5pt}\,\gyoung(#1)}}
\newcommand{\LeadingB}[1]{%
    {\Ylinecolour{BrickRed}\gyoung(#1)}%
}

\newcommand{\Wfirst}[1]{%
    {\Ylinecolour{Orchid}\gyoung(#1)}%
}
\newcommand{\Wsecond}[1]{%
    {\Ylinecolour{Dandelion}\gyoung(#1)}%
}
\newcommand{\Wthird}[1]{%
    {\Ylinecolour{Black}\gyoung(#1)}%
}
\newcommand{\Wfourth}[1]{%
    {\Ylinecolour{NavyBlue}\gyoung(#1)}%
}
\newcommand{\LeadingF}[1]{%
    {\Ylinecolour{NavyBlue}
    \gyoung(#1)_{\color{NavyBlue}\nicefrac12}}%
}
\newcommand{\NoGSO}[1]{%
    {\Ylinecolour{Gray}\gyoung(#1)}
}

\newcommand{\md}{\mathrm{d}}
\newcommand{\ov}{\overline} 
\newcommand{\depth}{{w}}
\newcommand{\pl}{\partial}
\newcommand{\Fock}{\mathfrak{F}}

\newcommand{\N}{\mathbb{N}}
\newcommand{\R}{\mathbb{R}}
\newcommand{\Y}{\mathbb Y}
\newcommand{\Z}{\mathbb{Z}}
\newcommand{\NS}{\textsf{NS}\,}
\newcommand{\GSO}{\textsf{GSO}\,}
\newcommand{\Ram}{\textsf{R}\,}
\newcommand{\Coeff}[1]{\mathtt{#1}}
\newcommand{\osp}{\mathfrak{osp}}
\renewcommand{\sp}{\mathfrak{sp}}
\newcommand{\so}{\mathfrak{so}}
\newcommand{\spLim}{{\color{BrickRed}\bullet}}
\newcommand{\oLim}{{\color{RoyalBlue}\bullet}}
\newcommand{\spRank}{{\color{BrickRed}\mathsf{N}}}
\newcommand{\oRank}{{\color{RoyalBlue}\mathsf{M}}}

\newcommand{\Diag}{{\color{RoyalPurple}\mathsf{K}}}
\newcommand{\BRST}{\mathcal{Q}}
\newcommand{\SUSY}{\mathbf{Q}}
\newcommand{\bone}{{\color{RoyalBlue}b_{-1}}}
\newcommand{\btwo}{{\color{RoyalBlue}b_{-2}}}
\newcommand{\boneh}{{\color{RoyalBlue}b_{-\nicefrac12}}}
\newcommand{\bthreeh}{{\color{RoyalBlue}b_{-\nicefrac32}}}
\newcommand{\bfiveh}{{\color{RoyalBlue}b_{-\nicefrac52}}}
\newcommand{\alphaone}{{\color{BrickRed}\alpha_{-1}}}
\newcommand{\alphatwo}{{\color{BrickRed}\alpha_{-2}}}
\newcommand{\alphathree}{{\color{BrickRed}\alpha_{-3}}}

\usetikzlibrary{positioning, calc, decorations.pathmorphing}
\counterwithin{equation}{section}
\catcode`\@=11
\def\marginnote#1{}
\def\crbig{\\\noalign{\vspace {3mm}}}

\relax

\begin{document}

\pagenumbering{gobble}
\hfill
\vskip 0.01\textheight

\begin{center}

{\Large\bfseries 
On the deep superstring spectrum}

\vspace{0.4cm}

\vskip 0.03\textheight
\renewcommand{\thefootnote}{\fnsymbol{footnote}}
Thomas \textsc{Basile} and Chrysoula \textsc{Markou}
\renewcommand{\thefootnote}{\arabic{footnote}}
\vskip 0.03\textheight

{\em Service de Physique de l'Univers, Champs et Gravitation, \\ Universit\'e de Mons, 20 place du Parc, 7000 Mons, 
Belgium}\\

\end{center}

\vskip 0.02\textheight

\begin{abstract}
    We propose a covariant method of constructing
    entire trajectories of physical states in superstring theory
    in the critical dimension. It is inspired by a recently
    developed covariant technology of excavating bosonic string 
    trajectories, that is facilitated by the observation
    that the Virasoro constraints
    can be written as linear combinations of lowering operators
    of a bigger algebra, namely a symplectic algebra,
    which is Howe dual to the spacetime Lorentz algebra.
    For superstrings, it is the orthosymplectic algebra
    that appears instead, with its lowest weight states forming
    the simplest class of physical trajectories in the \NS sector.
    To construct the simplest class in the \Ram sector,
    the lowest weight states need to be
    supplemented with other states, which we determine.
    Deeper trajectories are then constructed by acting
    with suitable combinations of the raising operators
    of the orthosymplectic algebra, which we illustrate
    with several examples. 
\end{abstract}

\newpage
\tableofcontents
\newpage

\fancyfoot[C]{\thepage\ of \pageref*{LastPage}}

\pagenumbering{arabic}
\setcounter{page}{1}

\section{Introduction}
\label{sec:intro}
The spectrum of superstring theory is organized
in supermultiplets of the $\mathcal{N}=1$ super--Poincar\'e group
in the critical dimension $d=10$. For open superstrings,
it contains a single massless supermultiplet,
composed of a spin--$1$ and a spin--$(\nicefrac12)$ state,
and \emph{infinitely many} massive ones.%
\footnote{The spectrum can be extracted
from the partition function given in terms
of $SO(9)$ characters and spelled out explicitly
up to level $10$ in \cite{Hanany:2010da}
(see also \cite{Curtright:1986di}).}
Each of these massive states is characterized
by a highest weight of the little group $SO(9)$,
i.e. a collection $\Y = (s_1, s_2, s_3, s_4)$
of \emph{ordered} non--negative numbers which are either
\emph{all integers} --- for bosonic states ---
or \emph{all half-integers} --- for fermionic states, which we may think of as the analogue of spin in four dimensions.
To construct these states explicitly
in a systematic manner (for low level examples, see e.g.
\cite{Sasaki:1985py, Tanii:1986ug,Ichinose:1986vj,Polchinski:1986qf, DHoker:1987rxo,Skliros:2011si,Skliros:2016fqs},
and more recently \cite{Biswas:2024unn,Pesando:2024lqa},
for the bosonic string, and \cite{Koh:1987hm,Giannakis:1998wi,Feng:2010yx,Feng:2011qc,Feng:2012bb,Lust:2021jps,Benakli:2021jxs,Benakli:2022edf,Benakli:2022ofz,Lust:2023sfk}
for the superstring), one can for example work
in the light cone,
or use the Del Giudice--Di Vecchia--Fubini (DDF)
construction \cite{DelGiudice:1971yjh, Brower:1972wj, Goddard:1972ky,
Brower:1973iz}, since both can generate 
the whole superstring spectrum. While both methods
have led to great success in the development
of (super)string theories (for instance, in the proof
of the no--ghost theorem \cite{Freeman:1986fx,
Figueroa-OFarrill:1988gxt, Figueroa-OFarrill:1989vhl}),
they also suffer from drawbacks: in the light cone, the output of
constructed  states is not immediately Lorentz covariant, and in the DDF formalism
 it is a superposition
of massive states with different spins.

For the bosonic string, a method for computing
whole \emph{trajectories} at once,
and in a Lorentz covariant manner, was introduced
in \cite{Markou:2023ffh}, based on a result
from representation theory known as Howe duality
\cite{Howe1989i, Howe1989ii}.
The idea of \cite{Markou:2023ffh}, which we extend
to the superstring in this paper, goes as follows.
First, one observes that a polarization tensor
$\varepsilon_\Y$ which is transverse, traceless
and has the symmetry of the Young diagram $\Y$,%
\footnote{The Young symmetry and tracelessness condition
ensure that the polarization tensor carries
the irrep $\Y$ of the Lorentz algebra $\so(d-1,1)$,
while the transversality constraint
reduces its $\so(d-1)$ content to a single irrep,
corresponding to the same Young diagram $\Y$
for the massive little algebra. In other words,
it allows one to describe an $\so(d-1)$--irrep
in a Lorentz--covariant manner. For more details
on this last point, see Appendix \ref{app:Young},
or \cite{Bekaert:2006py}).}
when contracted in a specific way
with $d$--dimensional oscillators of the string Fock space,
defines a simple solution to the Virasoro constraints,
i.e. gives rise to physical states
that were called \emph{principally embedded states}.
They are characterized by the fact that they minimize
the level for a fixed diagram $\Y$, i.e. they constitute
the first occurrence of the diagram $\Y$ 
when going through the bosonic string spectrum
in the direction of increasing mass level.
For instance, the leading Regge trajectory,
namely the set of highest--spins per level,
consists of principally embedded states; 
in the bosonic string, its interactions
were thoroughly explored in \cite{Ademollo:1974kz, Cremmer:1974jq,Sagnotti:2010at} and
 its transversality and tracelessness constraints were put in differential form in \cite{Sagnotti:2010at}, while its interactions in the superstring were investigated
in \cite{Hornfeck:1987wt, Schlotterer:2010kk}.

The second observation of \cite{Markou:2023ffh}
is to recognize the Virasoro algebra as a subalgebra
of $\sp(\bullet)$, the limit of $\sp(2K,\R)$
when $K \longrightarrow \infty$.
This algebra acts on the \emph{modes} of the oscillators,
and importantly, commutes with the Lorentz algebra.
As a consequence, the Fock space of the bosonic string
decomposes into a direct sum of tensor products of irreps
of $\so(d-1,1)$ and $\sp(\bullet)$ wherein each irrep
of one algebra appears together with a unique irrep
of the other. Put differently, the bosonic string Fock space
can be decomposed in subspaces, in which each state
carries a given Lorentz representation $\Y$, 
and also forms an irreducible module of $\sp(\bullet)$.
One can therefore consider this $\sp(\bullet)$--module,
often called ``Howe dual'' to $\Y$,
as its \emph{multiplicity space}. Upon working 
in the transverse subspace,\footnote{This restriction
is possible since the BRST cohomology of the bosonic string
is included in the transverse subspace
\cite{Kato:1982im, Henneaux:1986kp, Manes:1988gz}.}
all states with $\so(d-1)$--irrep $\Y$ can be found
in the $\sp(\bullet)$--module, and hence one can focus
on this module to look for physical states of spin $\Y$.

The last point is that the Howe dual representation 
to $\Y$ is a \emph{lowest weight module} of $\sp(\bullet)$,
whose lowest weight vector is the principally embedded state
discussed previously. Consequently, one can reach 
any other state with spin $\Y$ by acting
on the principally embedded one with raising operators
of $\sp(\bullet)$. Building physical states of spin $\Y$
therefore amounts to finding polynomials in $\sp(\bullet)$
raising operators which are compatible with the physicality
conditions, meaning that they commute with the positive--modes
Virasoro generators, up to lowering operators of $\sp(\bullet)$,
as the latter act trivially on principally embedded states.
In doing so, the dependence of the algorithm on $\Y$ is fairly minimal:
the lengths of its rows merely appear as parameters
in the polynomials in raising operators, thereby allowing one
to derive whole \emph{trajectories of states}
\cite[Sec. 4.3]{Markou:2023ffh}. In fact, a more relevant
parameter for defining the notion of trajectory
in the bosonic string seems to be the number of rows
of a Young diagram. The other important quantity
is the difference between the level
of the principally embedded state with diagram $\Y$,
and the level (necessarily higher) at which a clone
of the same diagram appears, called \emph{depth}.

In this work, we extend this method to the case
of the superstring in the Ramond--Neveu--Schwarz 
(RNS) formulation. The reasoning behind the construction
is almost identical to the bosonic string case,
upon replacing the symplectic algebra $\sp(\bullet)$
with its orthosymplectic counterpart $\osp(\oLim|\spLim)$,
and identifying the multiplicity space of a Lorentz irrep
with a particular lowest weight module of it.
There are however, a few noteworthy differences,
owing to the specificities of the superstring
compared to the bosonic one.
\begin{itemize}
\item To begin with, one should distinguish 
between the Neveu--Schwarz (\NS\!) and the Ramond (\Ram\!) sectors,
the former consisting of spacetime bosons while the latter
of spacetime fermions. While, in both sectors,
the $\osp(\oLim|\spLim)$--module dual to a given
Lorentz irrep $\Y$ is of lowest weight type,
its lowest weight vector is also the principally embedded
state of diagram $\Y$ \emph{only} in the \NS sector.
In the \Ram sector, we find that one should consider
a linear combination of the lowest weight vector 
together with other states, as elaborated on in subsection \ref{sec:principal}. This is related to the fact
that, in the $\Ram$ sector, some $\osp(\oLim|\spLim)$
raising operators do not change the level, and therefore
the subspace of states at minimal level in the lowest weight
$\osp(\oLim|\spLim)$--module is not reduced to
the lowest weight vector as in the \NS sector, 
but is of finite dimension (greater than $1$).
This also leads to the appearance of non--trivial multiplicities
for principally embedded fermions of mixed--symmetry, which are confirmed by the results of \cite{Hanany:2010da} for low levels.
The situation is illustrated in figure \ref{fig:NS-R} below.

\item The spectrum of superstring theory in the RNS formalism
becomes spacetime supersymmetric only after implementing
the Gliozzi--Scherk--Olive (\GSO\!) projection that is necessary for modular invariance
\cite{Gliozzi:1976qd}. In the \NS sector, this means
projecting out all half--integer levels. Our construction
is formulated before the \GSO projection and
we find that it is both possible to uplift physical states
that are projected out to ones that are not, and vice--versa.

\item Finally, we observe that the number of boxes
of the principal \textit{diagonal} of Young diagrams seems
to define a sensible notion of a superstring trajectory,
instead of the number of rows in the bosonic string.
\end{itemize}

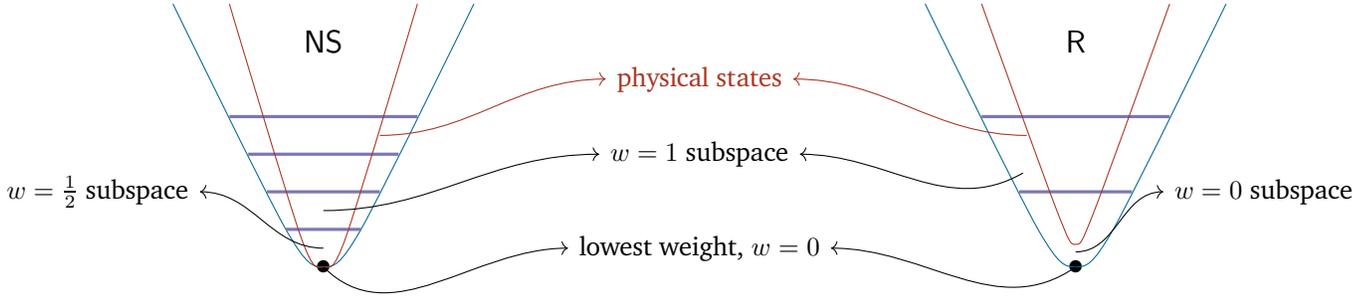
\begin{figure}[ht!]
    \centering\label{fig:NS-R}
    \begin{tikzpicture}
        \node (NS) at (0,3) {$\mathsf{NS}$};
        \node (R) at (10,3) {$\mathsf{R}$};
        \node (lwNS) at (0,0) {$\bullet$};
        \node (lwleg) at (5,0.25)
                {\footnotesize lowest weight, $w=0$};
        \node (w12legNS) at (-3,1)
                {\footnotesize $w=\tfrac12$ subspace};
        \node (w1leg) at (5,1.5)
                {\footnotesize $w=1$ subspace};
        \node (physleg) at (5,2.5) 
                {\footnotesize\color{Mahogany} physical states};
        \draw[MidnightBlue] plot [smooth] coordinates
        {(-2,3.5) (-0.5,0.5) (0,0) (0.5,0.5) (2,3.5)};
        \draw[Periwinkle, very thick]
                        (-0.5,0.5) -- (0.5,0.5);
        \draw[Periwinkle, very thick]
                        (-0.75,1) -- (0.75,1);
        \draw[Periwinkle, very thick]
                        (-1,1.5) -- (1,1.5);
        \draw[Periwinkle, very thick]
                        (-1.25,2) -- (1.25,2);
        \draw[Mahogany] plot [smooth] coordinates
        {(-1.25,3.5) (-0.3,0.4) (0,0) (0.3,0.4) (1.25,3.5)};
        \draw (0,0) edge[->, out=-45, in=180, 
                        decoration={zigzag}] (lwleg);
        \draw (0,0.25) edge[->, out=180, in=0, 
                        decoration={zigzag}] (w12legNS);
        \draw (0,0.75) edge[->, out=0, in=180, 
                        decoration={zigzag}] (w1leg);
        \draw (0.75,1.75) edge[Mahogany, ->, out=0, in=180,
                        decoration={zigzag}] (physleg);
        \node (lwR) at (10,0) {$\bullet$};
        \node (w0legR) at (12.5,1)
                {\footnotesize $w=0$ subspace};
        \draw[MidnightBlue] plot [smooth] coordinates
        {(8,3.5) (9.5,0.5) (10,0) (10.5,0.5) (12,3.5)};
        \draw[Periwinkle, very thick] (9.25,1) -- (10.75,1);
        \draw[Periwinkle, very thick] (8.75,2) -- (11.25,2);
        \draw[Mahogany] plot [smooth] coordinates
        {(8.75,3.5) (9.75,0.75) (10,0.3) (10.25,0.75) (11.25,3.5)};
        \draw (10,0) edge[->, out=215, in=0, 
                        decoration={zigzag}] (lwleg);
        \draw (10,0.2) edge[->, out=0, in=180, 
                        decoration={zigzag}] (w0legR);
        \draw (9.3,1.25) edge[->, out=210, in=0, 
                        decoration={zigzag}] (w1leg);
        \draw (9.35,1.75) edge[Mahogany, ->, out=180, in=0,
                        decoration={zigzag}] (physleg);
    \end{tikzpicture}
    \caption{Schematic representation of an $\osp(\spLim|\oLim)$
    lowest weight module, Howe dual to an irrep $\Y$ 
    of the Lorentz algebra $\so(d-1,1)$,
    in the \textcolor{MidnightBlue}{blue} outline.
    The \textcolor{Periwinkle}{purple} horizontal lines
    separate subspaces of different depths $w$,
    and the \textcolor{Mahogany}{red} outline the subspace
    of \emph{physical states}.}
\end{figure}

This paper is organized as follows.
In Section \ref{sec:superstring}, we give a brief summary
of superstring theory in the Ramond--Neveu--Schwarz 
formalism, focusing on the construction of physical
vertex operators and the corresponding physical states.
In Section \ref{sec:Howe}, we quickly review Howe duality
for the bosonic string, and  discuss its extension relevant
for the superstring. In Section \ref{sec:spectrum_sections},
we start by deriving the physicality conditions
in the \NS sector and discuss their counterpart
in the \Ram sector, before identifying
principally embedded states in both sectors. 
We then examine operators of small values of the depth,
present various solutions for those and use them
to derive examples of physical trajectories.
We conclude in Section \ref{sec:conclusion}.
Appendices \ref{app:lcRS} and \ref{app:Young}
contain a review of the construction
of some of the lightest \Ram states
in the light cone gauge and our conventions
for Young diagrams respectively.

\paragraph{Conventions.}
Spacetime indices are denoted by Greek letters,
$\mu, \nu, \ldots = 0, \ldots, d-1\,$
and we use the mostly--plus metric signature.
We work in the critical dimension of the superstring,
namely $d=10$ implicitly. The dot product ``$\cdot$''
denotes the Minkowski product in flat spacetime.
For the open string, we set $\alpha'=\tfrac12$
as is customary (and equivalently $\alpha'=2$
for the closed string), which means
that the string field $X^\mu$ has $0$ mass dimension.

\section{Superstring elements}
\label{sec:superstring}

\subsection{Ingredients of the worldsheet SCFT}

We will be employing the Ramond--Neveu--Schwarz
(RNS) formulation of the superstring, which we briefly
review in this subsection following
\cite{Friedan:1985ge, Blumenhagen:2013fgp}
(see also \cite{Schlotterer:2011psa}
for a pedagogical introduction). We focus
on the open superstring, namely holomorphic fields
on the string worldsheet, and recall that
it is straightforward to obtain closed strings
by introducing their antiholomorphic counterparts
and imposing the level--matching condition.
Fundamental ingredients are the ``matter'' spacetime vectors
$\partial X^\mu$ and $\psi^\mu$, that are superpartners
and primary fields of weight $1$ and $\tfrac12$
respectively with respect to
the worldsheet superconformal field theory (SCFT);
they are functions of the (complex) worldsheet coordinate $z$.
The corresponding worldsheet energy--momentum tensor
and supercurrent read
\begin{align}\label{EMsc1} 
    T^{X,\psi}(z) =  -\frac{1}{2}\, \partial X \cdot \partial X\,(z)
    -\frac{1}{2}\, \psi \cdot \partial \psi\,(z)\,, \quad 
    T^{X,\psi}_{\textrm{F}}(z)
    = \frac{i}{2}\, \psi \cdot \partial X\,(z)\,,
\end{align}
and are naturally superpartners. In the BRST quantization,
the conformal $(b,c)$ and superconformal $(\beta,\gamma)$
ghost systems, with weights equal to $(2,-1)$
and $\big( \frac{3}{2}, -\frac{1}{2}\big)$ respectively,
have to be introduced. The ghost energy--momentum tensor
and supercurrent read respectively
\begin{align}
    T^{\textrm{gh}}(z) &= T^{b,c}(z) + T^{\beta, \gamma}(z) 
    = -2\,b\partial c(z) - (\partial b)c(z)
    -\frac{3}{2}\,\beta \partial \gamma(z)
    -\frac{1}{2}\,(\partial \beta) \gamma(z)\,, \crbig
    T_{\textrm{F}}^{\textrm{gh}}(z)
    &= \frac{1}{2}\, b\gamma(z) - (\partial \beta) c(z)
    -\frac{3}{2}\, \beta \partial c(z) \,.
\end{align}
The (nilpotent) worldsheet BRST charge then takes the form
\begin{align}
    \BRST = \BRST_0 + \BRST_1 + \BRST_2\,,
\end{align}
with
\begin{subequations}
\begin{align}\label{charge1}
    \BRST_0 &= \oint \frac{\md z}{2\pi i} \, c \,
    \Big[T^{X,\psi} +T^{\beta,\gamma} +(\partial c) b \Big]\,,
    \\ \label{charge2}
    \BRST_1 &= -\oint \frac{\md z}{2\pi i} \, e^{-\chi} 
    e^{\varphi} \, T^{X,\psi}_{\textrm{F}}\,,
    \\ \label{charge3}
    \BRST_2 &= \frac{1}{4} \oint \frac{\md z}{2\pi i} \, b \, 
    e^{-2\chi} e^{2\varphi}\,,
\end{align}
\end{subequations}
where the superghost system is bosonized by the chiral bosons
$\varphi$ and $\chi$ as in
\begin{align}
    \beta = e^{-\varphi } e^\chi \partial \chi\,
    \qquad\text{and}\qquad
    \gamma = e^{-\chi} e^\varphi\,,
\end{align}
so that the index $i$ of $\mathcal Q_i$ stands
for its superghost charge.\footnote{The superghost
charges of $\beta$ and $\gamma$ are $-1$ and $+1$
respectively.} Note that $b$ and $c$  are Grassmann--odd,
while $\beta$ and $\gamma$ are Grassmann--even.
The worldsheet $\mathcal{N}=1$ superconformal algebra reads
\begin{subequations}
\begin{align}\label{SCF1}
    T(z) \, T(w) & \sim \tfrac12\,\frac{\mathsf c}{(z-w)^4}
    + \frac{2\,T(w)}{(z-w)^2} + \frac{\partial T(w)}{z-w}\,,\\ 
    \label{SCF2}
    T(z) \, T_F(w) & \sim \tfrac32\,\frac{T_F(w)}{(z-w)^2}
    + \frac{\partial T_F(w)}{z-w}\,, \\
    \label{SCF3}
    T_F(z) \, T_F(w) & \sim \tfrac16\,\frac{\mathsf c}{(z-w)^3}
    + \tfrac12\,\frac{T(w)}{z-w}\,,
\end{align}
\end{subequations}
where the symbol ``$\sim$'' stands for equality up to regular terms
and the central charge $\mathsf c$  is equal to $\pm 15$
for the matter fields and superghosts respectively.
Note that the CFTs of $X$, $\psi$, $(b,c)$ and $(\beta, \gamma)$
are free and thought of as being decoupled. 

Due to the two possibilities for the periodicity
of $\psi^\mu$, the open superstring comprises two sectors, 
the Neveu--Schwarz ($\mathsf{NS}$) and the Ramond
($\mathsf{R}$), in which $\psi$ is periodic and antiperiodic
on the complex plane respectively, which is mapped here
to the two--dimensional disk. The boson OPEs read
\begin{align} \label{boson2point}
    X^\mu(z) X^\nu(w) 
    \sim -  \eta^{\mu \nu} \, \ln |z - w|\,,
\end{align}
and
\begin{align}\label{boson2pointgh}
    \chi(z) \chi(w) \sim \ln |z - w|\,,
    \qquad\qquad
    \varphi(z) \varphi(w) \sim - \ln |z - w|\,,
\end{align}
whereas the fermion OPE reads
\begin{align} \label{fermion2pointNS}
    \psi^\mu(z) \psi^\nu(w) \sim \frac{\eta^{\mu \nu}}{z-w}\,,
\end{align}
in both sectors. 
The $\mathsf R$ sector requires the introduction
of a primary field $S_A$ of weight $\tfrac{d}{16}$, 
which is a Dirac spinor of $\mathfrak{so}(d-1,1)$.
Upon decomposing $S_A$ into left-- and right--handed Weyl spinors
$S_\alpha$ and $S^{\dot\alpha}$, the relevant OPE read
\begin{align}\label{weirdOPEs}
    S_\alpha(z) S_\beta(w) &\sim \frac{(\gamma^\mu C)_{\alpha \beta}\,\psi_\mu(w)}{\sqrt2\,(z-w)^{3/4}}
\,,\\ \label{fermion_mixed}
     \psi^\mu(z) S_\alpha(w) &=  \frac{ (\gamma^\mu)_{\alpha\dot\beta}\,S^{\dot\beta}(w)}{\sqrt{2}\,(z-w)^{1/2}}+(z-w)^{1/2}\Big[\tfrac{\sqrt{2}}{d-2}\psi^\mu \slashed{\psi}-\tfrac{1}{\sqrt{2}(d-2)(d-1)}\gamma^\mu \slashed{\psi}\slashed{\psi}\Big]_{\alpha \dot{\beta}} S^{\dot{\beta}}(w)+\dots\,,
 \end{align}
where we used the conventions of \cite[App. A \& B]{Feng:2012bb}
for the gamma and charge conjugation matrices,
denoted by $\gamma^\mu$ and $C$ respectively.
The subleading term in \eqref{fermion_mixed}
was first given in terms of contractions
of $\psi^\mu \psi^\nu S^{\dot{\beta}}$ in \cite{Kostelecky:1986xg}
(see also \cite{Koh:1987hm}) and much later obtained
in the covariant form in terms of $S^\mu_\alpha$
and $\partial S^{\dot{\beta}}$ \cite{Feng:2012bb},
with $S^\mu_\alpha$ being a vector--spinor
of weight $\nicefrac{(d+16)}{16}\,$.
Finally, we will also need the OPE
\begin{equation}\label{superghopeexp}
    \begin{aligned}
        e^{q_1\varphi(z)}e^{q_2\varphi(w)} & =  (z-w)^{-q_1 q_2} \\
        & \quad \times \bigg[1+ (z-w)\,q_1 \partial \varphi
        + \frac{1}{2}(z-w)^{2}\,q_1 \big[\partial^2\varphi
        + q_1 (\partial \varphi)^2\big]+\ldots \bigg] 
        e^{(q_1+q_2)\varphi(w)}\,.
    \end{aligned}
\end{equation}

\paragraph{Physical states.}
A generic vertex operator creating asymptotic states
of momentum $p^\mu$ takes the form
\begin{equation}\label{eq:Fdefin}
    V_F^{(q)}(p,z) = g_{\textrm{o}} \, T^a \,  e^{q\,\varphi}\, F
    (\partial^k X, \partial^r \psi)\,
    \begin{cases}
        e^{i\,p \cdot X}\,, & \textrm{ in the } \mathsf{NS} \textrm{ sector} \\
        S_A \, e^{i\,p \cdot X}\,,  & \textrm{ in the } \mathsf{R} \textrm{ sector,}
    \end{cases}
\end{equation}
where $g_{\textrm{o}}$ is the open string coupling constant
and $T^a$ the brane gauge group generators, namely the Chan--Paton factors \cite{Paton:1969je};
the factor $g_{\textrm{o}} \, T^a$ will always be implicit
from now on. 
Here, $F$ is a polynomial in
$\partial X^\mu$, $\psi^\mu$ and their derivatives,
and $q$ is called the ghost picture of $V_F^{(q)}$.
The factor $e^{q\,\varphi}$ and the momentum eigenstate
$e^{i\,p \cdot X}$ contribute $-\frac{1}{2}q(q+2)$
and $\tfrac12\,p^2$ units of conformal weight respectively,
while the descendants $\partial^k X$ and $\partial^r \psi$ 
have weights $k$ and $r+\tfrac12$ respectively.
Vertex operators are spacetime scalars,
so the polynomial $F$ further contains a priori
arbitrary tensors (tensor--spinors) in the $\mathsf{NS}$
($\mathsf{R}$) sector, which are introduced to construct
a Lorentz invariant quantity out of the previous ingredients.
Superstring states belonging to the $\mathsf{NS}$
($\mathsf R$) sector are thus spacetime bosons (fermions).
In the BRST quantization, the physical state condition
takes the form
\begin{align}\label{gen_phys}
    [\mathcal Q, V_F^{(q)}] = \textrm{tot. deriv.}\,,
\end{align}
and implies separately
\begin{equation}\label{physical_three}
    \begin{aligned}
        \relax [\mathcal Q_0,V_F^{(q)}(z)] =\textrm{tot. deriv.}\,,
        \qquad
        [\mathcal Q_1,V_F^{(q)}(z)] = \textrm{tot. deriv.}\,,
        \qquad
        [\mathcal Q_2,V_F^{(q)}(z)] = +\,\textrm{tot. deriv.}\,,
    \end{aligned}
\end{equation}
since the three parts $\mathcal{Q}_0$, $\mathcal{Q}_1$
and $\mathcal{Q}_2$ have different superconformal ghost charges. 

Moreover, the vertex operator
creating a physical state can be written
in infinitely many different ways, which correspond
to different values of the picture $q$.
Different pictures are obtained via the operation
\begin{align}\label{picture}
    V_F^{(q+1)}(w)
    &=2\Big[\mathcal Q, e^{\chi(w)}\, V_F^{(q)}(w)\Big]\,,
\end{align}
valid in both sectors, $\mathsf{NS}$ and $\mathsf{R}$.
$F$ is simplest in what is called the canonical ghost picture,
which corresponds to $q=-1$ and $q=-\tfrac12$
(or $-\tfrac32$) in the $\mathsf {NS}$
and $\mathsf R$ sector respectively.\footnote{In the \NS (\Ram) sector, it holds that $q\in \mathbb{Z}$ ($q\in \mathbb{Z}+\tfrac12$) for the allowed pictures.
} Notice that, while \eqref{picture} takes the form of a spurious state by construction, it \textit{does} contribute to scattering amplitudes, namely all allowed pictures are physical. In addition, because of the first of the constraints \eqref{physical_three},
every physical vertex operator has to have conformal weight
equal to $1$. Indeed, assuming that $V^{(q)}_F(w)$ has conformal weight $h$ with respect to the total energy--momentum tensor, $T=T^{X,\psi} + T^{b,c} + T^{\beta,\gamma}$, namely
\begin{align} \label{opeEMvo}
    T(z) V^{(q)}_F(w) \sim \frac{h \, V^{(q)}_F(w) }{(z-w)^2} + \frac{\partial V^{(q)}_F(w)}{z-w}\,,
\end{align}
the first of \eqref{physical_three} yields $h=1$ for any vertex operator in any picture $q$. In the canonical ghost picture,
this yields the conformal weight $h_F$ of its corresponding polynomial as
\begin{equation}
    h_F = \phi - \tfrac12 p^2\,,
\end{equation}
where
\begin{align}
    \phi=\begin{cases}
       \tfrac12\,, & \textrm{in the $\mathsf{NS}$ sector} \\  0 \,, & \textrm{in the $\mathsf{R}$ sector.} 
    \end{cases}
\end{align}

\subsection{Fock space and oscillators}

The various fields of the worldsheet CFTs can be expanded
in Fourier modes, for example
\begin{equation} \label{field_expansions}
    i\partial X^\mu(z)
    = \sum_{n \in \Z} \alpha^\mu_n\,z^{-n-1} \,,
    \qquad 
    \psi^\mu(z)
    = \sum_{r \in \Z+\phi} b_r^\mu\,z^{-r-\frac12}\,,
\end{equation}
and
\begin{equation}
    T^{X,\psi}(z) = \sum_{n \in \Z} L_n\,z^{-n-2} \,,
    \qquad 
    T^{X,\psi}_{\textrm{F}}(z)
    = \frac{1}{2}\sum_{r \in \Z+\phi} G_r\,z^{-r-\frac32}\,,
\end{equation}
where
\begin{equation}
    \alpha^\mu_0 = p^\mu \,. 
\end{equation}
The level number operator reads
\begin{align} \label{level}
    N := \sum_{m=1}^\infty \alpha_{-m} \cdot \alpha_m
    + \sum_{r \in \Z+\phi>0}^\infty r\,b_{-r} \cdot b_r\,,
\end{align}
while the modes of the energy--momentum tensor and supercurrent
are given by the bosonic and fermionic oscillators via%
\footnote{Note that, in the \Ram sector,
the $L_0$ has been shifted by $-\tfrac{d}{16}$
so as to have the standard realization
of the super--Virasoro algebra \eqref{superVis1}--\eqref{superVis3}.
This shift is cancelled by the \Ram vacuum's conformal weight
upon action of $L_0$ in the physicality condition \eqref{masscond}.}
\begin{align}
    L_n &= \tfrac12 \sum_{m\in \mathbb{Z}}
    :\alpha_{-m} \cdot \alpha_{m+n}:
    +\, \tfrac12 \sum_{r\in \mathbb{Z}+\phi}
    \big(r+\tfrac{n}{2}\big) :b_{-r} \cdot b_{n+r}:
    -\delta_{n,0}\,\tfrac{d}8(\tfrac12-\phi)\,,\\
    G_r & =\sum_{m\in \mathbb{Z}} :\alpha_{-m} \cdot b_{m+r}:\,,
\end{align}
where $:(\dots):$ denotes the normal ordering of oscillators.
Explicitly separating the positive and negative modes,
and implementing the normal ordering, the super--Virasoro
generators read
\begin{equation}
    \begin{aligned}
        L_{n>0} & = p \cdot \alpha_n
        + \sum_{m\geq1} \alpha_{-m} \cdot \alpha_{m+n}
        + \tfrac12\,\sum_{m=1}^{n-1} \alpha_m \cdot \alpha_{n-m}\\
        & \hspace{80pt} + \sum_{r\geq1} (r+\tfrac{n}2-\phi)\,
        b_{-r+\phi} \cdot b_{r-\phi+n}
        + \tfrac12\,\sum_{r=2\phi}^n (-r+\tfrac{n}2+\phi)\,
        b_{r-\phi} \cdot b_{n-r+\phi}\,,
    \end{aligned}
\end{equation}
and 
\begin{equation} \label{supervis_gen_osc}
    G_{r-\phi>0} = p \cdot b_{r-\phi}
    + \sum_{k\geq1} \alpha_{-k} \cdot b_{k+r-\phi}
    + \sum_{k\geq2\phi} b_{-k+\phi} \cdot \alpha_{k+r-2\phi}
    + \sum_{k=1}^{r-1} \alpha_k \cdot b_{r-\phi-k}\,,
\end{equation}
for the positive modes, and the negative ones
can be obtained by conjugation. Note that some sums
in the above expressions start at $2\phi$,
i.e. $1$ in the \NS sector and $0$ in the \Ram sector,
due to the existence of \emph{zero--modes} for the fermionic
oscillators in the \Ram sector. Finally, one finds
\begin{equation}
    L_0 = \tfrac12\,p^2
    + \sum_{k\geq1} \big(\alpha_{-k} \cdot \alpha_k
    + (k-\phi)\,b_{-k+\phi} \cdot b_{k-\phi}\big)
    - \tfrac{d}8\,(\tfrac12-\phi)
    \equiv \tfrac12\,p^2 + N - \tfrac{d}8\,(\tfrac12-\phi)\,,
\end{equation}
where $N$ is the level number operator introduced above.
Requiring states to be eigenvectors of $L_0$
therefore fixes their mass $m^2=-p^2$ in terms
of the level $N$.

The analogue of the boson OPE \eqref{boson2point} is the bosonic oscillator algebra 
\begin{equation}\label{eq:bos_osc}
   [\alpha_m^\mu, \alpha_n^\nu]     = m\,\delta_{m+n,0}\,\eta^{\mu\nu}\,,
    \qquad 
    (\alpha_m^\mu)^\dagger=\alpha_{-m}^\mu\,,
    \qquad 
    m, n\, \in  \Z \,,
\end{equation}
while the analogue of the fermion OPE \eqref{fermion2pointNS} is the fermionic oscillator algebra
\begin{equation}\label{eq:ferm_osc}
    \{b_m^\mu, b_n^\nu\} = \delta_{m+n,0}\,\eta^{\mu\nu}\,,
    \qquad 
    (b_m^\mu)^\dagger = b_{-m}^\mu  \,, \qquad m, n \in \Z+\phi\,,
\end{equation}
and the analogue of \eqref{SCF1}--\eqref{SCF3}
is the super--Virasoro algebra
\begin{subequations}
    \begin{align} \label{superVis1}
        [L_m, L_n] & = (m-n)\, L_{m+n}
        + \tfrac{d}{8}\, m(m^2-1)\, \delta_{m+n}\,, \\ 
        \label{superVis2}
        [L_m, G_r] & = \big(\tfrac{m}{2} - r\big) G_{m+r}\,,\\ 
        \label{superVis3}
        \{G_r, G_s\} &= 2\, L_{r+s}
        +\tfrac{d}{2} \big(r^2-\tfrac14\big) \delta_{r+s}\,.
    \end{align}
\end{subequations}
The Fock space is generated by the action
of the oscillators with negative modes
on a vacuum state $\ket0$, whose defining property
is to be annihilated by positive--mode oscillators
\begin{align}
    \alpha_m^\mu\,\ket{0} &= 0 = b_r^\mu\,\ket{0}\,,
    \qquad \forall\, m,r>0\,.    
\end{align}
The physicality conditions \eqref{physical_three}
on a general state $\ket{\textrm{phys}}$, built out of non--positive--mode oscillators acting on the vacuum, now take
the form
\begin{align} \label{masscond}
    \big( L_0-\phi \big) |\textrm{phys} \rangle &= 0\,, \\ \label{conosc1}
    L_m |\textrm{phys} \rangle &= 0\,, \quad m>0\,, \\ \label{conosc2}
    G_r |\textrm{phys} \rangle &= 0\,, \quad   r  \geq \phi\,.
\end{align}
Moreover, sufficient are the following constraints 
\begin{align}\label{suff}
    \mathsf{NS}\, &:\,
    \Big( L_0- \frac{1}{2} \Big)\ket{\textrm{phys}} = 0\,,
    \quad G_{\nicefrac12} \ket{\textrm{phys}} = 0\,,
    \quad G_{\nicefrac32} \ket{\textrm{phys}} = 0\,, \\
    \mathsf{R}\, & :\, G_0 \ket{\textrm{phys}} = 0\,,
    \quad G_1 \ket{\textrm{phys}} = 0\,,
\end{align}
as a consequence of the super--Virasoro algebra
\eqref{superVis1}--\eqref{superVis3}.

\subsection{Building the spectrum and the lightest states}

Using the creation operators $\alpha_{-n}$ and $b_{-r+\frac12}$
in the \NS sector and $\alpha_{-n}$ and $b_{-r}$ in the \Ram sector,
a general candidate for a physical state can be written as
\begin{align}\label{genNS}
   \varepsilon_{\mu_1(m_1)|\dots|\mu_I(m_I)\pmb|\nu_1[n_1]|\dots|\nu_J[n_J]}^{\NS}\, \alpha_{-k_1}^{\mu_1(m_1)}\dots \alpha_{-k_I}^{\mu_I(m_I)}\, b^{\nu_1[n_1]}_{-r_1+\frac12}\dots b^{\nu_J[n_J]}_{-r_J+\frac12}\, \ket{p;0}_\NS \,, \\ \label{genR}
    \varepsilon_{\mu_1(m_1)|\dots|\mu_I(m_I)\pmb|\nu_1[n_1]|\dots|\nu_J[n_J]}^{\Ram, A} \, \alpha_{-k_1}^{\mu_1(m_1)}\dots \alpha_{-k_I}^{\mu_I(m_I)}\, b^{\nu_1[n_1]}_{-r_1}\dots b^{\nu_J[n_J]}_{-r_J}\,  \ket{p; A;0}_\Ram\,,
\end{align}
in the \NS and in the \Ram sector respectively,
where we have used the notation
\begin{equation}
    \alpha_{-k}^{\mu(m)} := \alpha_{-k}^{(\mu_1}
    \dots \alpha_{-k}^{\mu_m)} \,,
    \qquad 
    b^{\nu [n]}_{-r+\phi} := b^{[\nu_1}_{-r+\phi}
    \dots b^{\nu_n]}_{-r+\phi} \,,
\end{equation}
both as a way to shorten the expression
and to highlight the commuting, respectively
anticommuting, nature of $\alpha$, respectively
$b$.  The polarization tensor $ \varepsilon^{\NS}$
and the polarization tensor--spinor $ \varepsilon^{\Ram, A}$
have $I$ groups of $m_i$ symmetric indices,
and $J$ groups of $n_j$ antisymmetric indices,
which a priori do not obey any particular symmetry condition
between one another --- requiring that a state be physical
will result in imposing certain symmetry
conditions on its polarization tensor. Using \eqref{level},
the respective levels at which the states \eqref{genNS}
and \eqref{genR} find themselves at are 
\begin{equation}
    N =  \sum_{i=1}^I k_i\,m_i + \sum_{j=1}^J (r_j-\phi)\,n_j\,.
\end{equation}

\paragraph{State--operator correspondence.}
By means of the state--operator correspondence,
string states can be mapped to vertex operators,
that is the language most suitable for the computation
of string scattering amplitudes,
as the latter can be written as vacuum expectation values
of the product of the vertex operators creating all external states
in a given diagram. To recall how the correspondence is realized,
let us start with the vacuum. The momentum eigenstate vacuum
in the \NS sector is created by
\begin{equation}
    \ket{p;0}_\NS = \lim_{z\to0} e^{i\,p \cdot X(z)} \ket0\,,
\end{equation} 
while in the \Ram sector, it is created by
\begin{equation}
    \ket{p; A;0}_\Ram = \lim_{z\to0} S_A\,e^{i\,p\cdot X}(z)\ket0\,.
\end{equation}
In fact, the interpolation between the two sectors
by means of the spin--field $S_A$ can be expressed via
\begin{align}
     \ket{p; A;0}_\Ram = \lim_{z\to0} S_A(z)  \ket{p;0}_\NS \,.
\end{align}

Moreover, the expansions \eqref{field_expansions}
can be inverted to yield, for example, the creation operators
as integrals of the fields $\partial X$ and $\psi$:
\begin{align}\label{bosint}
    \alpha^\mu_{-n} =\oint \frac{\md z}{2 \pi i}\frac{1}{z^n}\,i\partial X^\mu(z)
    \qquad\text{and}\qquad
    b^\mu_{-r+\phi}&=\oint \frac{\md z}{2 \pi i}
    \frac{1}{z^{r+\nicefrac12-\phi}}\,\psi^\mu(z)\,.
\end{align}
Consequently, the bosonic dictionary in both sectors
is easily derived from \eqref{bosint} as
\begin{equation}\label{dictionarybos}
    \alpha^\mu_{-n} \ket0
    \quad\longleftrightarrow\quad
    \frac{1}{(n-1)!}\, i\partial^n X^\mu(0) \,,
    \qquad
    \alpha^\mu_{n}
    \quad\longleftrightarrow\quad
    n! \, \eta^{\mu \nu} 
    \frac{\delta}{i \delta \partial^n X^{\nu}} \,,
    \qquad n \geqslant 1 \,,
\end{equation}
as is the fermionic dictionary in the \NS sector:
\begin{align} \label{dictionaryNS}
    b^\mu_{-r+\frac12} \ket{0}
    \quad\longleftrightarrow\quad
    \frac{1}{(r-1)!}\,\partial^{r-1} \psi^\mu(0) \,,
    \qquad
    b^\mu_{r-1/2}
    \quad\longleftrightarrow\quad
    (r-1)! \, \eta^{\mu \nu}
    \frac{\delta}{ \delta \partial^{r-1} \psi^{\nu}} \,,
    \qquad r \geqslant 1 \,.
\end{align}
Applying the dictionary \eqref{dictionarybos}
and \eqref{dictionaryNS} on the creation operators
of \eqref{genNS}, the vertex operator for a general \NS state
is given by
\begin{equation}\label{eq:word}
    \varepsilon_{\mu_1(m_1)|\dots|\mu_I(m_I)\pmb|\nu_1[n_1]|\dots|\nu_J[n_J]}^{\NS}\, i\partial^{k_1} X^{\mu_1(m_1)} \dots 
    i\partial^{k_I} X^{\mu_I(m_I)}\,
    \partial^{r_1-1} \psi^{\nu_1[n_1]} \dots 
    \partial^{r_J-1} \psi^{\nu_J[n_J]}\,,
\end{equation}
where we have absorbed the factors $(n_i-1)!$ and $(r_j-1)!$ in the polarization tensor $\varepsilon$.

To obtain the fermionic dictionary in the \Ram sector,
we have to calculate
\begin{align}\label{partialdictionaryR}
    b_{-r}^\mu  \ket{p; A;0}_\Ram
    = \lim_{w\rightarrow 0}  \oint \frac{\md z}{2 \pi i}
    \frac{1}{z^{r+\nicefrac12}} \psi^\mu(z) S_A(w) \ket{p;0}_\NS \,.
\end{align}
To perform this calculation for any $r$, knowledge of the OPE of $\psi^\mu(z) S_A(w)$ to \textit{arbitrary} order is required to obtain an integrand with non--vanishing integral, so, with the means presented so far, we cannot present a closed formula that relates the action of the creation operators $b_{-r}$, for any $r$, on the vacuum to $\psi$ and its descendants. However, once a certain value of $r$ is chosen, the computation is feasible at finite time. For example,
 \begin{align}\label{dictionaryR1}
    b_{-1}^\mu  \ket{ p;\alpha;0}_\Ram 
    \quad\longleftrightarrow\quad
    \Big[ \tfrac{\sqrt{2}}{d-2}\psi^\mu \slashed{\psi}
    - \tfrac{1}{\sqrt{2}(d-2)(d-1)} \gamma^\mu \slashed{\psi} \slashed{\psi} \Big]_{\alpha\dot{\beta}} \,S^{\dot{\beta}}(0) \ket{p;0}_\NS \,,
\end{align}
where we needed the first subleading term in the OPE
$\psi^\mu(z) S_A(w)$, given in \eqref{fermion_mixed}.
Nevertheless, knowledge of \eqref{dictionaryR1}
is not sufficient once more than one $b$ are present,
since for example
 \begin{align}
    b_{-1}^\mu b_{-1}^\nu \ket{p; A;0}_\Ram
    = \lim_{w\rightarrow 0}  \oint \frac{\md z}{2 \pi i} \frac{\md y}{2 \pi i}   \frac{1}{(zy)^{3/2}} \psi^\mu(z) \psi^\mu(y) S_A(w) \ket{p;0}_\NS \,,
\end{align}
which cannot be calculated with the standard version
of Wick's theorem.\footnote{Here we use the term ``standard'' to distinguish Wick's theorem for non--interacting fields from the generalized Wick's theorem for interacting fields described for example in \cite[Chap. 6.B]{DiFrancesco:1997nk}.} To avoid having to find the dictionary
for the states/trajectory in question,
we will be concentrating on the Fock space realization
of \Ram states.

Instead, in the \NS sector, writing the most general Ansatz
for a vertex operator at level $N$ boils down
to enumerating the half--integer partitions
of $N-\tfrac12$ (i.e. all possible ways
of obtaining $N-\tfrac12$ as a sum of positive
integers \emph{and half--integers}). Indeed,
such partitions encode monomials of the form \eqref{eq:word}.
Now let us recall how the string spectrum is built
using the traditional methods. In the covariant language
of vertex operators, to build states at a level $N$,
one writes all possible terms with conformal weight $N$
contributing to the respective polynomial $F$ that appears in the generic expression \eqref{eq:Fdefin}
and imposes the BRST constraints \eqref{physical_three}.
For example, at $N=\frac32$ namely $p^2=-\tfrac12$
in the $\mathsf{NS}$ sector, there are three possible monomials
that can contribute,
\begin{equation}
    F_{\nicefrac32}
    = \varepsilon_{\mu\pmb|\nu}\,i\partial X^\mu\,\psi^\nu
    + \varepsilon_\mu\,\partial \psi^\mu
    + \varepsilon_{\mu\nu\lambda}\,
    \psi^\mu \psi^\nu \psi^\lambda\,,
\end{equation}
where $\varepsilon_{\mu\pmb|\nu}$, $\varepsilon_\mu$
and $\varepsilon_{\mu\nu\lambda}$ are a priori arbitrary
(the latter is by construction totally antisymmetric).
Imposing \eqref{physical_three} yields \cite{Koh:1987hm, Feng:2010yx} 
the lightest massive spin--2 and the lightest spin--$(1,1,1)$
of the open superstring spectrum, with vertex operators given by 
\begin{align}\label{lightestspin2}
    V^{(-1)}(p,z) =   e^{-\varphi} \, \varepsilon_{\mu\nu}(p)\,i\partial X^\mu\,\psi^\nu \, e^{ip\cdot X} \,, \quad p^\mu  \varepsilon_{ \mu \nu} = 0 \,, \quad  \varepsilon_{[\mu \nu]} = 0\,,
\end{align}
and
\begin{align}
    V^{(-1)}(p,z) = e^{-\varphi}\,
    \varepsilon_{\mu \nu \lambda}(p)\,
    \psi^{\mu} \psi^{\nu}  \psi^{\lambda}\, e^{ip\cdot X}\,,
    \quad p^\mu \varepsilon_{ \mu \nu \lambda} = 0\,,
\end{align}
respectively. At level $2$, one finds for instance 
the lightest hook,
\begin{align}\label{eq:hook}
    V^{(-1)}(p,z) = e^{-\varphi}\,
    \varepsilon_{\mu\pmb|\nu\rho}(p)\,
    i\partial X^\mu\, \psi^\nu \psi^\rho\,e^{i\,p \cdot X}\,,
    \qquad 
    p^\mu\,\varepsilon_{\mu\pmb|\nu\rho} = 0\,,
    \qquad \varepsilon_{[\mu\pmb|\nu\rho]} = 0\,,
\end{align}
where $\varepsilon$ is also traceless.
The same procedure can be employed to construct levels
of the $\mathsf{R}$ sector. Alternatively,
one can apply the light cone technology,
namely use the transverse creation oscillators $\alpha^i_{-n}$
and $b^i_{-r}\,$, $i=0,\dots, d-2$ to construct
all possible $\so(d-2)$ irreps, which are then to be recombined
into $\so(d-1)$ irreps, on a level--by--level basis.
For the example of the lightest levels
of the $\mathsf{R}$ sector, this procedure is illustrated
in Appendix \ref{app:lcRS}. The lightest physical states
in both sectors are displayed in table \ref{lightest}.

\paragraph{\GSO projection and spacetime supersymmetry.}

The previously described methodology leads to a spectrum
which is not supersymmetric and contains a tachyonic state.
A way of treating these pathologies is via the \GSO projection,
which we can think of essentially as projecting out
half the states, namely all half--integer \NS levels,
displayed in gray in table \ref{lightest},
and half of the ``generalized'' chirality of all \Ram states,
see Appendix \ref{app:lcRS} for a very brief review.
The number of propagating d.o.f. after
the $\mathsf{GSO}$ projection is given in the last column
of table \ref{lightest}. $\mathcal{N}=1$ spacetime
super--Poincar\'e symmetry is generated by 
\begin{align}
    P^\mu = \oint \frac{\md z}{2 \pi i}\, i \partial X^\mu \,,
    \qquad
    J^{\mu \nu} = \oint \frac{\md z}{2 \pi i}\,
    \Big[iX^{[\mu} \pl X^{\nu]} + \psi^{[\mu} \psi^{\nu]}\Big]\,,
\end{align}
and 
\begin{align}
    \SUSY_\alpha^{(-\nicefrac12)}
    = 2^{1/4} \oint \frac{\md z}{2 \pi i}\, S_\alpha \,
    e^{-\varphi/2} \,,
    \qquad 
    \SUSY_\alpha^{(+\nicefrac12)}
    = \frac{1}{2^{1/4}} \oint \frac{\md z}{2 \pi i}\, 
    i\partial X \cdot (\gamma \, S) _\alpha \, e^{+\varphi/2} \,,
\end{align}
with algebra
\begin{align}
    \Big\{\SUSY_\alpha^{(-\nicefrac12)},
    \SUSY_\beta^{(+\nicefrac12)} \Big\}
    = (\gamma \, C)_{\alpha \beta} \cdot P\,,
    \quad \textrm{etc}\,.
\end{align}
After the $\mathsf{GSO}$ projection, the open superstring 
states at every mass level form multiplets of $\mathcal{N}=1$
supersymmetry, with the supersymmetry generators
creating $\mathsf R$ from $\mathsf {NS}$ states
(at the same level) and vice versa according to
\begin{subequations}
    \begin{align}\label{susy1}
        V^{(-\nicefrac12)}_{\mathsf{R}}(p,z)
        & = \Big[\eta^\alpha \, \SUSY_{\alpha}^{(+\nicefrac12)},
        V_{\mathsf{NS}}^{(-1)}(p,z) \Big]\,, \\ \label{susy2}
        V^{(-1)}_{\mathsf{NS}}(p,z)
        & = \Big[\eta^\alpha \, \SUSY_{\alpha}^{(-\nicefrac12)},
        V_{\mathsf{R}}^{(-\nicefrac12)}(p,z) \Big]\,.
    \end{align}
\end{subequations}
The existence of these operations suggests that,
once a physical state or trajectory is constructed
in one sector, they can be used to find its superpartner
in the other sector.

The lowest states form the massless vector supermultiplet,
whose vector's vertex operator in the canonical ghost picture reads
\begin{align}
    V^{(-1)}(p,z) = e^{-\varphi}\,
    \varepsilon_\mu(p)\, \psi^\mu\, e^{ip\cdot X}\,.
\end{align}
Its superpartner is given by
\begin{align} \label{fermionsuper}
    V^{(-\nicefrac12)}(p,z)
    = \Big[\eta^\alpha\, \SUSY_{\alpha}^{(+\nicefrac12)},
    V^{(-1)}(p,z) \Big]
    = e^{-\varphi/2}\,\upsilon^\alpha(p)\,
    S_\alpha\, e^{ip\cdot X}\,,
\end{align}
where we have used the OPEs \eqref{superghopeexp}
and \eqref{fermion_mixed} to show that the integrand
has a single pole and the spinor wavefunction $\upsilon^\alpha$
is identified with
\begin{align}
    \upsilon^\alpha(p) = \frac{1}{2^{3/4}}\,\varepsilon_\nu(p)\,
    p_\mu\,\eta^\beta\,\gamma^\mu_{\beta \dot{\beta}}\,
    \ov{\gamma}^{\nu \dot{\beta}\alpha}\,.
\end{align}
Notice that the vertex operator \eqref{fermionsuper}
of the spin--$(\nicefrac12)$ is produced in the canonical ghost picture
for the $\mathsf R$ sector with this method. More generally,
the spectrum can be visualised in terms of Regge trajectories,
as in the bosonic string. The leading Regge trajectory
consists of the highest spins of every level. In the NS sector,
it thus contains the states with symmetric rank--$s$
polarisation tensors $\varepsilon_{\mu_1 \dots \mu_s}$
that find themselves at level $s-1$, namely $\tfrac12p^2 = 1-s$,
with vertex operator \cite{Schlotterer:2010kk}
\begin{align} \label{leadingvo}
    V^{(-1)}_{\mathsf{NS}}(p,z) = e^{-\varphi}\,
    \varepsilon_{\mu_1 \dots \mu_s}(p)\,
    i\partial X^{\mu_1} \dots i\partial X^{\mu_{s-1}}\,
    \psi^{\mu_s}\, e^{ip\cdot X}\,,
    \quad p^\mu \varepsilon_{\mu \mu_2 \dots \mu_s} = 0\,,
    \quad \varepsilon^{\mu}_{\hphantom{\mu} \mu \mu_3 \dots \mu_s} = 0\,, 
\end{align}
and is highlighted in red in table \ref{lightest}.
Its superpartner trajectory, namely the leading Regge
in the \Ram sector that is highlighted in blue
in table \ref{lightest}, can be found via
\begin{align} \label{leadingsuper}
\begin{aligned}
    V^{(-\nicefrac12)}_{\mathsf{R}}(p,z)
    & = \Big[\eta^\alpha \, \SUSY_{\alpha}^{(+\nicefrac12)},
    V_{\mathsf{NS}}^{(-1)}(p,z) \Big] \crbig
    & = e^{-\varphi/2}\,
    \bigg[\upsilon^\alpha_{\mu_1 \dots \mu_{s-1}}(p)\,
    i\partial X^{\mu_1} \dots i\partial X^{\mu_{s-1}} \crbig
    & \qquad + \rho_{\dot{\beta}\, \mu_1 \dots \mu_{s-1}}(p)\, 
    i\partial X^{\mu_1} \dots i\partial X^{\mu_{s-2}}\, 
    \psi^{\mu_{s-1}}\,\slashed{\psi}^{\dot{\beta} \alpha}\bigg]\, S_\alpha\, e^{ip\cdot X}\,,
\end{aligned}
\end{align}
see also \cite{Koh:1987hm, Giannakis:1998wi, Schlotterer:2010kk}, where we have used the identity $\gamma_{\alpha \dot \beta}^{(\mu} \,\ov{\gamma}^{\nu) \, \dot \beta \gamma} = - \eta^{\mu \nu} \delta_\alpha^\gamma$
and that $\varepsilon_{\mu_1 \dots \mu_s}$ is symmetric
and traceless and we have defined
\begin{align}
   \upsilon^\alpha_{\mu_1 \dots \mu_{s-1}}(p)
   = \frac{1}{2^{3/4}} \, \varepsilon_{\mu_1 \dots \mu_s}(p)\, 
   p_\mu\,\eta^\beta \,\gamma^\mu_{\beta \dot{\beta}} \, 
   \ov{\gamma}^{\mu_s\, \dot{\beta}\alpha} \,,\quad
   \rho_{\dot{\beta}\, \mu_1 \dots \mu_{s-1}}(p)
   =\frac{1}{2^{1/4}}\,(s-1)\,\varepsilon_{\mu_1 \dots \mu_s}(p) \, \eta^\alpha \, \gamma^{\mu_s}_{\alpha\dot{\beta}}\,.
\end{align}
Notice that, for $s=1$, the second term in \eqref{leadingsuper} 
disappears and we recover \eqref{fermionsuper}.

\begin{table}
\centering 
\renewcommand{\arraystretch}{1.5}
  \begin{tabular}{ c || l | l | c }
   $m^2$ & $\mathsf{NS}$ & $\mathsf{R}$ & propagating d.o.f. \\ \hline \hline
   $-\tfrac12$ & $\textcolor{Gray}{\bullet}$ & &  \\
   $0$ & $\LeadingB{;}^{\textcolor{BrickRed}{so(d-2)}}$
   & $\textcolor{NavyBlue}{\bullet^{so(d-2)}_{\nicefrac12}}$
   & $\textbf{8}_{\textrm{B}}+ \textbf{8}_{\textrm{F}}$ \\
   $\tfrac12$ & $\NoGSO{;,;}$ & &  \\
   $1$ & $\LeadingB{;;} \oplus\, \gyoung(;,;,;)$
   & $\LeadingF{;}$ &  $\textbf{128}_{\textrm{B}}
   + \textbf{128}_{\textrm{F}}$ \\
   $\tfrac32$ & $\NoGSO{;;}\, \oplus\, {\color{Gray}\bullet}\,
   \oplus\, \NoGSO{;,;,;,;}\, \oplus\, \NoGSO{;;,;}$ && \\
   $2$ & $\LeadingB{;;;}\, \oplus\, \gyoung(;)\, 
   \oplus\, \gyoung(;;,;)\, \oplus\, \gyoung(;,;)\,
   \oplus\, \gyoung(;;,;,;)\, \oplus\, \gyoung(;,;,;,;,;)$ 
   & $\LeadingF{;;}\, \oplus\, \gyoung(;)_{\nicefrac12}\, 
   \oplus\, \bullet_{\nicefrac12}
   \oplus\, \gyoung(;,;)_{\nicefrac12}$
   & $\textbf{1152}_{\textrm{B}}
   + \textbf{1152}_{\textrm{F}}$ 
  \end{tabular}
\renewcommand{\arraystretch}{1}
\caption{Open superstring, physical content of the first few levels. All Young diagrams appearing
should be understood as irreducible representations
of the massive little group $SO(d-1)$ in $d$--dimensions,
except at mass level $m^2=0$ where they denote
irreps of the massless little group $SO(d-2)$.
The bosonic part of the leading Regge trajectory is indicated
in \textcolor{BrickRed}{red} and its fermionic superpartner
in \textcolor{NavyBlue}{blue}.}
\label{lightest}
\end{table}

At this point, it is clear that, as in the bosonic string,
the construction of physical superstring states on a level--by--level
basis can quickly become cumbersome as the level increases.
In the following, we will extend to the superstring the methodology
recently developed in \cite{Markou:2023ffh} using Howe duality.
To this end, we will first recall how Howe duality appears
in the bosonic string spectrum, namely for the bosonic oscillators
$\alpha_n^\mu$, and proceed to analyse how it can be realised
once the fermionic oscillators $b_n^\mu$ are added.

\section{Bosonic and fermionic versions
of Howe duality}
\label{sec:Howe}

Let us consider a generic Fock space $\Fock$ generated by 
pairs of creation--annihilation operators.
$\Fock$ can be thought of as the oscillator representation
of a group $G$, which is either the symplectic group $Sp$
if they are all bosonic, the orthogonal group $O$
if they are all fermionic, or the orthosymplectic group $OSp$
if both are present. A pair of reductive groups $H$
and $\widetilde H$, both of which are subgroups of $G$
and which are each other's centralisers%
\footnote{This last condition means that,
not only $H$ and $\widetilde H$ commute, but also
that $H$ is the maximal subgroup commuting
with $\widetilde H$ in $G$, and vice versa.},
is then called a ``reductive dual pair''.
The reductive dual pair correspondence,
also known as ``Howe duality'' \cite{Howe1989i, Howe1989ii},
see also \cite{Rowe:2012ym, Basile:2020gqi},
refers then to a ``1--1'' correspondence
between the irreducible representations of $H$ and $\widetilde H$.
In particular, the important property of the decomposition
of $\Fock$ under $H \times \widetilde H$ is that
it is \emph{strongly multiplicity--free},
meaning that not only pairs of representations
of $H$ and $\widetilde H$ appear only once
in the decomposition, but also that any irrep,
appears in one such pair at most. In other words,
as a $H \times \widetilde H$ representation,
the Fock space $\Fock$ is a direct sum
of the form\footnote{The symbol $\boxtimes$
denotes the \emph{external} tensor product 
of a $H$--irrep and a $\widetilde H$--irrep,
resulting in a $(H \times \widetilde H)$--irrep.}
\begin{equation}
    \Fock\big|_{H \times \widetilde H}\,
    \cong \bigoplus_\lambda \pi^H_\lambda
    \boxtimes \pi^{\widetilde H}_{\theta(\lambda)}\,,
\end{equation}
where the sum runs over a subset of irreps
$\pi^H_\lambda$ of $H$  (those appearing in $\Fock$)
labelled by $\lambda$, and $\theta(\lambda)$ is the label
for the corresponding $\widetilde H$ representation
$\pi^{\widetilde H}_{\theta(\lambda)}$.
More importantly, the map $\theta$ between $H$--irreps
and $\widetilde H$--irreps appearing in $\Fock$
is a \emph{bijection}, meaning that each representation
$\pi_\lambda^H$ of $H$ appears only once in the above
decomposition of $\Fock$, accompanied by a unique irrep
$\pi_{\theta(\lambda)}^{\widetilde H}$ of $\widetilde H$.

In the remainder of this section,
we will briefly review \cite{Markou:2023ffh},
and recall how the bosonic version of Howe duality
is tied to the description of arbitrary symmetry
Lorentz tensors \emph{in the symmetric basis}.
Next, we will discuss the fermionic version of Howe duality,
i.e. the duality between irreps of dual pairs
in fermionic Fock spaces and argue that it is related
to the description of mixed--symmetry Lorentz tensors 
\emph{in the antisymmetric basis}.
Finally, as both bosonic and fermionic oscillators
are needed for the construction of superstring states,
we will put the two versions together
and explain how Howe duality in a Fock space
generated by both types of oscillators
allows one to describe Lorentz tensors
in a mixture of the symmetric and antisymmetric basis.

\subsection{Bosonic version and the symmetric basis}

To discuss physical states in the superstring spectrum,
we will only need to use Howe duality between
\emph{Lie algebra representations}, so that
we will work at the level of algebras from now on.

For the bosonic string, the relevant Howe dual pair
is the one formed by the Lorentz algebra $\so(d-1,1)$
and the symplectic algebra $\sp(2\spRank,\R)$,
in the oscillator representation of $\sp(2\spRank d,\R)$,
i.e. the bosonic Fock space defined by $\spRank \times d$ pairs
of creation and annihilation operators.
Howe duality was first observed in the bosonic string spectrum
in \cite{Markou:2023ffh} and allows one to characterise
and construct tensors of $\so(d-1,1)$
with the symmetry of a $\spRank$--row Young diagram
in terms of lowest weight irreps of $\sp(2\spRank,\R)$,
thereby constructing entire physical string trajectories
efficiently and covariantly.

Let us recall the bosonic version
of Howe duality, following \cite[sec. 3.3]{Markou:2023ffh}.%
\footnote{Bosonic Howe duality has also been used
in the context of higher spin gravity,
see e.g.\cite{Vasiliev:2003ev, Alkalaev:2008gi, 
Boulanger:2008kw, Alkalaev:2009vm, Bekaert:2009fg, 
Alkalaev:2011zv, Bekaert:2013zya}
and \cite[Sec. 3]{Bekaert:2004qos} for a concise 
review of its use to treat arbitrary tensors.}
We will work on the subspace of the Fock space 
generated by the negative modes
of the bosonic oscillators \eqref{eq:bos_osc}, i.e. the space
of linear combinations of states of the form 
\begin{equation}
    \ket{\varepsilon}
    := \varepsilon_{\mu_1(\ell_1)|\cdots|\mu_p(\ell_p)}\,\alpha_{-i_1}^{\mu_1(\ell_1)} \dots \alpha_{-i_p}^{\mu_k(\ell_p)} \ket0\,,
    \qquad
    1 \leq i_j \leq \spRank\,, \quad i_j \neq i_k\,,
\end{equation}
where $\varepsilon_{\mu_1(\ell_1)|\cdots|\mu_l(\ell_p)}$
is a tensor with $p$ groups of \emph{symmetric} indices
(with $0 \leq p \leq \spRank$), without additional
symmetries among indices of different groups. $\spRank$ is an integer that enumerates the different types of creation operators excited; its value depends on the level or trajectory in question, while it can become arbitrarily large in principle, reflecting the infinite size of the string spectrum --- it will be thought of as implicitly finite throughout this work. The action of the Lorentz algebra $\so(d-1,1)$
commutes with that of $\sp(2\spRank,\R)$, with the latter being
generated \emph{on this subspace} by
\begin{equation}\label{eq:sp}
    T^{mn} := \tfrac1{m\,n}\,
    \alpha_{-m} \cdot \alpha_{-n}\,,
    \qquad 
    T^m{}_n := \tfrac1m\,\alpha_{-m} \cdot \alpha_n
    + \tfrac{d}2\,\delta_{mn}\,,
    \qquad 
    T_{mn} := \alpha_m \cdot \alpha_n\,,
\end{equation}
where $m,n=1,2,\dots,\spRank$.
The $\sp(2\spRank,\R)$ indices 
correspond to the various groups of symmetric
spacetime indices carried by a given type
of creation operator.\footnote{Our convention
for the prefactors in the definition of the generators 
\eqref{eq:sp} or \eqref{eq:sp:fields}
of the $\mathfrak{sp}(\spLim)$ algebra is slightly
different from the one used in \cite{Markou:2023ffh}.}
Consequently, the $\sp(2\spRank,\R)$ generators
encode various operations on these groups of indices,
namely:
\begin{itemize}
\item The generators $T^k{}_k$, without implicit
summation, count how many indices
are in the $k$th group,
\begin{equation}
    T^k{}_k \ket\varepsilon
    = (\ell_k+\tfrac{d}2)\,\ket\varepsilon\,,
\end{equation}
modulo a constant shift by $\tfrac{d}2$.
\item The generators $T^m{}_n$, with $m \neq n$,
symmetrize an index of the $n$th group
with \emph{all} indices of the $m$th group. 
In particular, it allows one to reformulate
the condition that $\varepsilon$ is irreducible
under $\mathfrak{gl}_d$. Let us assume that $\varepsilon$ 
is of the form $\ket\varepsilon
=\varepsilon_{\mu_1(s_1)|\cdots|\mu_l(s_l)}\,
\alpha_{-1}^{\mu_1(\ell_1)} \dots \alpha_{-p}^{\mu_p(\ell_p)} \ket0$ with
\begin{equation}
    \ell_1 \geq \ell_2 \geq \dots \geq \ell_p > 0\,,
\end{equation}
and $0 \leq p \leq \spRank$. Then, in the symmetric basis,
irreducibility of $\varepsilon$ reads
\begin{equation}
    T^m{}_n \ket{\varepsilon} = 0\,,
    \quad 1 \leq m<n \leq \spRank
    \qquad\Longleftrightarrow\qquad 
    \varepsilon_{\cdots|\mu_m(\ell_m)|\cdots|\mu_m\mu_n(\ell_n-1)|\cdots}=0\,,
\end{equation}
or in plain words, the symmetrization of all indices
in the $m$th group with one index of the \emph{subsequent},
$n$th group, should vanish.
\item The generators $T_{mn}$ take a trace between
one index in the $m$th group and another one
in the $n$th group. Requiring that $\varepsilon$
be traceless therefore translates into
\begin{equation}
    T_{mn} \ket{\varepsilon} = 0
    \qquad\Longleftrightarrow\qquad
    \eta^{\alpha\beta}\,\varepsilon_{\cdots,\mu_m(\ell_m-1)\alpha,\cdots,\mu_n(\ell_n-1)\beta,\cdots}=0\,,
\end{equation}
for all pairs $1 \leq m, n \leq \spRank$.
\item Finally, the generators $T^{mn}$ multiply
the tensor by a metric $\eta$, and symmetrizes
one of its indices with all those of the $m$th group,
and the other index with all indices of the $n$th group.
\end{itemize}

The action of these generators gives us a way to define
an irreducible tensor $\varepsilon$ of the Lorentz algebra
in a bosonic Fock space by 
a vector $\ket\varepsilon$, verifying
\begin{equation}\label{lowestwB}
    \varepsilon\ \simeq\
    \begin{tikzpicture}[scale=0.6, baseline=(ref.base)]
    \coordinate (ref) at (0.5,1.25);
    \coordinate (toprow) at (-3,2.5);
    \edef\s{\dimexpr9cm\relax}
    \edef\h{\dimexpr0.5cm\relax}
    \edef\g{\dimexpr2cm\relax}
    
    \draw[thick] (toprow) rectangle ++(\s,\h);
    \draw[thick] ($(toprow)-(0,\h)$)
                    rectangle ++(0.8*\s,\h);
    \draw[thick] ($(toprow)-(0,2*\h)$)
                    rectangle ++(0.65*\s,\h);
    \draw[thick] ($(toprow)-(0,2*\h+\g)$)
                    rectangle ++(0.25*\s,\h);
    \draw[thick, dashed] ($(toprow)+(0.55*\s,-2.5*\h)$)
                    -- ($(toprow)+(0.3*\s,-0.5*\h-\g)$);
    \draw[thick] ($(toprow)+(0.55*\s,-2.5*\h)$)
        -- ++(0,0.5*\h) -- ($(toprow)-(0,2*\h)$)
        -- ++(0,-\g+\h) -- ++(0.3*\s,0) -- ++(0,0.5*\h);
    \node at ($(toprow)+(0.4*\s,0.5*\h)$) {\scriptsize$\ell_1$};
    \node at ($(toprow)+(0.4*\s,-0.5*\h)$) {\scriptsize$\ell_2$};
    \node at ($(toprow)+(0.4*\s,-1.5*\h)$) {\scriptsize$\ell_3$};
    \node at ($(toprow)+(0.12*\s,-1.5*\h-\g)$) {\scriptsize$\ell_N$};
    \end{tikzpicture}
    \quad\Longleftrightarrow\quad\left\{
    \begin{aligned}
        T^k{}_k \ket\varepsilon
        & = (\ell_k+\tfrac{d}2)\ket\varepsilon\,,
        & k=1,\dots,\spRank\,,\\
        T^m{}_n \ket\varepsilon & = 0\,,
        & 1 \leq m < n \leq \spRank\,,\\
        T_{mn} \ket\varepsilon & = 0\,,
        & m,n=1,\dots,\spRank\,,
    \end{aligned}
    \right.
\end{equation}
i.e. $\ket\varepsilon$ is a \emph{lowest weight vector}
for $\sp(2\spRank,\R)$.\footnote{Indeed,
the first equation fixes the $\sp(2\spRank,\R)$ weight
of $\ket\varepsilon$ --- its eigenvalues
under the Cartan subalgebra generators
$T^k{}_k$ --- while the other two equations
mean that  $\ket\varepsilon$ is annihilated
by all lowering operators.}
This is due to the fact that the Lorentz
and symplectic algebras form a reductive dual pair,
and hence Howe duality tells us that the subspace 
of states generated by the first $\spRank$ pairs of modes
of $\partial X$ can be decomposed
into irreducible representations of their direct sum,
$\so(d-1,1) \oplus \sp(2\spRank,\R)$, such that
each representation of a given algebra appearing
in this decomposition does so together with a \emph{unique}
irrep of the other algebra. In other words,
the infinitely many disparate and physical apparitions
of a given $\so(d-1,1)$ irrep in the bosonic string spectrum
correspond to different irreps of  $\sp(2\spRank,\R)$
and can be reached by acting with the raising operators,
$T^{mn}$ and $T^{m>n}{}_n\,$, of the latter
on lowest weight states as defined in \eqref{lowestwB}.

\subsection{Fermionic version
and the antisymmetric basis}
\label{sec:fermionic_Howe}
Now let us turn to the fermionic
version of Howe duality (see e.g.
\cite{Rowe:2011zz, Rowe:2012ym} for reviews). 
In this case, the Fock space in question can be thought of as the oscillator representation of the orthogonal group, in the sense
that ``fermionic dual pairs'' are found as mutual
centralisers in it. We will be interested in
the dual pairs composed of the Lorentz algebra
$\so(d-1,1)$ and another orthogonal algebra,
either $\mathfrak{o}(2\oRank)$
or $\mathfrak{o}(2\oRank+1)$. The action
of these algebras on a fermionic Fock space
is generated by  bilinears in the oscillators,
as in the bosonic case. There are,
however, a couple of specificities to the fermionic
case: $(i)$ the Fock space is finite--dimensional
which restricts the type of irreps of the dual pairs
that can appear in the decomposition,
and $(ii)$ whether the dual algebra is
$\mathfrak{o}(2\oRank)$ or $\mathfrak{o}(2\oRank+1)$
determines the bosonic or fermionic nature
of the corresponding Lorentz representations.%
\footnote{See also, e.g.,
\cite[Sec. VI]{Gunaydin:1987hb} for a similar
discussion in the context of constructing
lowest weight representation of $\osp(2m+1|2n,\R)$.} 

\paragraph{\NS sector.}
Let us start with the subspace of the \NS sector
of the Fock space that is generated by
the negative modes half--integer
$-\tfrac12, -\tfrac32, \dots, -\oRank+\tfrac12$
of the fermionic oscillators \eqref{eq:ferm_osc},
i.e. the space of linear combinations of states
of the form
\begin{equation}
    \ket{\varepsilon}
    := \varepsilon_{\mu_1[h_1]|\dots|\mu_M[h_l]}\,
    b^{\mu_1[h_1]}_{-r_1+\frac12}
    \dots b^{\mu_l[h_l]}_{-r_l+\frac12} \ket0_{\sf NS}\,,
    \qquad 
    1 \leq r_j \leq \oRank\,, \quad r_j \neq r_k\,,
\end{equation}
where $\varepsilon_{\mu_1[h_1]|\dots|\mu_l[h_l]}$
is a tensor of $l$ groups of antisymmetric indices
(with $0 \leq l \leq \oRank$),
without additional symmetries among indices
of different groups. As in the bosonic case,
$\oRank$ is implicitly thought of as a finite integer. 
The Lorentz algebra $\so(d-1,1)$ , generated by the bilinears 
 \begin{equation} \label{LorentzR}
     J^{\mu\nu} = 2\,\sum_{r\geq 1}
     b_{-r+\frac12}^{[\mu} b^{\nu]}_{r-\frac12}\,,
 \end{equation}
on this subspace commutes with the action
of the orthogonal algebra $\mathfrak{o}(2\oRank)$,
generated by
\begin{equation}\label{eq:o}
    M^{rs} := b_{-r+\frac12} \cdot b_{-s+\frac12}\,,
    \qquad 
    M^r{}_s := b_{-r+\frac12} \cdot b_{s-\frac12}
    - \tfrac{d}2\,\delta_{rs}\,,
    \qquad 
    M_{rs} = b_{r-\frac12} \cdot b_{s-\frac12}\,,
\end{equation}
where $r,s=1,2,\dots,\oRank$. For the sake 
of completeness, let us recall the commutation
relation of $\mathfrak{o}(2\oRank)$ in the above basis,
\begin{subequations}
    \label{eq:o_commutation}
    \begin{align}
        [M^r{}_s, M^{tu}] & = 2\,\delta^{[u}_s\,M^{t]r}\,,
        &&& [M^r{}_s, M^t{}_u]
        & = \delta^t_s\,M^r{}_u - \delta^r_u\,M^t{}_s\,,\\
        [M^r{}_s, M_{tu}] & = -2\,\delta_{[u}^r\,M_{t]s}\,,
        &&& [M^{rs}, M_{tu}]
        & = -4\,\delta^{[r}_{[t}\,M^{s]}{}_{u]}\,.
    \end{align}
\end{subequations}
Now in parallel with the bosonic case,
the $\mathfrak{o}(2\oRank)$ indices correspond
to the various groups of \emph{antisymmetric}
Lorentz indices, and hence the $\mathfrak{o}(2\oRank)$
generators encode operations on these groups. 
More precisely:
\begin{itemize}
\item The generators $M^r{}_r$, without summation
implied, returns the number of indices in the $r$th group,
\begin{equation}
    M^r{}_r \ket{\varepsilon}
    = (h_r - \tfrac{d}2) \ket{\varepsilon}\,,
\end{equation}
modulo a shift by $-\tfrac{d}2$.
\item The generators $M^r{}_s$ with $r \neq s$
antisymmetrize an index of the $s$th group
with \emph{all} indices of the $r$th group. 
Here again, it allows one to reformulate
the condition that $\varepsilon$ is irreducible
under $\mathfrak{gl}_d$. Assuming that the groups
of indices of $\varepsilon$ are ordered by decreasing
length, i.e. $h_1 \geq h_2 \geq \dots \geq h_M$,
irreducibility of $\varepsilon$ reads
\begin{equation}
    M^r{}_s \ket{\varepsilon} = 0\,,
    \quad 1 \leq r < s \leq \oRank
    \qquad\Longleftrightarrow\qquad 
    \varepsilon_{\cdots|\mu_r[h_r]|\cdots|\mu_r\mu_s[h_s-1]|\cdots}=0\,,
\end{equation}
in the \emph{antisymmetric basis}.
\item The generators $M_{rs}$ take a trace between an index
of the $r$th group and another one of the $s$th group.
In particular, tracelessness of $\varepsilon$ reads
\begin{equation}
    M_{rs} \ket{\varepsilon} = 0
    \qquad\Longleftrightarrow\qquad
    \eta^{\alpha\beta}\,\varepsilon_{\cdots,\mu_r[h_r-1]\alpha,\cdots,\mu_s[h_s-1]\beta,\cdots}=0\,,
\end{equation}
for any $1 \leq r, s \leq \oRank$.
\item Finally, the generators $M^{rs}$ act by multiplying
the tensor $\varepsilon$ with a metric $\eta$,
and antisymmetrizing one of its index with all those
of the $r$th group of indices in $\varepsilon$,
and the other one with all indices of the $s$th group.
\end{itemize}

In complete parallel with the bosonic case, we can single out
an irreducible Lorentz tensor in $\Fock^b_{\sf NS}$ from
a state $\ket{\varepsilon}$ that verifies
\begin{equation}\label{eq:lw_o(2M)}
    \varepsilon\simeq\
    \begin{tikzpicture}[scale=0.6, baseline=(ref.base)]
    \coordinate (top) at (0,0);
    \coordinate (ref) at (0, -2);
    \edef\h{\dimexpr5cm\relax}
    \edef\l{\dimexpr0.5cm\relax}
    \edef\g{\dimexpr3cm\relax}
    
        \draw[thick] (top) rectangle ++(\l,-\h);
        \draw[thick] ($(top)+(\l,0)$)
                        rectangle ++(\l,-0.8*\h);
        \draw[thick] ($(top)+(2*\l,0)$)
                        rectangle ++(\l,-0.65*\h);
        \draw[thick] ($(top)+(\g+2*\l,-0.35*\h)$)
        -- ++(0,0.35*\h) -- ++(-\g+\l,0) -- ++(0,-0.6*\h) -- ++(\l,0) -- ++(0,0.5*\l); 
        \draw[thick, dashed] ($(top)+(4*\l,-0.55*\h)$) -- ++(\g-2*\l,0.2*\h);
        \draw[thick] ($(top)+(\g+2*\l,0)$)
                        rectangle ++(\l,-0.3*\h);
        \node at ($(top)+(0.5*\l,-0.5*\h)$) 
                                {\scriptsize$h_1$};
        \node at ($(top)+(1.5*\l,-0.4*\h)$) 
                                {\scriptsize$h_2$};
        \node at ($(top)+(2.5*\l,-0.3*\h)$) 
                                {\scriptsize$h_3$};
        \node at ($(top)+(\g+5*\l,-0.2*\h)$) 
                                {\footnotesize$h_M$};
        \draw ($(top)+(\g+2.5*\l,-0.15*\h)$)
        edge[in=100,out=75,->] ($(top)+(\g+5*\l,-0.15*\h)$);
    \end{tikzpicture}
    \qquad\Longleftrightarrow\qquad\left\{
    \begin{aligned}
        M^r{}_r \ket{\varepsilon}
        & = (h_r-\tfrac{d}2) \ket{\varepsilon}\,,\
        & r=1,\dots,\oRank\,,\\
        M^r{}_s \ket{\varepsilon} & = 0\,,\
        & 1 \leq r < s \leq \oRank\,,\\
        M_{rs} \ket{\varepsilon} & = 0\,,\
        & r, s = 1, \dots, \oRank\,,
    \end{aligned}
    \right.
\end{equation}
which is to say that $\ket{\varepsilon}$ is a lowest weight
state for $\mathfrak{o}(2\oRank)$. This is again in accordance
with the previous claim that the Lorentz and the orthogonal
algebras form a dual pair, in $O(d \times 2\oRank)$ here,
and hence their irreps appear in mutually distinct
couples. Here we have seen that the Lorentz irrep corresponding
to a Young diagram with columns of height $[h_1,\dots,h_M]$
is dual to the lowest weight irrep of $\mathfrak{o}(2\oRank)$
with lowest weight $[h_1-\tfrac{d}2,\dots,h_M-\tfrac{d}2]$.

\paragraph{\Ram sector.}
Now let us turn our attention to the subspace
of the Ramond sector generated by fermionic
oscillators \eqref{eq:ferm_osc} of modes
$0,-1, \dots, -\oRank$. The zero modes
form a Clifford algebra and it can be shown
that they can be defined as the $\gamma$--matrices
in $d$--dimensions
\begin{equation}
    b_0^\mu  := \tfrac{1}{\sqrt2}\,\gamma^\mu \,,
    \qquad 
    \{\gamma^\mu, \gamma^\nu\} = 2\,\eta^{\mu\nu}\,,
\end{equation}
which implicitly carry Dirac indices $A,B,\dots$ and which we suppress. Let us also keep in mind that they anticommute
with the other fermionic oscillators,
$\{\gamma^\mu, b_{\pm r}^\nu\} = 0\,,$
for $r=1,2,\dots,\oRank$. Correspondingly,
the generators of the Lorentz algebra
take the form
\begin{equation}
    J^{\mu\nu} = \tfrac12\,[\gamma^\mu, \gamma^\nu]
    + 2\,\sum_{r\geq1} b_{-r}^{[\mu}\,b_{+r}^{\nu]}\,,
\end{equation}
on this subspace. These $\gamma$--matrices
can be split into creation and annihilation operators
as well, as explained in, e.g.,
\cite[Chap. 8.1]{Blumenhagen:2013fgp}.
Assuming that the spacetime dimension $d=2n$
is even, we can define the creation and annihilation
operators
\begin{equation}
    \mathrm{b}_\pm^I := \tfrac1{2\sqrt2}\,
    (\sigma_I\gamma^I \pm i\,\gamma^{2n-1-I})\,,
    \qquad 
    I=0,\dots,n-1\,,
    \qquad 
    \sigma_I := \left\{
    \begin{aligned}
        i & \quad \text{if}\quad I=0\,,\\
        1 & \quad \text{otherwise}\,,
    \end{aligned}
    \right.
\end{equation}
where \emph{no summation} is implied 
in the term $\sigma_I\,\gamma^I$, and which verify
\begin{equation}
    \{\mathrm{b}^I_+, \mathrm{b}^J_-\} = \delta^{IJ}\,,
    \qquad 
    \{\mathrm{b}^I_+, \mathrm{b}^J_+\} = 0
    = \{\mathrm{b}^I_-, \mathrm{b}^J_-\}\,.
\end{equation}
Requiring that the \Ram vacuum 
be annihilated by half of the zero--modes,
say $\mathrm{b}_-^I\ket0 = 0$, we obtain a subspace
generated by the action of the other half
of the zero--modes, $\mathrm{b}_+^I$ here, 
subspace of states which are all annihilated
by the positive modes $b_r^\mu$ with $r=1,2,\dots,\oRank$, 
and that is left invariant by the action 
of the Lorentz algebra, thereby defining
a finite--dimensional representation of it.
Note that this representation is reducible,
due to the fact that the Lorentz
generators $\gamma_{[\mu}\,\gamma_{\nu]}$
are quadratic in $\mathrm{b}_\pm^I$ so that
they will leave invariant the subspaces
with an even or odd number of creation operators.
These two invariant subspaces form the spin--$(\nicefrac12)$
positive and negative chirality representations
of $\so(1,2n-1)$.

The Fock space generated by the creation operators
$b_{-r}^\mu$ and $\mathrm{b}_+^I$ is then a linear combination
of states of the form
\begin{equation}
    \ket{\varepsilon}
    := \varepsilon^A_{\mu_1[h_1]|\dots|\mu_l[h_l]}\,b^{\mu_1[h_1]}_{-r_1} \dots b^{\mu_l[h_l]}_{-r_l} 
    \ket{A;0}_\Ram\,,
    \qquad 
    1 \leq r_i \leq \oRank\,,
    \qquad r_i \neq r_j\,,
\end{equation}
where $\varepsilon$ is a tensor--spinor with $l$ groups
of antisymmetric indices (again, $0 \leq l \leq \oRank$),
a priori without particular symmetries between indices
of different groups and and a spinor index.
The Lorentz algebra commutes with bilinears of the form\footnote{The bilinears $M^r$ of course also carry Dirac indices $A,B,\dots$ due to the gamma matrices they contain.}
\begin{equation}
    M^r := \tfrac1{\sqrt2}\,\gamma \cdot b_{-r}\,,
    \qquad 
    M_r := \tfrac1{\sqrt2}\,\gamma \cdot b_r\,,
    \qquad
    r = 1,2,\dots,\oRank\,,
\end{equation}
as well as with the previously defined $M^{rs}$,
$M^r{}_s$ and $M_{rs}\,$; note that the Lorentz
and $\mathfrak{o}(2\oRank)$ generators
are still given by the expressions \eqref{LorentzR}
and \eqref{eq:o} respectively,
where now the fermionic oscillators have integer modes.
Altogether, they define a representation
of $\mathfrak{o}(2\oRank+1)$,
whose commutation relations read
\begin{subequations}
    \begin{align}
        [M^r{}_s, M^t] & = \delta_s^t\,M^r\,,
        &&& [M^r{}_s, M_t] & = -\delta_t^r\,M_s\,,
        &&& [M^r, M_s] & = -\,M^r{}_s\,,\\
        [M^{rs}, M_t] & = 2\,\delta_t^{[s}\,M^{r]}\,,
        &&& [M_{rs}, M^t] & = 2\,\delta^t_{[s}\,M^{\,}_{r]}\,,
    \end{align}
\end{subequations}
on top of those recalled in \eqref{eq:o_commutation}.
These new generators also encode operations
on the groups of antisymmetric Lorentz indices,
and on the spinor index now carried by the ``vacuum space''.
To be explicit:
\begin{itemize}
\item The generators $M_r$ act on $\varepsilon$
by contracting the spacetime index of $\gamma_\mu$
with an index in the $r$th group. In particular,
$\gamma$--tracelessness of $\varepsilon$ amounts to
\begin{equation}\label{eq:gamma_trace}
    M_r \ket{\varepsilon} = 0
    \qquad\Longleftrightarrow\qquad 
    \gamma^{\nu\,A}{}_B\,
    \varepsilon^B{}_{\cdots,\nu\mu_r[h_r-1],\cdots} = 0\,,
\end{equation}
for any $1 \leq r \leq \oRank$.
\item The generators $M^r$ act by multiplying
the tensor--spinor $\varepsilon$ by $\gamma_\mu$
and antisymmetrizing its index with those of $r$th group
of $\varepsilon$.
\end{itemize}
So now we have a way to characterize irreducible
Lorentz tensor--spinors, namely by imposing
the conditions \eqref{eq:lw_o(2M)} together with
the $\gamma$--tracelessness \eqref{eq:gamma_trace}.

In summary, the lowest weight conditions for
$\mathfrak{o}(2\oRank)$ or $\mathfrak{o}(2\oRank+1)$
can be thought of as a way to glue columns generated
by fermionic creation operators of a given species
(meaning with a fixed value of the $\mathfrak{o}(2\oRank)$
or $\mathfrak{o}(2\oRank+1)$ index),
and well as imposing ($\gamma$--)tracelessness. A couple of
differences with respect to the bosonic case
are worth repeating. First, both bosonic and fermionic irreps
of the Lorentz algebra appear, and the distinction
between the two is tied to the nature of the dual algebra,
$\mathfrak{o}(2\oRank)$ or $\mathfrak{o}(2\oRank+1)$,
and therefore to the dimension of the fermionic
Fock space under consideration (upon restricting to certain levels or trajectories). Indeed, we have seen
that to go from the former to the latter, 
one needs to add one family of gamma matrices,
which reflects the fact that $\so(d-1,1)$
form a dual pair with $\mathfrak{o}(2\oRank)$
in $\mathfrak{o}\big(d \times 2\oRank\big)$,
and with $\mathfrak{o}(2\oRank+1)$
in $\mathfrak{o}\big(d \times (2\oRank+1)\big)$.
The relevant fermionic Fock space then requires
$d \times \oRank$ pairs of oscillators
in the first case, and $d \times \oRank + \tfrac{d}2$
in the second case. Second, the use of fermionic oscillators
naturally leads us to describing tensors(--spinors)
in the antisymmetric basis. As a consequence, 
the types of Young diagrams that can appear
is restricted in both the horizontal
and the vertical: each column cannot be higher
than $d$ as in the bosonic case, but the length
of each row is also bounded by $\oRank$
--- the maximal number of columns that we can create.

\subsection{Super--Howe duality}
\label{sec:super-Howe}
Finally, let us consider the ``super--version''
of Howe duality,\footnote{We will refrain
from referring to this duality as ``supersymmetric'',
since it does not relate bosonic and fermionic
representations of some spacetime isometry
algebra, but instead relates 
irreps of the Lorentz algebra to irreps
of a dual superalgebra (at least
in the cases of interest for us in this work).}
by which we mean the duality between representations
of dual pairs in an orthosymplectic algebra
(for more details, see e.g. 
\cite{Cheng2000,Cheng:2009sz,Cheng2010,Cheng2012}
and references therein).\footnote{See also, e.g.
\cite{Bars:1982ep, Nishiyama1991, Nishiyama1994}
for early use of it to build lowest weight modules
of the orthosymplectic algebra.}
More specifically,
we will be interested in the pair formed by
the Lorentz algebra $\so(d-1,1)$
and the orthosymplectic algebras
$\osp(2\oRank|2\spRank,\R)$
or $\osp(2\oRank+1|2\spRank,\R)$.

\paragraph{\NS sector.}
Let us start with the former pair, which acts
on the Fock space generated by \emph{both}
the bosonic and fermionic oscillators in the \NS sector.
As before, we will focus on the subspace 
generated by the first $\spRank$ bosonic modes
and $\oRank$ fermionic modes, so that a generic state
in this subspace will be a linear combination
of elements of the form
\begin{equation}\label{eq:state_osp}
    \ket\varepsilon = \varepsilon_{\mu_1(\ell_1)|\cdots|\mu_p(\ell_p)\pmb|\nu_1[h_1]|\cdots|\nu_l[h_l]}\,
    \alpha_{-i_1}^{\mu_1(\ell_1)} \dots 
    \alpha_{-i_p}^{\mu_p(\ell_p)}\, b_{-r_1+\frac12}^{\nu_1[h_1]} \dots 
    b_{-r_l+\frac12}^{\nu_l[h_l]} \ket{0}_{\mathsf{NS}}\,,
\end{equation}
where $\varepsilon$ is a tensor with $p$ groups
of symmetric indices \emph{and} $l$ groups
of antisymmetric indices (with $0 \leq p \leq \spRank$
and $0 \leq l \leq \oRank$), without any symmetry
properties between indices of difference groups
a priori. The algebras $\sp(2\spRank,\R)$
and $\mathfrak{o}(2\oRank)$ act on such states,
via the bilinears in bosonic and fermionic
oscillators spelled out in \eqref{eq:sp} and \eqref{eq:o}.
The important difference is that now
the direct sum of these two algebras
form the bosonic subalgebra of the superalgebra
$\osp(2\oRank|2\spRank,\R)$, whose fermionic generators
are represented by
\begin{equation}\label{eq:Q_osp}
    Q^{nr} := \tfrac1n\,\alpha_{-n} \cdot b_{-r+\frac12}\,,
    \qquad 
    Q^n{}_r := \tfrac1n\,\alpha_{-n} \cdot b_{r-\frac12}\,,
    \qquad 
    Q_n{}^r := b_{-r+\frac12} \cdot \alpha_n\,,
    \qquad 
    Q_{nr} := \alpha_n \cdot b_{r-\frac12}\,,
\end{equation}
i.e. all possible bilinears made out of one bosonic
and one fermionic oscillator, and whose Lorentz indices
are contracted. The anticommutator of the fermionic
generators gives back the bosonic ones,
\vspace{-15pt}
\begin{subequations}
    \begin{align}
        \{Q^m{}_r, Q^{ns}\} & = \delta^s_r\,T^{mn}\,,
        & \{Q_m{}^r, Q^{ns}\} & = \delta^n_m\,M^{rs}\,,\\
        \{Q^m{}_r, Q_{ns}\} & = -\delta^m_n\,M_{rs}\,,
        & \{Q_m{}^r, Q_{ns}\} & = \delta^r_s\,T_{mn}\,,\\
        \{Q^m{}_r, Q_n{}^s\} & = \delta_r^s\,T^m{}_n
        + \delta^m_n\,M^s{}_r\,,
        & \{Q^{mr}, Q_{ns}\} & = \delta^r_s\,T^m{}_n
        - \delta^m_n\,M^r{}_s\,,
    \end{align}
\end{subequations}
and fermionic generators transform under the action
of $\sp(2\spRank,\R)$ ones as
\begin{subequations}
    \begin{align}
        [T^k{}_l, Q^m{}_r] & = \delta^m_l\,Q^k{}_r\,,
        & [T^{kl}, Q_m{}^r] & = -2\,\delta^{(k}\,Q^{l)r}\,,\\
        [T^k{}_l, Q_m{}^r] & = -\delta^k_m\,Q_l{}^r\,,
        & [T^{kl}, Q_{m r}] & = -2\,\delta^{(k}_m\,Q^{l)}{}_r\,, \\
        [T^k{}_l, Q^{m r}] & = \delta^m_l\,Q^{k r}\,,
        & [T_{kl}, Q^m{}_r] & = 2\,\delta_{(k}^m\,Q_{l)r}\,, \\
        [T^k{}_l, Q_{m r}] & = -\delta^k_m\,Q_{l r}\,,
        & [T_{kl}, Q^{m r}] & = 2\,\delta_{(k}^m\,Q_{l)}{}^r\,,
    \end{align}
\end{subequations}
and under $\mathfrak{o}(2\oRank)$ as
\begin{subequations}
    \begin{align}
        [M^r{}_s, Q^m{}_t] & = -\delta^r_t\,Q^m{}_s\,,
        & [M^{rs}, Q^m{}_t] & = -2\,\delta^{[r}_t\,Q^{ms]}\,,\\
        [M^r{}_s, Q_m{}^t] & = \delta_s^t\,Q_m{}^r\,,
        & [M^{rs}, Q_{m t}] & = -2\,\delta^{[r}_t\,Q_m{}^{s]}\,,\\
        [M^r{}_s, Q^{mt}] & = \delta_s^t\,Q^{mr}\,,
        & [M_{rs}, Q_m{}^t] & = -2\,\delta^t_{[r}\,Q_{m s]}\,,\\
        [M^r{}_s, Q_{mt}] & = -2\,\delta^r_t\,Q_{ms}\,,
        & [M_{rs}, Q^{m t}] & = -2\,\delta^t_{[r}\,Q^m{}_{s]}\,.
    \end{align}
\end{subequations}
These commutation relations simply reflect the fact
that the fermionic generators are in the tensor product
of the fundamental irrep of $\sp(2\spRank,\R)$
and the fundamental irrep of $\mathfrak{o}(2\oRank)$.

We have discussed previously
how the generators of $\sp(2\spRank,\R)$ encode 
symmetrization and trace operations on rows
of Young diagrams associated with different
types of bosonic creation operators,
and how the generators of $\mathfrak{o}(2\oRank)$
encode antisymmetrization and trace operations
on columns of Young diagrams associated with 
different types of fermionic operators.
In particular, lowering operators of $\sp(2\spRank,\R)$
allow us to express how to ``glue'' rows together,
while lowering operators of $\mathfrak{o}(2\oRank)$
allow us to express how to ``glue'' columns together.
Now that we have both bosonic and fermionic 
oscillators together, it is only natural that
these new generators encode operations involving
both rows and columns of Lorentz Young diagrams:
\begin{itemize}
\item The generators $Q^m{}_r$ symmetrize an index
of the $r$th group of antisymmetric indices
with all those of the $m$th group of symmetric indices.

\item The generators $Q_m{}^r$ antisymmetrize an index
of the $m$th group of symmetric indices with all those
of the $r$th group of antisymmetric indices.

\item The generators $Q_{mr}$ takes a trace between
an index of the $m$th group of symmetric indices
and one of the $r$th group of antisymmetric indices.

\item Finally, the generators $Q^{mr}$ multiply
the tensor $\varepsilon$ with a metric $\eta$,
and symmetrize one of its indices with those
of the $m$th group of symmetric indices,
and anitsymmetrize the other index
with those of the $r$th group of antisymmetric indices.
\end{itemize}
The condition that $\varepsilon$ be traceless
is therefore equivalent to 
\begin{equation}
    T_{mn} \ket\varepsilon = 0
    = M_{rs} \ket\varepsilon
    \qquad\text{and}\qquad
    Q_{mr} \ket\varepsilon = 0\,,
\end{equation}
since $T_{mn}$ take traces between two groups
of symmetric indices, $M_{rs}$ between two groups
of antisymmetric indices, and $Q_{mr}$ between
a group of symmetric indices and a group
of antisymmetric ones. We also know from
the two previous sections that imposing
\begin{equation}
    T^m{}_n \ket\varepsilon = 0
    \quad\text{for}\quad m<n
    \qquad\text{and}\qquad
    M^r{}_s \ket\varepsilon = 0
    \quad\text{for}\quad r<s
\end{equation}
will respectively glue together rows (corresponding
to the groups of symmetric indices of $\varepsilon$),
and columns (corresponding to the groups of antisymmetric
indices of $\varepsilon$), in decreasing length order.
It remains to understand how the block of rows
and the block of columns are glued together.

To do so, let us have a closer look at the action
of $Q^m{}_r$ and $Q_m{}^r$, starting with the simplest 
case $\oRank=1=\spRank$. This means that the tensor 
$\varepsilon$ has one group of antisymmetric indices,
and one of symmetric indices. In terms of Young
diagram, it corresponds to the tensor product
of a column with a row. As a representation
of $\mathfrak{gl}_d$, this tensor product
will generally contain two irreducible pieces,
one obtained by gluing the column on the left
of the row, and another one by gluing it below.
Requiring that the generators $Q^1{}_1$ or $Q_1{}^1$
annihilates the state $\ket\varepsilon$, selects
one of these two $\mathfrak{gl}_d$ irreducible tensors,
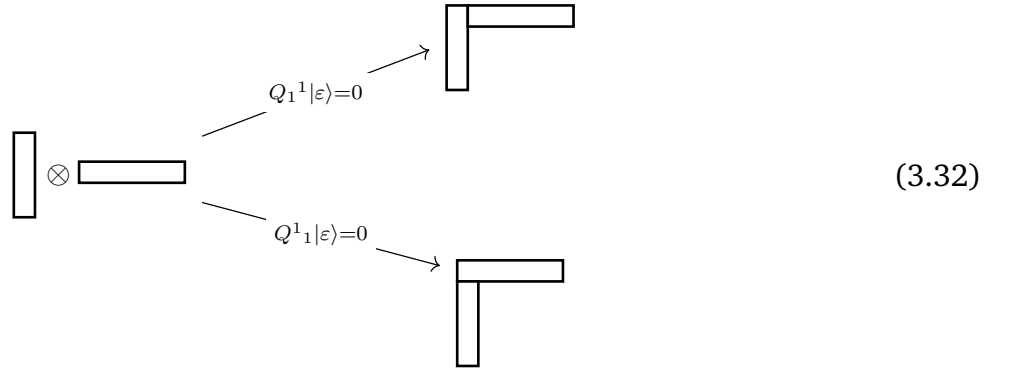
\begin{equation}
    \begin{tikzcd}[row sep=tiny, column sep=large]
        && \gyoung(|4_5)\\
        \parbox{10pt}{\gyoung(|4)} \otimes \gyoung(_5) 
        \ar[rru, "Q_1{}^1\ket\varepsilon=0" description]
        \ar[rrd, "Q^1{}_1\ket\varepsilon=0" description] && \\
        && \gyoung(_5,|4)
    \end{tikzcd}
\end{equation}
so that, together with the tracelessness conditions
$T_{11}\ket\varepsilon=M_{11}\ket\varepsilon
=Q_{11}\ket\varepsilon=0$, either one of the conditions
$Q_1{}^1\ket\varepsilon=0$ or $Q^1{}_1\ket\varepsilon=0$
define an irreducible Lorentz tensor.

Now let us try to add, for example, one row, i.e. we consider the case
$\oRank=1$ and $\spRank=2$. We therefore have four odd
generators which implement anti/symmetrization
between indices of a column and a row,
namely $Q_1{}^1$, $Q_2{}^1$, $Q^1{}_1$ and $Q^2{}_1$
(and four odd generators adding/removing traces).
Choosing a pair of generators among those four
allows us to select one of three possible ways
we can glue these rows and column, 
\begin{equation}
    \begin{tikzcd}[row sep=tiny, column sep=large]
        && \gyoung(|5_7,/1_5)\\
        \parbox{10pt}{\gyoung(|5)} \otimes\left(
        \begin{aligned}
            \gyoung(_7)\\[-5pt]
            \otimes\qquad\,\ \\[-3pt]
            \gyoung(_5)\quad\
        \end{aligned}
        \right)\,
        \ar[rru, "Q_1{}^1\ket\varepsilon=0
            =Q_2{}^1\ket\varepsilon" description]
            \ar[rr, "Q^1{}_1\ket\varepsilon=0
                =Q_2{}^1\ket\varepsilon"]
        \ar[rrd, "Q^1{}_1\ket\varepsilon=0
            =Q^2{}_1\ket\varepsilon" description,
            end anchor=west]
        && \gyoung(_7,|5_5) \\
        && \gyoung(_7,_5,|5)
    \end{tikzcd}
\end{equation}
assuming that we also impose the rows to be glued
one below the other\footnote{Note that the pair
of generators $Q_1{}^1$ and $Q^2{}_1$ does not appear
here, as it would glue the second row on top 
of the first one, thereby violating the convention
for the construction of a Young diagram, 
whose rows should be of decreasing lengths.}
--- i.e. that we also imposed $T^1{}_2\ket\varepsilon=0$.
Imposing that $\ket\varepsilon$ is also annihilated by
the trace--type operators $T_{11}$, $T_{12}$, $T_{22}$,
$M_{11}$, $Q_{11}$ and $Q_{21}$ makes $\varepsilon$
into a Lorentz irreducible tensor as before.

These different options correspond to different
choices of positive/negative roots for
the orthosymplectic algebra $\osp(2\oRank|2\spRank,\R)$.
Indeed, the odd generators $Q$
are all ladder operators with respect to the Cartan
subalgebra of the bosonic subalgebra, and hence
should be divided into two groups, of raising
and of lowering operators. Having in mind
that $Q_{mr}$ take traces, we would prefer to think
of them as lowering operators, like $T_{mn}$
and $M_{rs}$. It therefore remains to split
the set of generators $Q_m{}^r$ and $Q^m{}_r$
into raising and lowering operators. The pairs
singled out in the previous example correspond
to different such splittings. For arbitrary ranks
$\oRank$ and $\spRank$ of the orthogonal and symplectic
subalgebras respectively, this freedom in the choice
of positive and negative roots associated with 
$Q_m{}^r$ and $Q^m{}_r$ will lead to different ways
of gluing columns and rows and, crucially, no choice can be favoured without additional criteria.

For the superstring, we know that additional criteria on the structure of string states do exist: they are the super--Virasoro constraints, enforcing that general states be physical. As we will see in the next section, their form in the \NS sector is consistent with the choice of
\begin{equation} \label{choice_low}
    Q_m{}^r \quad\text{with}\quad m \geq r
    \qquad\text{and}\qquad
    Q^m{}_r \quad\text{with}\quad m < r\,,
\end{equation}
as being part of the lowering operators. For the moment,
let us point out that the lowest weight conditions
associated with this choice corresponds to constructing
a Young diagram out of columns and rows
as in the following example,
\begin{equation}
    \gyoung(|7_9,/1|5_6,/1/1|3_3)
\end{equation}
namely putting the $i$th row on the top right
of the $i$th column so as to form a ``hook'',
and then inserting this $i$th hook in the wedge 
of the $(i-1)$th hook, thereby creating a diagram with a ``\textit{nested--hook--like}'' structure.

To summarize, we found that a tensor $\varepsilon$
is irreducible for the Lorentz algebra if and only if
the state $\ket\varepsilon$ built out of it
as in \eqref{eq:state_osp} is a lowest weight
for $\osp(2\oRank|2\spRank,\R)$ with respect
to the previously identified root system, i.e.
\begin{equation}\label{eq:lw_osp}
    \varepsilon\ \simeq\ 
    \begin{tikzpicture}[scale=0.6, baseline=(ref.base)]
    \coordinate (ref) at (0,0);
    \coordinate (1stcolumn) at (-3,3);
    \coordinate (1strow) at (-2.5,2.5);
    \edef\s{\dimexpr7.5cm\relax}
    \edef\h{\dimexpr6cm\relax}
    \edef\g{\dimexpr2cm\relax}
    \edef\l{\dimexpr0.5cm\relax}
    
    \draw[thick] (1stcolumn) rectangle ++(\l,-\h);
    \draw[thick] (1strow) rectangle ++(\s,\l);
    \draw[thick] ($(1stcolumn)+(\l,-\l)$)
                    rectangle ++(\l,-0.7*\h);
    \draw[thick] ($(1strow)+(\l,-\l)$)
                    rectangle ++(0.65*\s,\l);
    \draw[thick] ($(1stcolumn)+2*(\l,-\l)$)
                    rectangle ++(\l,-0.5*\h);
    \draw[thick] ($(1strow)+2*(\l,-\l)$)
                    rectangle ++(0.4*\s,\l);
    \node at ($(1strow)+(0.5*\s,0.5*\l)$) {\scriptsize$\ell_1$};
    \node at ($(1strow)+(0.4*\s,-0.5*\l)$) {\scriptsize$\ell_2$};
    \node at ($(1strow)+(0.3*\s,-1.5*\l)$) {\scriptsize$\ell_3$};
    \node at ($(1stcolumn)+(0.5*\l,-0.6*\h)$) {\scriptsize$h_1$};
    \node at ($(1stcolumn)+(1.5*\l,-0.5*\h)$) {\scriptsize$h_2$};
    \node at ($(1stcolumn)+(2.5*\l,-0.4*\h)$) {\scriptsize$h_3$};
    \node at (ref) {$\ddots$};
    \end{tikzpicture}
    \quad\Longleftrightarrow\quad
    \left\{
    \begin{aligned}
        T^m{}_m \ket\varepsilon
        & = (\ell_m+\tfrac{d}2) \ket\varepsilon\,, &\\
        M^r{}_r \ket\varepsilon
        & = (h_r-\tfrac{d}2) \ket\varepsilon\,, &\\
        Q^m{}_r \ket\varepsilon & = 0\,, & m < r\,,\\
        Q_m{}^r \ket\varepsilon & = 0\,, & m \geq r\,,\\
        Q_{mr} \ket\varepsilon & = 0\,, &\\
        T_{mn} \ket\varepsilon & = 0 = M_{rs} \ket\varepsilon\,, &\\
        T^m{}_n \ket\varepsilon & = 0\,, & m < n\,,\\
        M^r{}_s \ket\varepsilon & = 0\,, & r < s\,,
    \end{aligned}
    \right.
\end{equation}
and with $m,n=1,\dots,\spRank$ and $r,s=1,\dots,\oRank$.
The pattern is similar
to what we have seen previously for the bosonic
and fermionic versions of Howe duality: conditions
fixing the lowest weight of $\osp(2\oRank|2\spRank,\R)$
determine the total number of boxes making up
the Young diagram labelling the dual $\so(d-1,1)$ irrep,
while the action of lowering operators impose 
that the tensor be traceless and have the symmetry
of the Young diagram displayed above. The main
difference is that now the boxes of the final Young
diagram are arranged both into column and rows,
so that on top of the symmetry and tracelessness
conditions on rows or columns only, which are encoded by
lowering operators of $\sp(2\spRank,\R)$ or
$\mathfrak{o}(2\oRank)$ respectively, one should
add similar conditions for both rows and columns
using the \emph{odd} lowering operators
of $\osp(2\oRank|2\spRank,\R)$. In particular,
some of these conditions tell us \emph{how to glue
rows with columns together}.

\paragraph{The Frobenius notation.}

Note that the relation between the Young diagram
labelling the Lorentz irrep and the lowest weight
of the dual $\osp(2\oRank|2\spRank,\R)$ irrep
is a bit more difficult to express than before,
since neither the symmetric nor the antisymmetric
basis for Young diagram is adapted.
To solve this problem, it is useful to turn to
what is called the \emph{Frobenius notation}
for Young diagrams. The idea is the following:
say that we are given a Young diagram denoted by
$(s_1, \dots, s_n)$ in the symmetric basis
(i.e. $s_i$ denotes the length of the $i$th row), 
and $[h_1, \dots, h_m]$ in the antisymmetric basis
(i.e. $h_i$ is the height of the $i$th column),
and whose diagonal contains $\Diag$ boxes. In the Frobenius
notation, this diagram will be denoted
\begin{equation} \label{Frobdef}
    (\ell_1, \dots, \ell_\Diag
    \mid \bar\ell_1, \dots, \bar\ell_\Diag)\,,
\end{equation}
where 
\begin{equation}
    \ell_i = s_i-i\,,
    \qquad\text{and}\qquad
    \bar\ell_i = h_i-i\,,
    \qquad i=1,\dots,\Diag\,.
\end{equation}
In plain words, the first $k$ entries correspond to
the length of the $i$th row \emph{on the right
of the diagonal}, and the next $\Diag$ entries to the height
of the $i$th column \emph{below the diagonal}.
For instance, the diagram
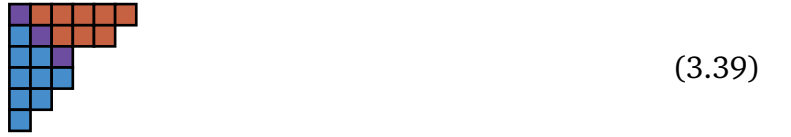
\begin{equation}\label{eq:ex_frobenius}
    \newcommand\blc{\Yfillcolour{RoyalPurple!90}}
    \newcommand\rd{\Yfillcolour{BrickRed!75}}
    \newcommand\bl{\Yfillcolour{RoyalBlue!70}}
    \gyoung(!\blc;!\rd;;;;;,!\bl;!\blc;!\rd;;;,!\bl;;!\blc;,!\bl;;;,;;,;)
\end{equation}
is denoted 
\begin{equation}
    (6,5,3,2,2,1)
    \qquad\text{and}\qquad
    [6,5,4,2,2,1]\,,
\end{equation}
in the symmetric and antisymmetric basis respectively,
and 
\begin{equation}
    (5,3,0|5,3,1)\,,
\end{equation}
in the Frobenius notation. This notation is 
convenient to describe the $\osp(2\oRank|2\spRank,\R)$
lowest weight states identified above,
since $\ell_i$ entries will correspond to the number
of $\alpha_{-i}^\mu$, and $\bar\ell_i+1$ to the number
of $b_{-i}^\mu$. This last shift by one unit is a consequence
of the fact that in the principal embedding, the diagonal
consists of fermionic oscillators $b$, and hence
are not taken into account in the ``Frobenius labels''.
We will therefore use a slightly modified version
of this notation, namely we will denote a diagram
whose diagonal is composed of $\Diag$ boxes 
by $(\ell_1,\dots,\ell_\Diag \mid h_1, \dots, h_\Diag)$,
where $\ell_i$ is the length of the $i$th row
on the right of the diagonal, and $h_i$ is the height
of the $i$th column \emph{starting from the diagonal},
i.e. including the $i$th box of the diagonal.
In this notation, the labels used to characterise 
a diagram are simply related to the components 
of the $\osp(2\oRank|2\spRank,\R)$ lowest weight
by a shift of $\pm\tfrac{d}2$, as in the bosonic
or fermionic version of Howe duality discussed previously.
For instance, the previous diagram \eqref{eq:ex_frobenius}
would be denoted $(5,3,0|6,4,2)$, or to put it visually,
this modified notation would correspond to the diagram,
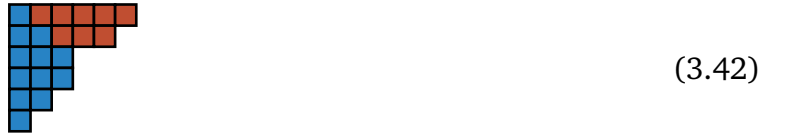
\begin{equation}\label{eq:ex_frobenius_mod}
    \newcommand\blc{\Yfillcolour{Black}}
    \newcommand\rd{\Yfillcolour{BrickRed!85}}
    \newcommand\bl{\Yfillcolour{RoyalBlue!80}}
    \gyoung(!\bl;!\rd;;;;;,!\bl;!\bl;!\rd;;;,!\bl;;!\bl;,!\bl;;;,;;,;)
\end{equation}
where the first three entries are the length of the rows
in red, and the next three entries the height of the columns
in blue.

\paragraph{\Ram sector.}
Let us now turn to the subspace
of the Ramond sector generated by the first $\spRank$
modes of the bosonic oscillators, and the first $\oRank$
modes of the fermionic ones, acting on the \Ram vacum
$\ket{A;0}_{\mathsf{R}}$. On top of the generators
of $\mathfrak{o}(2\oRank+1) \oplus \sp(2\spRank,\R)$,
as well as the odd generators \eqref{eq:Q_osp}
introduced previously (upon making the implicit
replacement $b_{\pm(r-\frac12)} \to b_{\pm r}$),
the additional odd generators\footnote{The generators $Q^m$ of course also carry Dirac indices $A,B,\dots$ due to the gamma matrices they contain.}
\begin{equation}
    Q^m := \tfrac1{\sqrt2\,m}\,\gamma \cdot \alpha_{-m}\,,
    \qquad 
    Q_m := \tfrac1{\sqrt2}\,\gamma \cdot \alpha_m\,,
\end{equation}
also act on this subspace, and commute with the Lorentz
algebra generators, and altogether these generators
make up a representation of $\osp(2\oRank+1|2\spRank,\R)$.
The last new addition corresponds to creating
or taking $\gamma$--traces \emph{in rows}
of a tensor-spinor. Naturally, the generators $Q_m$
which take $\gamma$--traces should be added
to the lowering operators previously identified.

For the sake of completeness, let us also recall
the anticommutation relations obeyed by
these new generators,
\begin{subequations}
\begin{align}
    \{Q^m,Q^n\} & = T^{mn}\,,
    & \{Q^m, Q_n\} & = T^m{}_n\,,
    & \{Q_m, Q_n\} & = T_{mn}\,,\\
   \{Q^m{}_r,Q_n\} & = \delta_n^m \,M_r\,,
   & \{ Q_m{}^r,Q^n\} & =-\,\delta_m^n\, M^r\,,\\
   \{Q^{mr}, Q_n\} & = \delta^m_n\,M^r\,,
   & \{Q_{mr}, Q^n\} & = -\delta^n_m\,M_r\,,
\end{align}
\end{subequations}
as well as the action of $\sp(2\spRank,\R)$,
\begin{subequations}
    \label{eq:sp-Q}
    \begin{align}
        [T^k{}_l, Q^m] & = \delta^m_l\,Q^k\,,
        & [T^k{}_l, Q_m] & = -\delta^k_m\,Q_l\,,\\
        [T_{kl}, Q^m] & = 2\,\delta^m_{(k}\,Q_{l)}\,,
        & [T^{kl}, Q_m] & = -2\,\delta^{(k}_m\,Q^{l)}\,,
    \end{align}
\end{subequations}
and of $\mathfrak{o}(2\oRank+1)$
\begin{subequations}
    \begin{align}
        [M^r, Q^m] & = -Q^{m r}\,,
        & [M^r, Q_m] & = -Q_m{}^r\,,\\
        [M_r, Q^m] & = -Q^m{}_r\,,
        & [M_r, Q_m] & = -Q_{mr}\,,
    \end{align}
\end{subequations}
on them. On top of that, the generators $M^r$ and $M_r$
of $\mathfrak{o}(2\oRank+1)$ act on the odd generators
$Q^m{}_r$, $Q_m{}^r$, $Q^{mr}$ and $Q_{mr}$ as
\begin{subequations}
\begin{align}
    [M^r,Q^m{}_s] & = \delta^r_s\, Q^m\,,
    & [M_r, Q_m{}^s] & = \delta_r^s\, Q_m\,,\\
    [M^r, Q_{ms}] & = \delta^r_s\,Q_m\,,
    & [M_r, Q^{ms}] & = \delta^s_r\,Q^m\,.
\end{align}
\end{subequations}
Again, the set of relations \eqref{eq:sp-Q}
merely expresses that the additional odd generators 
$Q^m$ and $Q_m$ are in the fundamental representation
of $\sp(2\spRank,\R)$.

Let us stress that, contrarily to the situation
in the \NS sector, the choice of lowering operators
given by \eqref{eq:lw_osp} together with $Q_m$
and $M_r$ is \emph{not} singled out by the super--Virasoro
constraints in the \Ram sector, as we will discuss 
in details in Section \ref{sec:principal}.

\section{Superstring spectrum sections}
\label{sec:spectrum_sections}

\subsection{Physicality conditions}

\paragraph{\NS sector.}
We begin with the $\mathsf{NS}$ sector. Let us introduce
the shorthand notation
\begin{subequations}
    \begin{align}\label{shorthand}
        X^{(n)}_\mu(z) & := \tfrac{i}{(n-1)!}\,\pl^n X_\mu(z)\,,
        & \psi^{(r)}_\mu(z) & := \tfrac1{(r-1)!}\,
        \pl^{r-1} \psi_\mu(z)\,, & \qquad n, r \geq 1\,,\\
        \delta_X^{(n)}(z) & := -i\,n!\,
        \frac{\delta}{\delta \partial^n X(z)}\,,
        & \delta_\psi^{(r)}(z) & := (r-1)!\,
        \frac{\delta}{\delta \partial^{r-1}\psi(z)}\,, &
    \end{align}
\end{subequations}
with $X^{(n)}$, $\delta^{(n)}_X$, $\psi^{(r)}$
and $\delta^{(r)}_\psi$ having conformal weights
equal to $n$, $-n$, $r-\tfrac12$ and $-r+\tfrac12$ respectively.
Note that all $X^{(n)}$ and all $\psi^{(r)}$ are treated
as independent variables. Using the dictionary 
\eqref{dictionarybos} and \eqref{dictionaryNS}, the generators \eqref{eq:sp}, 
\eqref{eq:o} and \eqref{eq:Q_osp}
of the $\mathfrak{osp}(\oLim|\spLim)$ algebra
now take the form
\begin{subequations}
\begin{equation}\label{eq:sp:fields}
    T^{kl} = \tfrac{1}{k\,l}\,X^{(k)} \cdot X^{(l)}\,,
    \qquad
    T^k{}_l = \tfrac1{k}\,X^{(k)} \cdot \delta_X^{(l)}\,,
    \qquad
    T_{kl} = \delta_X^{(k)} \cdot \delta_X^{(l)}\,,
\end{equation}
with conformal weights equal to $k+l$, $k-l$
and $-k-l$ respectively,
\begin{equation}\label{eq:o_fields}
    M^{rs} = \psi^{(r)}\cdot \psi^{(s)}\,,
    \qquad
    M^r{}_s = \psi^{(r)} \cdot \delta_\psi^{(s)}\,,
    \qquad
    M_{rs} = \delta_\psi^{(r)} \cdot \delta_\psi^{(s)}\,,
\end{equation}
with conformal weights equal to $r+s-1$, $r-s$ 
and $-r-s+1$ respectively, and
\begin{equation}\label{eq:osp_fields}
    Q^{nr} = \tfrac1n\, X^{(n)} \cdot  \psi^{(r)}\,,
    \qquad
    Q^n{}_r = \tfrac1n\, X^{(n)} \cdot \delta_\psi^{(r)}\,, 
    \qquad
    Q_n{}^r = \psi^{(r)} \cdot \delta_X^{(n)}\,,
    \qquad
    Q_{nr} = \delta_X^{(n)} \cdot \delta_\psi^{(r)}\,,
\end{equation}
with conformal weights equal to $n+r-\tfrac12$,
$n-r+\tfrac12$, $-n+r-\tfrac12$
and $-n-r+\tfrac12$ respectively.
\end{subequations}

In the above, we have omitted the terms
proportional to $\tfrac{d}{2}$ in the definition
of the generators \eqref{eq:sp} and \eqref{eq:o},
as it is the structures without the terms
in question that will naturally appear
inside the Virasoro constraints, as we will see.
This modification affects no (anti)commutators
of the orthosymplectic algebra
other than those involving one trace--creation
and one trace--annihilation operator,
\begin{subequations}
\begin{align}\relax \label{alg:mod1}
    [T^{mn}, T_{kl}] & = - 4\,\delta^{(m}_{(k}\,T^{n)}{}_{l)}
    -2 d \, \delta^m_{(k} \delta^n_{l)}\,, 
    & [M^{rs}, M_{tu}] & = -4\,\delta^{[r}_{[t}\,M^{s]}{}_{u]}
    + 2 d \, \delta^r_{[t} \delta^s_{u]}\,, \\
    \label{alg:mod2}
    \{Q^{mr}, Q_{ns}\} & =  \delta^r_s\,T^m{}_n
    - \delta^m_n\,M^r{}_s + d\, \delta^m_n \delta^r_s\,,\\
    \label{alg:mod3}
    \{Q^{m}, Q_{n}\} & = T^m{}_n + \tfrac{d}2\, \delta^m_n \,,
    & [M^{r}, M_{s}] & = -M^r{}_s + \tfrac{d}2\, \delta^r_s \,,
\end{align}
\end{subequations}
which obtain a contribution containing
the number of spacetime dimensions. 
These contributions to \eqref{alg:mod1} will not influence
the derivation of physical solutions to the (necessary
and sufficient) Virasoro constraints given by $G_\phi$ and $G_{\phi+1}$,
as the latter will only involve the lowering operators 
\eqref{choice_low} and not $T_{mn}$ neither $M_{rs}$, as we will see. 
The $d$--dependent contribution to \eqref{alg:mod2}
will also not influence the construction of the physical states 
considered in this work, unlike the first of \eqref{alg:mod3},
as we will see in an example in the \Ram sector later.

Let us now consider the most general (integrated)
vertex operator in the canonical ghost picture
of the \NS sector
\begin{align} \label{vo_all}
   \mathbb{V}_F^{(-1)}(p,z) = e^{-\varphi}\,
   F_\NS\big(X^{(1)}, X^{(2)}, \dots,
   \psi^{(1)}, \psi^{(2)}, \dots\big)\,e^{i\,p \cdot X}\,,
\end{align}
which we may think of as an operator
that can create \emph{any} state or trajectory
of the open superstring. We will derive
the consequences of physicality conditions \eqref{physical_three}
on the general $\mathbb{V}_F^{(-1)}$. Let us first recall that
\begin{align} \label{zeroc}
    \big[\BRST_0, \mathbb{V}_F^{(-1)}\big]
    = \partial \big(c\mathbb{V}_F^{(-1)}\big)\,,
\end{align}
for $\mathbb{V}_F^{(-1)}$ of conformal weight $h=1$.
Now let us plug in the general form \eqref{vo_all}
and employ the fact that all CFTs in question are decoupled.
Since the energy--momentum tensor $T^{X,\psi}$ given in \eqref{EMsc1}
is the sum of two terms, 
\begin{equation}
    T^X(z) = -\tfrac12\,\partial X \cdot \partial X(z)\,,
    \qquad
    T^\psi(z) = -\tfrac12\,\psi \cdot \partial \psi(z)\,,
\end{equation}
we can then use the result of \cite{Markou:2023ffh}
for the contribution of the bosonic part $T^X(z)$
(see also \cite{Nergiz:1993gw} for an early,
and partial, version in the language of oscillators)
and focus only on the contribution of $T^\psi(z)$, 
given by
\begin{equation}
    \label{general_ope_f}
    T^\psi(z)\,F_\NS(w) \sim \tfrac12 \sum_{r,s\geq1}
    \bigg[ \frac{(r+s-1)}{(z-w)^{r-s+2}}\,
    \psi^{(s)} \cdot \delta_\psi^{(r)}
    + \frac{s}{(z-w)^{r+s+1}}\,\delta_\psi^{(r)}
    \cdot \delta_\psi^{(s)} \bigg]\,F_\NS(w)\,.
\end{equation}
Gathering this result with the contributions from $T^X$
and $T^{\beta,\gamma}$, we can compare the coefficients
of the poles of every order with \eqref{opeEMvo} for $h=1$. 
At order $1$, we find the coefficient
\begin{align}
    \bigg( \sum_{n\geq1} \Big[ X^{(n+1)} \cdot \delta_X^{(n)}
    + n\,\psi^{(n+1)} \cdot \delta_\psi^{(n)} \Big]
    -\partial \varphi + ip \cdot \partial X \bigg)\,
    \mathbb V^{(-1)}_F = \partial \mathbb V^{(-1)}_F\,,
\end{align}
so agreement is achieved identically. At order $2$,
we obtain the on--shell condition
\begin{equation}
    \bigg( \sum_{n\geq1} \Big[ X^{(n)} \cdot \delta_X^{(n)}
    + (n-\tfrac12)\, \psi^{(n)} \cdot \delta_\psi^{(n)} \Big]
    + \tfrac12\,(p^2 -1) \bigg)\, F_\NS(w) = 0\,,
\end{equation}
which we can rewrite as
\begin{equation}\label{onshell}
    \big(L_0 - \tfrac12\big) F_\NS
    = \bigg( \sum_{n\geq1} \Big[ n\,T^n{}_n
    + (n-\tfrac12)\,M^n{}_n \Big]
    + \tfrac12\,(p^2 -1) \bigg)\, F_\NS = 0\,,
\end{equation}
i.e. in terms of the Cartan subalgebra generators
of the algebra $\osp(\oLim|\spLim)$.
At order $n+2>2$, we find the differential constraints
\begin{equation}
    \begin{aligned}
        \bigg( p \cdot \delta_X^{(n)}
        & + \sum_{m\geq1} \Big[ X^{(m)} \cdot \delta_X^{(n+m)}
        + (m+\tfrac{n-1}2)\,
        \psi^{(m)} \cdot \delta_\psi^{(n+m)} \Big] \\
        & \hspace{80pt} + \tfrac12 \sum_{m=1}^{n-1} 
         \delta_X^{(m)} \cdot \delta_X^{(n-m)}
         + \tfrac12 \sum_{m=1}^n (\tfrac{n+1}2-m)\,
         \delta_\psi^{(m)}
         \cdot \delta_\psi^{(n+1-m)}\Big] \bigg) F_\NS = 0\,,
    \end{aligned}
\end{equation}
which we may rewrite as
\begin{align}\label{Virasoro1_gen}
    \begin{aligned}
        L_n F_\NS & \equiv \bigg( T^0{}_n
        + \sum_{m\geq1} \Big[ m\, T^m{}_{n+m}
        + (m+\tfrac{n-1}2)\, M^m{}_{n+m} \Big] \\
        & \hspace{80pt} + \tfrac12 \sum_{m=1} \Big[ T_{m,n-m}
        + (\tfrac{n+1}2-m)\,M_{m,n-m+1} \Big] \bigg) F_\NS =0\,, 
    \end{aligned}
    \qquad \forall\, n > 0\,,
\end{align}
where we have introduce the notation
$T^0{}_n = p \cdot \delta_X^{(n)}$, following
\cite[Sec. 3.3]{Markou:2023ffh}.
Next, let us turn to the action of $\mathcal Q_1$.
We first have that
\begin{align}\label{result2}
    T^{X,\psi}_{\textrm{F}}(z)\, \big(F_\NS\,
    e^{i\,p \cdot X}\big)(w) & \sim \frac1{2}\,
    \bigg( \sum_{n,m\geq1} \frac{1}{(z-w)^{n-m+1}} 
    \Big[ \psi^{(m)} \cdot \delta_X^{(n)}
    + X^{(m)} \cdot \delta_\psi^{(n)} \Big] \\
    & \hspace{20pt} -\frac{1}{z-w}\, p \cdot \psi
    + \sum_{m,n\geq1} \frac1{(z-w)^{n+m+1}}\,
    \delta_X^{(n)} \cdot \delta_\psi^{(m)} \bigg)\, \big(F_\NS\, e^{i\,p \cdot X}\big)(w)\,.
    \qquad\nonumber
\end{align}
Now let us recall that requiring $F_\NS\,e^{i\,p \cdot X}$
to be a primary field of weight $h$ with respect to the $\mathcal N=1$
super--Virasoro algebra imposes that it has a superpartner
$\widetilde{F}_\NS$ of weight $h+\frac12$
with respect to worldsheet supersymmetry,
and that the two are related by the supercurrent, 
in the sense that
\begin{subequations}
\begin{align}
    \label{proprim}
    T^{X,\psi}_{\textrm{F}}(z)\,\big(F_\NS\,e^{i\,p \cdot X}\big)(w) & \sim \frac{\nicefrac12}{z-w}\,
    \big(\widetilde{F}_\NS\,e^{i\,p \cdot X}\big)(w)\,, \\
    T^{X,\psi}_{\textrm{F}}(z)\, \big(\widetilde{F}_\NS\,e^{i\,p \cdot X}\big)(w) &\sim \bigg[ \frac{h}{(z-w)^2} + \frac{\nicefrac12}{z-w} \partial \bigg]\big(F_\NS\,e^{i\,p \cdot X}\big)(w)\,,
\end{align}
\end{subequations}
with $h=\nicefrac12$ here. Comparing \eqref{result2}
and \eqref{proprim}, we find that the worldsheet superpartner
of $F_\NS$ is given by
\begin{equation}\label{Virasoro2_gen_1}
    \widetilde{F}_\NS = \Big[ Q_0{}^1
    + \sum_{n\geq1} (Q_n{}^{n+1} + n\, Q^n{}_n)\Big]\,
    F_\NS= G_{-\nicefrac12}F_\NS\,,
\end{equation}
where we introduced $Q_0{}^1 := p \cdot \psi^{(1)}$,
and that the following constraints
\begin{subequations}
\begin{align}\label{Virasoro2_gen_2}
    G_{\nicefrac12} F_\NS &= \tfrac12\bigg(Q_{01}
    + \sum_{k\geq1}\, \Big[ k\,  Q^k{}_{k+1}
    + Q_{k}{}^k \Big]\bigg) F_\NS = 0\,, \\ 
    \label{Virasoro2_gen_2bis}
    G_{\nicefrac32} F_\NS &= \tfrac12\,\bigg( Q_{02}
    + Q_{11} + \sum_{k=1} \Big[ k\,Q^k{}_{k+2}
    + Q_{k+1}{}^k \Big] \bigg)\, F_\NS = 0\,,\\
    \label{Virasoro2_gen_3}
    G_{r-\,\frac12} F_\NS
    &= \tfrac12\,\bigg(Q_{0\,r} + \sum_{k=1}^{r-1} Q_{r-k,k}
    + \sum_{k\geq1} \Big[ k\, Q^k{}_{k+r}
    + Q_{k+r-1}{}^k \Big] \bigg) F_\NS = 0\,,
    \qquad \forall\, r > 2\,,
\end{align}
\end{subequations}
are necessary in order to ensure the absence
of poles of order $r>1$ in the OPE \eqref{result2}.
Note that we introduced $Q_{0\,r} := p \cdot \delta_\psi^{(r)}$.
At this point, let us observe that the form
of \eqref{Virasoro2_gen_2} and \eqref{Virasoro2_gen_3}
justifies the choice \eqref{choice_low}
of lowering operators. In addition, we find that
\begin{align}
    \big[\BRST_1, \mathbb{V}_F^{(-1)}\big] = 0\,, \quad \big[\BRST_2, \mathbb{V}_F^{(-1)}\big] = 0\,,
\end{align}
as there are no poles in the integrands (where we have used \eqref{superghopeexp} for the former), 
so the second and third of the conditions \eqref{physical_three}
are also satisfied. Finally, since only $L_0$, $G_{\nicefrac12}$
and $G_{\nicefrac32}$ are sufficient,
we will only need \eqref{onshell}, \eqref{Virasoro2_gen_1},
\eqref{Virasoro2_gen_2} and \eqref{Virasoro2_gen_2bis}.

\paragraph{Picture changing.}
Applying the picture--changing operation \eqref{picture}
on \eqref{vo_all} gives
\begin{align}
    \Big[\BRST_0, e^{\chi} \,\mathbb V_F^{(-1)}\Big]
    = \partial \big( c\, e^{\chi} \, \mathbb V_F^{(-1)} \big) \,, 
    \quad 
    \Big[\BRST_1,e^{\chi} \,\mathbb V_F^{(-1)} \Big]
    = -\tfrac12\, \widetilde{F}_{\NS}\, e^{ip\cdot X} \,,
    \quad
    \Big[\BRST_2, e^{\chi} \,\mathbb V_F^{(-1)}\Big] = 0\,,
\end{align}
so only $\BRST_1$ contributes to
\begin{equation}
    \mathbb V_F^{(0)}
    = 2\,\big[\BRST, e^\chi\,\mathbb V^{(-1)}_F\big]
    = -\widetilde{F}_{\NS}\, e^{ip\cdot X} \,,
\end{equation}
with $\widetilde{F}_{\NS}$ given by \eqref{Virasoro2_gen_1}.
We can now verify physicality on $\mathbb{V}_F^{(0)}$, and find
\begin{align} 
    \big[\BRST_0, \mathbb{V}_F^{(0)}\big]
    = \partial \big(c\mathbb{V}_F^{(0)}\big)\,,
\end{align}
which is a total derivative as should be the case,
with the on--shell condition now taking the form
\begin{align}
    \big( L_0 - \tfrac12 \big) \widetilde{F}_\NS
    = \bigg\{ \sum_{n\geq1} \Big[ n\,T^n{}_n
    + \big( n-\tfrac12 \big) M^n{}_n \Big]
    + \tfrac12\big(p^2 -2\big) \bigg\}\,
    \widetilde{F}_\NS = 0\,,
\end{align}
and the form of the differential constraints \eqref{Virasoro1_gen}
is unaffected, but now act on $\widetilde{F}_\NS$.
We also find that 
\begin{align}\label{nonvan1}
    \big[\BRST_1, \mathbb{V}_F^{(0)}\big]
    = -\tfrac12 \partial \big( e^{-\chi} e^{\varphi}\, F_{\NS}   e^{ip\cdot X} \big) \,,
    \qquad
    \big[\BRST_2, \mathbb{V}_F^{(0)}\big] = 0\,,
\end{align}
with the form of \eqref{proprim} and \eqref{Virasoro2_gen_1}
along with the constraints
\eqref{Virasoro2_gen_2}--\eqref{Virasoro2_gen_3} is unaffected.
Notice that the two superghost pictures, $(-1)$ and $(0)$,
are sufficient to calculate scattering amplitudes of \NS states,
as the sum of the pictures of all external states has to be equal
to $-2$ in order to cancel the background superghost charge of $2$.

\paragraph{\Ram sector.}
To derive the super--Virasoro conditions
on a general vertex operator,
\begin{align} \label{vo_allR}
   \mathbb{V}_F^{(-1)}(p,z) = e^{-\varphi/2}\,
   F_\Ram \big(X^{(1)}, X^{(2)}, \dots,
   \psi^{(1)}, \psi^{(2)}, \dots\big)\,
   S_\alpha\, e^{i\,p \cdot X}\,,
\end{align}
in the \Ram sector, two (equivalent) possibilities appear a priori:  impose the physicality condition \eqref{physical_three} on \eqref{vo_allR} or \textit{supersymmetrize}, via \eqref{susy1}, a general \textit{physical} vertex operator in the \NS sector, namely one of the form  \eqref{vo_all} which also \textit{satisfies} the constraints \eqref{Virasoro2_gen_2}--\eqref{Virasoro2_gen_3}. These procedures involve OPEs of the form
\begin{align}
    \psi(z) \cdot \partial \psi(z)\, 
    F_\Ram(w)S_A(w)\,e^{i\,p \cdot X(w)}\,,
    \quad  \psi(z) \cdot \partial X(z)\, 
    F_\Ram(w)S_A(w)\,e^{i\,p \cdot X(w)}\,,
\end{align}
and
\begin{align}
    i\partial X(z) \cdot \big(\gamma \, S(z)\big) _A \, F_\NS (w) \, e^{i\,p \cdot X(w)}\,,
\end{align}
respectively. However, as indicated by the form
of their $2$--point functions \eqref{weirdOPEs},
$\psi$ and $S$ are \textit{interacting} fields:
in the \Ram sector the fermion CFT is \textit{not} free,
unlike in the \NS sector. Consequently,
the standard version of Wick's theorem
cannot be applied to calculate OPEs
that involve $\psi$ and $S$, a problem
that can in principle be bypassed by means
of the technique of bosonization for $\psi$ and $S$
in particular \cite{Friedan:1985ge,Kostelecky:1986xg}. 
More specifically, $\psi$ and $S$ can be bosonized
by means of a system of free chiral bosons
$H_i\,, i = 1, \dots, 5\,$, with $2$--point function 
naturally given by
\begin{equation}
    \big\langle H_k(z)\, H_l(w) \big\rangle
    = - \delta_{kl}\, \ln |z - w|\,,
\end{equation}
via
\begin{align}\label{bosonization}
    \psi^\mu = e^{ i \mu \cdot H} \,,
    \qquad S_A = e^{i A \cdot H}\,,
\end{align}
(up to cocycle factors à la Jordan--Wigner, which guarantee that all components of $\psi$ and $S$ anticommute), where now $\mu$ and $A$ stand both for vector/spinor spacetime indices and for the vectors
\begin{align}
    \mu=\big(0,\dots,0,\pm1,0,\dots,0 \big) \,,
    \qquad
    A=\big(\pm \tfrac12, \pm \tfrac12, \pm \tfrac12, \pm \tfrac12, \pm \tfrac12 \big)\,,
\end{align}
where $\mu$ is a $5$--component vector,
and the previous notation is meant to highlight
that only one of its components is non--vanishing 
(and equal to $\pm1$).
Let us then attempt to calculate the OPE of $S_A$
with a general function of $\psi$ and its derivatives,
$F \big(\psi^{(1)}, \psi^{(2)}, \dots\big)$.
Since we need to bosonize all $S$ and $\psi$,
we must first calculate the OPE of $S_A$
with the transform $F_{\mathsf{bos}} \big( \mu\cdot H^{(1)}, \mu\cdot H^{(2)}, \dots\big)$ of $F$,
where
\begin{align}
     \mu\cdot H^{(n)} := \mu\cdot \partial^{n-1} H\,.
\end{align}
We thus have
\begin{eqnarray} \nonumber
        S_A(z)\, F_{\mathsf{bos}}(w)
    &\sim& \sum_{l\geq0}\frac{(z-w)^l}{l!}\partial^{l}S_A(w)\,  \bigg\{ \pm  \tfrac{i}{2} \bigg[ -\ln|z-w| \frac{\delta }{\delta \big( \mu \cdot H^{(1)}\big) }+\sum_{n\geq2}  \frac{(n-2)! }{(z-w)^{n-1}}\frac{\delta }{\delta \big( \mu \cdot H^{(n)}\big) }\bigg] \\ \nonumber
    && \qquad  -\tfrac14  \bigg[\ln^2|z-w|\frac{\delta^2 }{\delta \big( \mu \cdot H^{(1)}\big) \delta \big( \mu \cdot H^{(1)}\big) } \\ \nonumber
    &&\qquad  \qquad  - 2\ln|z-w| \sum_{n=2}  \frac{(n-2)! }{(z-w)^{n-1}}\frac{\delta^2 }{\delta \big( \mu \cdot H^{(1)}\big) \delta \big( \mu \cdot H^{(n)}\big) } \\
    && \qquad   \qquad  + \sum_{n,m\geq2}  \frac{(n-2)!(m-2)! }{(z-w)^{n+m-2}}\frac{\delta^2 }{\delta \big( \mu \cdot H^{(n)}\big)  \delta \big( \mu \cdot H^{(m)}\big) }\bigg] + \dots \bigg\}F_{\mathsf{bos}}\,.
\end{eqnarray}
For the leading Regge trajectory, there appears
only $\psi$ in $F$ according to \eqref{leadingvo},
so $F_{\mathsf{bos}}=e^{i\mu\cdot H}$
and only (an infinite series in) the derivatives
w.r.t. $\mu\cdot H^{(1)}$ survive,
that can be resummed to the result \eqref{leadingsuper}.
For general subleading trajectories,
there appears a double infinity:
one can act with infinitely many derivatives
w.r.t. to each of the infinitely
many $\mu\cdot H^{(n)}$. 

Instead of attempting to transform such a general 
operator acting on  $F_{\mathsf{bos}}$ into its form, in terms of $\psi$
and its derivatives, acting on $F$, we can turn to the Fock space
description of states: in this language, we can easily 
derive the super--Virasoro constraints from \eqref{supervis_gen_osc}:
\begin{subequations}
\begin{align}\label{eq:Vir_R1}
    G_0 \, F_{\mathsf{R}}
    & = \Big[ \tfrac1{\sqrt{2}}\,\slashed{p}
    + \sum_{k=1}^\infty \Big(k\,Q^k{}_{k}
    + Q_k{}^k\Big) \Big] F_{\mathsf{R}} =0 \,,\\ 
    \label{eq:Vir_R2}
    G_1\, F_{\mathsf{R}} & = \Big[ Q_{01} + Q_1
    + \sum_{k=1}^\infty \Big(k\,Q^k{}_{k+1}
    + Q_{k+1}{}^k\Big) \Big] F_{\mathsf{R}}=0 \,,
\end{align}
\end{subequations}
with the operators $Q$ of the $\osp(\spLim|\oLim)$ algebra
given by \eqref{eq:Q_osp} in terms of the oscillators
$\alpha_{\pm n}$ and $b_{\pm r}$. While the former
can be mapped to the field $\partial X$
and its descendants by means of the dictionary
\eqref{dictionarybos} in the constraints \eqref{eq:Vir_R1}--\eqref{eq:Vir_R2},
as explained around equation \eqref{partialdictionaryR},
we have no closed formula available that maps
$b_{-r}$ to $\psi$ and its descendants. Consequently,
at this moment there is no obvious way of writing
the constraints \eqref{eq:Vir_R1}--\eqref{eq:Vir_R2}
in the language most adapted to constructing
physical vertex operators, which are necessary
for the calculation of string scattering amplitudes;
we wish to revisit this point in the future.

\paragraph{Transverse subspace simplification.}
As argued for example in \cite{Manes:1988gz}
for the bosonic string
(see also \cite{Kato:1982im, Henneaux:1986kp}
for arguments obtained in the BRST formalism),
restricting to the transverse subspace, with metric
\begin{align}\label{transv_m}
    \eta_{\mu \nu}^\perp
    =\eta_{\mu \nu} -\frac{p_\mu p_\nu}{p^2}\,,
\end{align}
is sufficient to construct physical states,
which we may understand intuitively as eliminating
spurious longitudinal modes. This argument can be extended
to the case of the superstring since the BRST cohomology
is known, and has been shown to consists
of only transverse states, generated by DDF operators,
see e.g. \cite{Ohta:1986wz, Freeman:1986fx, Henneaux:1987ux}.
Since one has simply $p^2_\perp=0=\slashed{p}_\perp$
in the transverse subspace, so the operators $p^2$
and $\slashed{p}$, which form the Laplacian and Dirac operators
respectively and which are necessary for the Klein--Gordon
and Dirac equations (in momentum space) to be satisfied
in the $\mathsf{NS}$ and $\mathsf{R}$ sectors respectively,
remain unchanged. The Virasoro constraints become then
\begin{subequations}
\begin{align}
    \label{VirTNS1}
    \Big(L_0-\tfrac12 \Big) F_{\mathsf{NS}} 
    & = \Big[  N_{\mathsf{NS}} 
    +\tfrac12(p^2 -1)  \Big] \, F_{\mathsf{NS}} = 0\,, \\ 
    \label{VirTNS2}
    G_{\nicefrac12} F_{\mathsf{NS}}
    & = \sum_{n\geq1}\, \Big[ n\,  Q^n{}_{n+1}
    + Q_{n}{}^n \Big] F_{\mathsf{NS}} = 0\,,  \\ \label{VirTNS3}
    G_{\nicefrac32} F_{\mathsf{NS}} & = \Big[ Q_{11}^\perp
    + \sum_{k\geq1} \Big( k\,Q^k{}_{k+2}
    + Q_{k+1}{}^k \Big) \Big]F_{\mathsf{NS}} = 0\,,
\end{align}
\end{subequations}
in the $\mathsf{NS}$ sector, and 
\begin{subequations}
\begin{align}
    \label{VirTR1}
    G_0 F_{\mathsf{R}} & = \Big[ \tfrac1{\sqrt{2}}\,\slashed{p}
    + \sum_{k\geq1} \Big(k\,Q^k{}_{k} + Q_k{}^k\Big) \Big] 
    F_{\mathsf{R}} = 0 \,,\\  \label{VirTR2}
    G_1 F_{\mathsf{R}} & = \Big[ Q_1 + \sum_{k\geq1}\Big(k\,Q^k{}_{k+1} 
    + Q_{k+1}{}^k\Big) \Big] F_{\mathsf{R}} = 0\,,
\end{align}
\end{subequations}
in the $\mathsf{R}$ sector, with the level being given by
\begin{equation}
    N_{\mathsf{NS}}  F_{\mathsf{NS}}
    = \sum_{n\geq1} \Big[ n \, T^n{}_n
    + \big( n-\tfrac12 \big) M^n{}_n \Big] F_{\mathsf{NS}}
    \qquad\text{and}\qquad
    N_{\mathsf{R}} F_{\mathsf{R}}
    = \sum_{n\geq1} n \, \Big( T^n{}_n
    +  M^n{}_n \Big)F_{\mathsf{R}} \,.
\end{equation}

\subsection{Principal embedding}
\label{sec:principal}
As in the bosonic string, a state characterized
by a given polarization tensor,
visualised as a certain Young diagram $\mathbb Y$,
can and will appear \emph{infinitely} many times in the spectrum,
but at \emph{different} mass levels,
up to a possible non--trivial multiplicity.
A given Young diagram $\mathbb Y$ can namely be embedded in infinitely many different ways in the string spectrum, starting from a lowest possible level $N_{\rm min}$, whose value depends on the former's symmetries. Using the terminology of \cite{Markou:2023ffh}, we will refer to the polynomial dressing $\mathbb Y$ at $N_{\rm min}$, such that the corresponding state be physical, as its \emph{principal embedding}. We will now determine its form in the two sectors of the superstring. In both sectors, the anticommuting
character of the fermionic modes $b_{-r}$ implies that, if $k$ copies of a certain mode
are used in the construction of a state,
the corresponding polarization will be labelled by
a diagram having at least $k$ rows.

\paragraph{Neveu--Schwarz sector.}
Let us start with the simplest example,
namely one--row Young diagrams of length $s$,
i.e. totally symmetric rank--$s$ tensors. To minimise
the level, we should restrict ourselves to the conformal
ingredients with the lowest conformal weights, i.e.
$\alpha^\mu_{-1}$ and $b^\mu_{-\nicefrac12}\,$.
Since the $b$ are anticommuting, there cannot be more than
one, since a second would create a second row and so on.
There are, therefore, only two possibilities, namely 
to use $s$ times $\alpha_{-1}$, or to use 
$s-1$ copies of $\alpha^\mu_{-1}$ and one $b_{-\nicefrac12}$.
Considering that $b_{-\nicefrac12}$ has weight $\tfrac12$
and $\alpha_{-1}$ has weight $1$, the second option
is the one minimising the level, with the latter being then equal
to $N_{\rm min}(s)=s-\tfrac12$. Schematically,
we can represent the polynomial in $b_{-\nicefrac12}$
and $\alpha_{-1}$, out of which the corresponding
state will be constructed, as
\begin{equation} \label{onerow}
    \Yboxdim{24pt}\scriptstyle
    \gyoung(\boneh;\alphaone_9{\cdots};\alphaone)
\end{equation}
where the position of $\textcolor{NavyBlue}{b_{-r+\frac12}}$ 
and $\textcolor{BrickRed}{\alpha_{-n}}$
in the diagram is meant to denote the index
of the polarization tensor with which
their spacetime index is contracted. Such vertex 
operators build the leading Regge trajectory, 
discussed in e.g. \cite{Schlotterer:2010kk}.

Now let us proceed to two--row diagrams $(s_1,s_2)$. The first row, of length $s_1\,$, will naturally be built according to \eqref{onerow}, so a single $b_{-\nicefrac12}$ and several $\alpha_{-1}$ have been used. For the second row, again a single $b_{-\nicefrac12}$ can be used, but no $\alpha_{-1}\,$, since powers of $\alpha_{-1}$
are automatically symmetric in their spacetime
indices and, as such, cannot be used to construct yet another row
in the Young diagram. Consequently, to fill the second row, we can use a single $b_{-\nicefrac12}$ as well as the next possible modes, namely $b_{-\nicefrac32}$ and $\alpha_{-2}$, of weight $\tfrac32$ and $2$ respectively. Again, only a single $b_{-\nicefrac32}$ can appear, but several $\alpha_{-2}\,$, so there are four options: $s_2$ copies of $\alpha_{-2}\,$, one $b_{-\nicefrac12}$ or $b_{-\nicefrac32}$
and $s_2-1$ copies of $\alpha_{-2}\,$, a pair of $b_{-\nicefrac12}$ and $b_{-\nicefrac32}$ and $s_2-2$ copies of $\alpha_{-2}$. Obviously, the last option is the one minimising the level, so that a two--row diagram in the principal embedding
can be schematically represented as
\begin{equation}
    \Yboxdim{24pt}\scriptscriptstyle
    \gyoung(\boneh;\alphaone_9{\cdots};\alphaone,%
    \boneh;\bthreeh;\alphatwo_6{\cdots};\alphatwo)
\end{equation}
from which we can read off the level,
\begin{equation}
    N_{\rm min}(s_1,s_2) = \left\{
    \begin{aligned}
        s_1 \qquad & \qquad\text{if}\quad s_2=1\,,\\
        s_1 + 2\,s_2 - \tfrac52
        & \qquad\text{if}\quad s_2 \geq 2\,.
    \end{aligned}
    \right.
\end{equation}
Note that we need to distinguish between two cases,
either the second row has only one box, or it has two
or more, in order to account for the presence or absence
of $b_{-\nicefrac32}\,$.

One can convince oneself, following the same line
of arguments, that for three row diagrams,
the principal embedding is given by
\begin{equation}
    \Yboxdim{24pt}\scriptscriptstyle
    \gyoung(\boneh;\alphaone_9{\cdots};\alphaone,%
    \boneh;\bthreeh;\alphatwo_6{\cdots};\alphatwo,%
    \boneh;\bthreeh;\bfiveh;\alphathree_3{\cdots};\alphathree)
\end{equation}
and at this point one can infer the following pattern:
the $i$th row of a Young diagram in the principal
embedding is created by one copy of $b_{-r+\frac12}$
for every $r=1,\dots,i\,$, and $s_i-i$ copies
of $\alpha_{-i}$. Taking into account that
$b_{-r+\frac12}$ has weight $r-\tfrac12$,
and $\partial^i X$ has weight $i$, we find that
the lowest possible level at which a given Young diagram can be embedded in the spectrum reads
\begin{equation}
    N^{{\mathbb Y}, \mathsf{NS}}_{\rm min} = \sum_{i=1}^{\Diag} i\,\ell_i
    + (i-\tfrac12)\,h_i\,,
\end{equation}
where we use the (modified) Frobenius notation introduced
in the previous section for the diagram, namely
$\mathbb Y = (\ell_1,\dots,\ell_\Diag \mid h_1, \dots, h_\Diag)\,,$
where $\Diag$ is the length of its diagonal. Let us highlight that the derived principal embedding automatically satisfies the lowest weight conditions \eqref{eq:lw_osp} of $\osp(2\oRank|2\spRank,\R)$.

\paragraph{Ramond sector.}
Things are slightly more subtle in the Ramond
sector, mainly due to the fact that the bosonic
and fermionic oscillators carry now the same amount of energy units.
Consider for instance the leading Regge trajectory,
consisting of totally symmetric tensor--spinors
of rank $s \in \N$: there are two ways of creating
such states at level $s$, either with $s$ bosonic
oscillators $\alpha_{-1}^\mu$, or with one fermionic
oscillator $b_{-1}^\mu$ and $s-1$ bosonic $\alpha_{-1}^\mu$, which can be schematically represented as
\begin{equation}
    \Yboxdim{20pt}\scriptstyle   \gyoung(\bone;\alphaone_4{\cdots};\alphaone)
    \qquad{\textstyle \textrm{ and }}\qquad
    \Yboxdim{20pt}\scriptstyle   \gyoung(\alphaone_5{\cdots};\alphaone)\,.
\end{equation}
Unlike in the \NS sector, where the analogue of the first option
was favoured as the one minimizing the level as in \eqref{leadingvo},
here the two possibilities cannot be distinguished
on this ground and \textit{both} contribute to the trajectory, as is clear from \eqref{leadingsuper}. For mixed--symmetric tensor--spinors,
the situation becomes more ambiguous: for example, for two--row diagrams, there are $4$ \textit{distinct} possibilities
\begin{equation}
    \begin{aligned}
        \Yboxdim{20pt}\scriptscriptstyle
        \gyoung(\bone;\alphaone_5{\cdots};\alphaone,%
        \bone;\btwo;\alphatwo_2{\cdots};\alphatwo)\,
        {\textstyle,} \hspace{45pt}
        \Yboxdim{20pt}\scriptscriptstyle
        \gyoung(\bone;\alphaone_5{\cdots};\alphaone,%
        \bone;\alphatwo_3{\cdots};\alphatwo)\, 
        {\textstyle,} \\\\
        \Yboxdim{20pt}\scriptscriptstyle
        \gyoung(\alphaone_6{\cdots};\alphaone,%
        \bone;\btwo;\alphatwo_2{\cdots};\alphatwo)\,
        {\textstyle,} \hspace{45pt}
        \Yboxdim{20pt}\scriptscriptstyle
        \gyoung(\alphaone_6{\cdots};\alphaone,%
        \bone;\alphatwo_3{\cdots};\alphatwo)
        {\textstyle,}
    \end{aligned}
\end{equation}
and for three--row diagrams, we can depict the $2^3$ possibilities as
\begin{equation}
    \Yboxdim{20pt}\scriptscriptstyle
    \gyoung(;;\alphaone_9{\cdots};\alphaone,%
    \bone;;\alphatwo_5{\cdots};\alphatwo,%
    \bone;\btwo;;\alphathree_2{\cdots};\alphathree)
    {\textstyle,}
\end{equation}
where the empty box $i$ in the diagonal is to be understood
as accommodating either an $\alpha_{-i}$ or a $b_{-i}$, with $i=1,2,3$. 
It is thus straightforward to see that, more generally,
for the principal embedding in the $\mathsf{R}$ sector
and a diagram with a diagonal of length $\Diag$,
there are $2^{\Diag}$ possible \textit{disparate} configurations
of the bosonic and fermionic oscillators,
which differ only along the diagonal and which all lead
to the \textit{same} lowest possible level, that is given by
\begin{equation} \label{lowestNR}
    {\mathsf N}^{\mathsf{R}}_{\mathbb Y}
    = \sum_{i=1}^{\Diag} i\, \big( \ell_i + \bar\ell_i + 1 \big)
    = \sum_{i=1}^{\Diag} i\, \big( \ell_i+ h_i \big) \,,
\end{equation}
where in the first and in the second equality we use the Frobenius \eqref{Frobdef} and the modified Frobenius notation respectively.

Consequently, the candidate physical states
belonging to the principal embedding
in the $\mathsf{R}$ sector are a priori
linear combinations of the $2^{\Diag}$ configurations.
Moreover, since the operators $Q_n{}^n$ and $Q^n{}_n$
interchange bosonic with fermionic oscillators
of the same conformal weight along the diagonal
of a general Young diagram, the $2^{\Diag}$ configurations
can be represented by starting from any among them
and then acting with $Q_n{}^n$ and $Q^n{}_n$
to obtain the rest. A simple way of constructing
such a representation is to start from the configuration
whose diagonal is filled by fermionic oscillators only.
This is a state with spin $\mathbb Y_{\nicefrac12}$,
namely a representation of the Lorentz group
carried by a ($\gamma$--)traceless tensor--spinor,
whose tensor part has the symmetry of the Young diagram
$\mathbb Y$ in precisely the same way as in the \NS sector.
That is to say, we consider a polarisation tensor--spinor
with symmetry $\mathbb Y_{\nicefrac12}$ and contract
its indices corresponding to the $k$th row (column)
above (below) the diagonal with the bosonic
(fermionic) oscillators $\alpha^\mu_{-k}$\,
($b^\mu_{-k}$, and its spinor index
with that of the Ramond vacuum). We will denote 
such a state by the symbol $\ket{\Y}_{\nicefrac12}$;
its level $N_\Y$ is given by \eqref{lowestNR}.
As discussed previously,
the irreducibility of this tensor--spinor
will be reflected in the fact that the state
constructed in this manner is annihilated by
the generators listed in \eqref{eq:lw_osp},
as well as $Q_n$ (encoding its $\gamma$--tracelessness),
namely 
\begin{align}
    Q^m{}_r \ket{\Y}_{\nicefrac12} = 0\,,\, m < r\,,
    \quad
    Q_m{}^r \ket{\Y}_{\nicefrac12} = 0\,,\, m \geq r\,, 
    \quad
    Q_n \ket{\Y}_{\nicefrac12} = 0\,,\, n>0\,.
\end{align}
The candidate physical states of the principal embedding
can then be represented as
\begin{align}\label{candidate}
    \ket{\textrm{cand}}_{\nicefrac12}
    := P_{\Y}\ket{\Y}_{\nicefrac12} := \bigg( \Coeff{b}_0+\sum_{n=1}^{\Diag} \tfrac1{n!}\,\Coeff{b}_{i_1 \dots i_n}\,
    Q^{i_1}{}_{i_1} \dots Q^{i_n}{}_{i_n} \bigg) \ket{\Y}_{\nicefrac12}\,,
\end{align}
where, because of the anticommuting nature
of the $Q^{i}{}_{i}$, the $\Coeff{b}$'s are a priori
arbitrary totally antisymmetric coefficients,
that account for $2^{\Diag}$ parameters.
By construction, $P_{\Y}$ has conformal weight equal to $0$.

With a candidate \eqref{candidate} available
for the principal embedding, let us look at
the super--Virasoro constraints
\eqref{VirTR1}--\eqref{VirTR2}, which are respectively
equivalent to imposing that physical states be annihilated
by $G_0$ and $G_1$. First let us notice that
the product of $G_1$ and $P_\Y$
has conformal weight equal to $-1$,
so it has to be proportional to the lowering operators
of $\osp(\oLim|\spLim)$; the latter annihilate
$\ket{\Y}_{\nicefrac12}$ by construction,
so the constraint \eqref{VirTR2} can only produce identities.
We thus turn to \eqref{VirTR1}:
$G_0$ contains the Dirac operator $\slashed{p}$,
so it is this constraint that will yield the Dirac equation,
namely
\begin{equation} \label{Diraceom}
    G_0 \ket{\textrm{phys}}_{\nicefrac12} = 0
    \qquad\Longleftrightarrow\qquad 
    (\slashed{p}-m)\,\ket{\textrm{phys}}_{\nicefrac12}=0\,,
\end{equation}
as is expected for physical $\mathsf{R}$ states.
This observation is suggestive to recasting \eqref{VirTR1}
as a mass--shell condition on the polarisation
tensor--spinor and imply that principally embedded states
in the \Ram sector are those eigenstates of the ``mass operator'',
\begin{equation}\label{eq:mass_op}
    \pmb{m} := G_0 - \tfrac1{\sqrt2}\,\slashed{p}
    \equiv \sum_{n\geq1} n\,Q^n{}_n + Q_n{}^n\,,
\end{equation}
which appear at the \emph{lowest possible level},
for a given spin. Since $G_0$ ``squares to'' $L_0$
according to the super--Virasoro algebra \eqref{superVis3},
we have that
\begin{align} \label{meigenstate}
    \pmb{m}^2 + \tfrac1{\sqrt2} \{\pmb{m},\slashed{p} \}
    = N_{\mathsf R}
    \quad \Rightarrow \quad
    \pmb{m}\ket{\textrm{phys}}_{\nicefrac12}
    = \pm \sqrt{N_{\mathsf R}}\ket{\textrm{phys}}_{\nicefrac12}\,,
\end{align}
so the eigenvalues of $\pmb{m}$ are square roots of the mass level, 
given by \eqref{lowestNR} for principally embedded eigenstates. 
Consequently, we have to consider the equation \eqref{meigenstate}
on the candidate \eqref{candidate}
and solve for the parameters $\Coeff{b}$.

To this end, we first consider the action of $\pmb{m}$
on the state $\ket{\Y}_{\nicefrac12}$. By definition,
the generators $Q_n{}^n$ annihilate such a state,
while the $Q^n{}_n$ do not, nor do they act diagonally on it,
so that 
\begin{equation}
    \pmb{m}\ket{\Y}_{\nicefrac12}
    = \sum_{n\geq1} n\,Q^n{}_n\,\ket{\Y}_{\nicefrac12}\,.
\end{equation}
Considering now the action of $\pmb{m}$ on the candidate \eqref{candidate},
and that $\{Q^m{}_m, Q^n{}_n\} = 0$, we can simply
replace $Q^m{}_m$ with anticommuting variables,
say $\theta^m$. We can also use these variables
to collect the $\Coeff{b}_{i_1 \dots i_n}$ coefficients as
\begin{equation}
    \Coeff{b}_{[n]} := \tfrac1{n!}\,
    \Coeff{b}_{i_1 \dots i_n}\,
    \theta^{i_n} \dots \theta^{i_n}\,,
\end{equation}
where summation is implied. Moreover,
\begin{equation}
    \{Q_m{}^m, Q^n{}_n\} = \delta_{mn}\,
    (T^n{}_n + M^n{}_n)\,,
    \qquad
    [T^k{}_k + M^k{}_k, Q^m{}_m] = 0
    = [T^k{}_k + M^k{}_k, Q_m{}^m]\,,
\end{equation}
so that $Q_m{}^m$ effectively acts as a derivation
on $P_\Y^\pm\ket{\Y}_{\nicefrac12}$,
and $T^k{}_k+M^k{}_k$ by multiplication 
with $\ell_k+h_k$. This means that we can replace 
these depth zero $\osp(\oLim|\spLim)$ generators by
\begin{subequations}
\begin{align}\label{eq:Q-theta}
    Q^m{}_m \quad & \longleftrightarrow \quad \theta_m\,,
    &&& Q_m{}^m \quad & \longleftrightarrow \quad 
    (\ell_m+h_m)\,\tfrac{\partial}{\partial \theta_m}\,,\\
    T^k{}_k \quad & \longleftrightarrow \quad
    \theta_k\,\tfrac{\partial}{\partial \theta_k} + \ell_k\,,
    &&& M^k{}_k \quad & \longleftrightarrow \quad
    -\theta_k\,\tfrac{\partial}{\partial \theta_k} + h_k\,,
\end{align}
\end{subequations}
when acting on the space of states
of the form \eqref{candidate}.
Introducing the operators
\begin{equation}
    \delta := \sum_{m\geq1} m\,\theta^m\,,
    \qquad
    \delta^* := \sum_{m\geq1} (\ell_m+h_m)\,
    \tfrac{\partial}{\partial\theta^m}\,,
\end{equation}
which obey
\begin{equation}
    \delta^2 = 0 = (\delta^*)^2\,,
    \qquad
    \{\delta,\delta^*\} = {\mathsf N}^{\mathsf{R}}_{\mathbb Y} \,,
\end{equation}
we can now rewrite the constraint \eqref{meigenstate}
on $\ket{\textrm{cand}}_{\nicefrac12}$ simply as
\begin{equation}\label{eq:system_R}
    \pm\,\sqrt{{\mathsf N}^{\mathsf{R}}_{\mathbb Y}}\,
    \Coeff{b}_{[n]} = \delta\Coeff{b}_{[n-1]}
    + \delta^*\Coeff{b}_{[n+1]}\,,
    \quad \forall\, 0 \leq n \leq \Diag\,.
\end{equation}
As the notation suggests, we may think
of $\delta$ and $\delta^*$ as a differential 
and a codifferential on the spaces of forms,
with basis given by monomials
in $\theta^k \leftrightarrow Q^k{}_k$.

The system \eqref{eq:system_R} contains as many equations
as there are of unknowns, which correspond
to the sum of the number of components of antisymmetric
tensors in $\Diag$ dimensions of all ranks,
i.e. $\sum_{n=0}^{\Diag} \binom{\Diag}{n}=2^{\Diag}$.
Note, however, that the system \eqref{eq:system_R}
contains some redundancy, i.e. not all equations
are independent. Indeed, suppose we are given
the equations
\begin{equation}\label{eq:pair_eq}
    \pm\,\sqrt{{\mathsf{N}}^{\mathsf{R}}_{\mathbb Y}}\,
    \Coeff{b}_{[n]} = \delta\Coeff{b}_{[n-1]}
    + \delta^*\Coeff{b}_{[n+1]}
    \qquad\text{and}\qquad 
    \pm\,\sqrt{{\mathsf N}^{\mathsf{R}}_{\mathbb Y}}\,
    \Coeff{b}_{[n+2]} = \delta\Coeff{b}_{[n+1]}
    + \delta^*\Coeff{b}_{[n+3]}\,,
\end{equation}
for some value of $n$, and let us act with $\delta$
on the first one. This yields
\begin{equation}
    \pm\sqrt{{\mathsf N}^{\mathsf{R}}_{\mathbb Y}}\
    \Coeff{b}_{[n+1]} = \delta\Coeff{b}_{[n]}
    \pm \tfrac1{\sqrt{{\mathsf N}^{\mathsf{R}}_{\mathbb Y} }}\,
    \delta^*\delta\Coeff{b}_{[n+1]}\,,
\end{equation}
which, upon using the second equation of \eqref{eq:pair_eq}
to replace $\delta\Coeff{b}_{[n+1]}$, leads to
\begin{equation}
    \pm \sqrt{{\mathsf N}^{\mathsf{R}}_{\mathbb Y}}\
    \Coeff{b}_{[n+1]} = \delta\Coeff{b}_{[n]}
    + \delta^*\Coeff{b}_{[n+2]}\,,
\end{equation}
i.e. we obtain the equation for $\Coeff{b}_{[n+1]}$.
One can check that, upon acting on the second equation
of \eqref{eq:pair_eq} with $\delta^*$ and using the first one
also produces the equation for $\Coeff{b}_{[n+1]}$.
Since this mechanism works for all values of $n$,
this tells us that roughly ``half'' of the equations
are independent. More precisely, the subsets of equations
\eqref{eq:system_R} with $n$ taking only \emph{even}
or \emph{odd} values are independent. For the sake 
of definiteness, let us say that we keep only the latter,
i.e. our system of equations is
\begin{equation}
    \pm\sqrt{{\mathsf N}^{\mathsf{R}}_{\mathbb Y}}\
    \Coeff{b}_{[2k+1]} = \delta\Coeff{b}_{[2k]}
    + \delta^*\Coeff{b}_{[2k+2]}\,,
    \quad \forall\, 0 \leq k \leq [\Diag/2]\,,
\end{equation}
which can be read as the \emph{definitions} for the unknowns
$\Coeff{b}_{[2k+1]}$ in terms of \emph{free} parameters, the
$\Coeff{b}_{[2k]}$. In other words, the above equation
is the most general solution of the defining equation
for $P_\Y$. The space of solutions is therefore parameterized
by the antisymmetric tensors $\Coeff{b}_{[2k]}$ of even ranks,
and hence of dimension
\begin{equation}\label{eq:mult_R}
    \sum_{n=0}^{[\Diag/2]} \binom{\Diag}{2n}
    = \tfrac12\,\sum_{n=0}^{\Diag} \binom{\Diag}{n}\,
    \big(1+(-1)^n\big) = 2^{\Diag-1}\,.
\end{equation}
For example, for $K=2$, up to an overall coefficient, the principal embedding takes the form
\begin{align}
\begin{aligned} \label{traj_ex_R_0}
  \bigg\{ \Coeff{b}_0 \, b_{-1}^{\nu_1} b_{-2}^{\nu_2} & \pm\tfrac{1}{\sqrt{\mathsf N}}\Big[h_1 \big(\Coeff{b}_0-(\ell_2+h_2)\Coeff{b}_{12}\big)     \,\alpha_{-1}^{\mu_1} b_{-2}^{\nu_2} \\ & \hspace{50pt}
    +h_2 \big(\Coeff{b}_0 +\tfrac12 (\ell_1+ h_1) \Coeff{b}_{12} \big) \, b_{-1}^{\nu_1} \alpha_{-2}^{\mu_2} \Big]
+\tfrac12 h_1 h_2 \Coeff{b}_{12} \, \alpha_{-1}^{\mu_1} \alpha_{-2}^{\mu_2} \bigg\}\\ & \hspace{100pt} \times \alpha_{-1}^{\mu_1(\ell_1)} \alpha_{-2}^{\mu_2(\ell_2)} b_{-1}^{\nu_1[h_1-1]} b_{-2}^{\nu_2[h_2-1]}\,\varepsilon_{\mu_1(\ell_1),\mu_2(\ell_2)\pmb|\nu_1[h_1], \nu_2[h_2]}^A  \ket{p;A;0}_\mathsf{R} \,,
\end{aligned}
\end{align}
with $\mathsf N=\ell_1+h_1 +2(\ell_2+h_2)$ and multiplicity
equal to $2$, namely there are two distinct such trajectories.

The multiplicity \eqref{eq:mult_R} may seem surprising at first glance,
considering that it concerns \emph{principally embedded} states,
as it contrasts drastically with the situation
in the \NS sector, or in the bosonic string
\cite{Markou:2023ffh}, where principally embedded states
all appear with multiplicity one. This unexpected feature
of the \Ram sector in fact seems to be in accordance
with the result of \cite{Hanany:2010da},
where the superstring spectrum
has been computed up to level 10. More specifically, 
using the \emph{factorized} partition function
(3.18)--(3.19) of \cite{Hanany:2010da}, one can compute
 the content of each supermultiplet
appearing at a given mass level. In the \Ram sector,
the first diagram with $\Diag=2$, namely the ``window''  $\gyoung(;;,;;)_{\nicefrac12}\,$, appears at level $N=5$ and in two \textit{distinct} supermultiplets,%
\footnote{Recall that any supermultiplet is obtained
by tensoring the little group irreps of the fundamental
supermultiplet with any other little group irrep,
so that the latter can serve as a label
for the supermultiplet in question.}
\begin{equation} \label{twosupermu}
    \Big(\gyoung(;;) \oplus \gyoung(;,;,;)
    \oplus \gyoung(;)_{\nicefrac12} \Big)
    \otimes \gyoung(;;)_{\nicefrac12}
    \quad \supset \quad \gyoung(;;,;;)_{\nicefrac12}
    \quad \subset \quad \Big(\gyoung(;;)
    \oplus \gyoung(;,;,;)
    \oplus \gyoung(;)_{\nicefrac12} \Big)
    \otimes \gyoung(;,;)\,,
\end{equation}
i.e. the first time this state appears,
it is with multiplicity two. This is precisely in agreement
with our result: the window in question is the lightest member
of the trajectory \eqref{traj_ex_R_0} and is given by
\begin{align}\label{lightestR}
\begin{aligned}
    & \varepsilon_{\mu\pmb{|}\nu_1 \nu_2,\nu}^A\,
    \alpha_{-1}^{\mu} \bigg\{\Coeff{b}_0\,b_{-1}^{[\nu_1}
    b_{-1}^{\nu_2]} b_{-2}^\nu \pm\tfrac{1}{\sqrt{5}}
    \Big[2 \big(\Coeff{b}_0-\Coeff{b}_{12}\big) \alpha_{-1}^{[\nu_1}  
    b_{-1}^{\nu_2]}b_{-2}^{\nu} \\
    & \hspace{150pt} + \big(\Coeff{b}_0 +\tfrac{3}{2} \Coeff{b}_{12} \big) b_{-1}^{[\nu_1} b_{-1}^{\nu_2]}\alpha_{-2}^{\nu}  \Big]
    + \Coeff{b}_{12} \, \alpha_{-1}^{[\nu_1} b_{-1}^{\nu_2]} \alpha_{-2}^{\nu} \bigg\} \,\ket{p;A;0}_\mathsf{R} \,,
 \end{aligned}
\end{align}
so that it indeed appears with multiplicity two,
as encoded in the two free parameters $\Coeff{b}_0$
and $\Coeff{b}_{12}$. Note, however, that, to associate
the two states with their respective supermultiplets,
we would have to determine those two linear combinations
of \eqref{lightestR} that close the spacetime 
supersymmetry algebra on the two supermultiplets
that appear in \eqref{twosupermu}. Similarly,
we find two other principally embedded diagrams
with $\Diag=2$ at level $N=6$,
\begin{equation}
    \Big(\gyoung(;;) \oplus \gyoung(;,;,;)
    \oplus \gyoung(;)_{\nicefrac12} \Big)
    \otimes \gyoung(;;;)_{\nicefrac12}
    \quad \supset \quad \gyoung(;;;,;;)_{\nicefrac12}
    \quad \subset \quad \Big(\gyoung(;;)
    \oplus \gyoung(;,;,;)
    \oplus \gyoung(;)_{\nicefrac12} \Big)
    \otimes \gyoung(;;;,;)\,,
\end{equation}
and 
\begin{equation}
    \Big(\gyoung(;;) \oplus \gyoung(;,;,;)
    \oplus \gyoung(;)_{\nicefrac12} \Big)
    \otimes \gyoung(;,;)_{\nicefrac12}
    \quad \supset \quad \gyoung(;;,;;,;)_{\nicefrac12}
    \quad \subset \quad \Big(\gyoung(;;)
    \oplus \gyoung(;,;,;)
    \oplus \gyoung(;)_{\nicefrac12} \Big)
    \otimes \gyoung(;;,;,;)\,,
\end{equation}
both of which also appear with multiplicity $2$,
in accordance with the previous discussion.
At level $N=7$, one finds principally embedded states
with spin
\begin{equation}
    \gyoung(;;;;,;;)_{\nicefrac12}\,,
    \qquad \gyoung(;;;,;;;)_{\nicefrac12}\,,
    \qquad \gyoung(;;;,;;,;)_{\nicefrac12}\,,
    \qquad \gyoung(;;,;;,;;)_{\nicefrac12}
    \qquad\text{and}\quad
    \qquad \gyoung(;;,;;,;,;)_{\nicefrac12}\,,
\end{equation}
and which all appear with multiplicity $2$ again.
This trend at relatively low levels seem to confirm
the degeneracy \eqref{eq:mult_R}, at least for $\Diag=2$.
In order to check this for $\Diag=3$, one would need
to push the computation of the superstring spectrum
as outlined in \cite{Hanany:2010da} up to level $N=14$
at the minimum.\footnote{The diagram with $\Diag=3$
and the lowest principal embedding level is
$\gyoung(;;;,;;;,;;;)_{\nicefrac12}$, which has
${\mathsf N} =14$.}

Let us conclude by remarking that there exists
a fairly simple solution for $P_\Y^\pm$ which is
only linear in $Q^n{}_n$, namely
\begin{equation}\label{eq:linear}
    P^\pm_\Y \propto \mathbf{1} \pm \tfrac1{\sqrt{{\mathsf N}^{\mathsf{R}}_{\mathbb Y} }}\,
    \sum_{n=1}^{\Diag} n\,Q^n{}_n\,,
\end{equation}
obtained by setting the free parameters $\Coeff{b}_{[2k]}=0$
for all $k>0$. Notice also that this solution exists
for any diagram $\Y$. For example, we find the principally embedded 
totally symmetric tensor--spinor as the general solution 
\eqref{eq:linear}, restricted to a one--row diagram of length $s$,
to be
\begin{equation}
    \ket{\varepsilon}_\pm := \Big( 1 \pm \tfrac{1}{\sqrt s}\,Q^1{}_1 \Big)\,\varepsilon_{\mu(s)}^{A\pm}\,
    b^\mu_{-1}\,\alpha^{\mu(s-1)}_{-1}\!
    \ket{A} = \varepsilon_{\mu(s)}^{A\pm}\,
    \big(b^\mu_{-1} \pm \tfrac1{\sqrt s}\,
    \alpha^\mu_{-1}\big)\,\alpha^{\mu(s-1)}_{-1}\!
    \ket{A}\,,
\end{equation}
where the polarization tensor--spinor $\varepsilon$
is traceless and $\gamma$--traceless. 
Now the generator $G_1$ does annihilate such a state,
while the generator $G_0$ produces the equation
\begin{align}
    \big(\slashed{p} \pm \sqrt{2s}\big) \,
    \varepsilon_{\mu(s)}^{A\pm} =0\,,
\end{align}
so after the \GSO projection we obtain
\begin{align}
    \slashed{p}_{\alpha\dot{\beta}}\,
    \varepsilon^{\dot{\beta}\pm}_{\mu(s)}
    \pm \sqrt{2s}\,\varepsilon^\pm_{\mu(s)\,\alpha}=0\,,
\end{align}
upon using the chiral representation of $\gamma$--matrices
available in even dimensions, which matches known results
for leading Regge states \cite{Koh:1987hm, Schlotterer:2010kk}.

\subsection{Non--principal embedding}

We can now extend the notion of ``\textit{depth}'' $w$ in both sectors, defined \cite{Markou:2023ffh} in the bosonic string as the difference between the level at which a certain diagram appears from the lowest possible level at which the same diagram can be physically embedded, to the case of the superstring, via:
\begin{align}
    w = N-  {\mathsf N}_{\mathbb Y} 
    = N - \sum_{i=1}^{\Diag} i\,\ell_i
    + (i - \phi)\,h_i\,.
\end{align}
In addition, while a bosonic string ``trajectory'' can be defined as the set of Young diagrams with a fixed number of rows that appear at a fixed value of $w$, for the case of the superstring the former integer should be replaced by the length $\Diag$ of their diagonal. We thus define a superstring ``\textit{trajectory}'' as the set of Young diagrams with a fixed pair $(\Diag,w)$. The analogue of the simplest, namely $1$--row trajectory, in the bosonic string is hence the hook
\begin{equation}
    \Yboxdim{10pt}
    \gyoung(|7{\scriptstyle h}_9{\scriptstyle \ell})
\end{equation}
in both sectors of the superstring, of the $2$--row
it is the ``nested'' hook of two rows and two columns and so on. 
\begin{table}
\centering 
\renewcommand{\arraystretch}{1.5}
  \begin{tabular}{ c || l | l  }
   $m^2$ & $\mathsf{NS}$ & $\mathsf{R}$  \\ \hline \hline
   $-\tfrac12$ & $\textcolor{BrickRed}{\bullet}$ &   \\
   $0$ & $\LeadingB{;}^{\textcolor{BrickRed}{so(d-2)}}$
   & $\textcolor{BrickRed}{\bullet^{so(d-2)}_{\nicefrac12}}$
    \\
   $\tfrac12$ & $\LeadingB{;,;}$ &   \\
   $1$ & $\LeadingB{;;} \oplus\, \LeadingB{;,;,;}$
   & $\LeadingB{;}_{\textcolor{BrickRed}{\nicefrac12}}$  \\
   $\tfrac32$ & $ \LeadingB{;,;,;,;}\, \oplus\, \LeadingB{;;,;} \,
   \oplus\, \Wfirst{;;}\, \oplus\, {\color{NavyBlue}\bullet}$ & \\
   $2$ & $\LeadingB{;;;}\, \oplus\, \LeadingB{;,;,;,;,;} \, \oplus\,  \LeadingB{;;,;,;}  \, 
   \oplus\, \Wfirst{;;,;}\, \oplus\, \Wthird{;,;} \,
   \oplus\, \Wfourth{;} $ 
   & $\LeadingB{;;}_{\textcolor{BrickRed}{\nicefrac12}}\, \oplus\, \LeadingB{;,;}_{\textcolor{BrickRed}{\nicefrac12}} \, \oplus\, \Wsecond{;}_{\textcolor{Dandelion}{\nicefrac12}}\, 
   \oplus\, \textcolor{NavyBlue}{\bullet_{\nicefrac12}}
  $
  \end{tabular}
\renewcommand{\arraystretch}{1}
\caption{Open superstring, physical content of the first few levels. 
The principal embedding is indicated
in \textcolor{BrickRed}{red}, while the embeddings of depth $w=\frac12,1,\frac32,2$ 
in \textcolor{Orchid}{orchid}, \textcolor{Dandelion}{dandelion}, black and \textcolor{NavyBlue}{blue} respectively.}
\label{lightestnew}
\end{table}
Consequently, we can view $1$--row and $1$--column diagrams
as a subclass of the hook trajectory. The principal embedding
of this first trajectory, consisting of all hooks at $w=0$,
is highlighted in red in table \ref{lightestnew}
and its states take the form
\begin{align}\label{princNS}
    \varepsilon_{\mu(\ell),\nu[h]}^{\NS}\, \alpha_{-1}^{\mu(\ell)}\, b^{\nu[h]}_{-\nicefrac12}\, \ket{p;0}_\NS \,, \\ \label{princR}
    \varepsilon_{\mu(\ell),\nu[h]}^{\Ram, A\pm} \, \big( \alpha_{-1}^{\mu(\ell)} \, b^{\nu[h]}_{-1} \pm\tfrac{h}{\sqrt{\ell+h}} \alpha_{-1}^{\mu(\ell)\nu}b_{-1}^{\nu[h-1]} \big)  \ket{p; A;0}_\Ram\,.
\end{align}
By applying the dictionary \eqref{dictionarybos},
\eqref{dictionaryNS} on \eqref{princNS},
the vertex operators of the principally embedded \NS hooks
are very easily found to be
\begin{align}
    \mathbb V^{(-1)}_{\depth=0}
    =  \varepsilon_{\mu(\ell),\nu[h]}^{\NS}\, e^{-\varphi} \, 
    i\partial X^{\mu(\ell)}\, \psi^{\nu[h]}\, e^{ip\cdot X} \,,
\end{align}
of which the leading Regge trajectory \eqref{leadingvo}
is a subcase. Notice also that all principally embedded trajectories 
are built \emph{before} the GSO projection, as is for example the hook \eqref{princNS}, on which the \GSO has to be applied; the leading Regge \eqref{leadingvo} is already given after the \GSO, as every time a box is added to the row, the level changes by an integer. On the other hand, constructing the vertex operators
of the principally embedded \Ram hooks would necessitate
the dictionary for $n$ copies of $b_{-1}$, which, as explained around \eqref{dictionaryR1}, requires knowledge of the OPE $\psi^\mu(z) S_A(w)$ to \textit{arbitrary} order, even though only one type of $b$ is present. For the subcase of the leading Regge trajectory, there appears only one copy of $b_{-1}$, so we can apply the dictionary \eqref{dictionaryNS}, \eqref{dictionaryR1} which yields \eqref{leadingsuper}.

In the rest of this work, we will restrict ourselves
to the Fock space description of \Ram states. By construction,
all trajectories of the principal embedding can be uplifted
to both half--integer and integer depths in the \NS
and only to integer depths in the \Ram sector.
We will be considering specific cases of small values
of nontrivial depth that are highlighted
in table \ref{lightestnew} and construct the Ans\"atze
for the respective dressing functions,
on which we will then apply the super--Virasoro constraints
to find the physical trajectories, and once obtained,
check whether they are spurious or not.
Proceeding to deeper trajectories is straightforward
by following the same algorithm.

\paragraph{\NS sector at depth $w=\tfrac12$.}
An operator of depth $w=\tfrac12$ is generically
of the form
\begin{equation}\label{eq:Ansatz_w=1/2}
    f = \sum_{k\geq1} \big(\tfrac1k\,\Coeff{u}_k\,
    Q_k{}^{k+1} + \Coeff{v}_k\,Q^k{}_k\big)\,,
\end{equation}
where the factor $\tfrac1k$ before the first term
has been added solely for later convenience and $\Coeff{u}_k$, $\Coeff{v}_k$ are a priori arbitrary coefficients. The
anticommutator of $f$ with $G_{\nicefrac12}$ is given by
\begin{equation}
    \{G_{\nicefrac12},f\}
    = \sum_{k\geq1} (\Coeff{u}_{k-1}+\Coeff{v}_k)\,M^k{}_k
    + (\Coeff{u}_k+\Coeff{v}_k)\,T^k{}_k\,,
\end{equation}
whereas the anticommutator
$\{G_{\nicefrac32}, f\}$ is of depth $-1$,
and hence necessarily proportional
to the lowering operators of the orthosymplectic algebra.
As a consequence, when evaluating these anticommutators
on a principally embedded state with Young diagram
$(\ell_1,\dots \ell_\Diag | h_1, \dots, h_\Diag)$,
one finds that the result vanishes provided that
\begin{equation}\label{ns:cons12}
    \sum_{k=1}^{\Diag} \Coeff{u}_k\,(h_{k+1} + \ell_k)
    + \Coeff{v}_k\,(h_k + \ell_k) = 0\,.
\end{equation}
Note that $h_k+\ell_k$ is the ``hook length''
at the $k$th box on the principal diagonal
of the diagram.

Let us consider specific examples that correspond
to explicit solutions of \eqref{ns:cons12} for the first trajectory.
\begin{itemize}
    \item For the leading Regge subcase, the solution is
    (up to an overall non--zero prefactor),
    \begin{align}
        f = Q^1{}_1 -\tfrac{\ell+1}{\ell} Q_1{}^2\,,
    \end{align}
    whose lightest member is the spin--2
    \begin{align}
        F = \varepsilon_{\mu\nu}(p)\,
        \big(\tfrac12\,i\partial X^\mu\,i\partial X^\nu
        +\psi^\mu \partial \psi^\nu \big)\,.
    \end{align}
    This is the lightest clone of \eqref{lightestspin2}.
    Notice that this trajectory does \emph{not} survive
    the $\mathsf{GSO}$ projection.
    \item For hook--type diagrams, the solution
    is (up to an overall non--zero prefactor),
    \begin{align}
        f = -\tfrac{\ell}{\ell+h}\,Q^1{}_1+Q_1{}^2\,.
    \end{align}
    For the smallest diagram of this type, $\sYng{;;,;}$\,, this yields the polynomial
    \begin{align}
        F = \varepsilon_{\mu\pmb|\nu\rho}(p)\,
        \big(-\tfrac23\,i\partial X^\mu\,i\partial X^\nu \psi^\rho
        + \partial \psi^\mu \psi^\nu \psi^\rho\big)
    \end{align}
    of the vertex operator in the $(-1)$ superghost picture
    of the lightest clone of the hook \eqref{eq:hook}
    in the principal embedding,
    highlighted in \textcolor{Orchid}{orchid}
    in table \ref{lightestnew}. Using \eqref{Virasoro2_gen_1},
    the polynomial of the vertex operator in the $(0)$ picture
    is found to be given by
     \begin{align}
        \widetilde{F} = \varepsilon_{\mu\pmb|\nu\rho}(p)\,
        \big(\tfrac43\,\partial \psi^\mu\, i \partial X^\nu \psi^\rho
        -\tfrac23\,i\partial X^\mu\partial \psi^\nu \psi^\rho
        + i\partial^2 X^\mu \psi^\nu \psi^\rho\big)\,.
    \end{align}
    \item For $p$--forms, namely when $\ell=0$ and $h$ is arbitrary,
    there is no non--trivial solution for the $w=\tfrac12$ dressing
    function, meaning that one cannot uplift principally
    embedded $p$--forms at level $\tfrac{p}2$
    to physical $p$--forms at level $\tfrac{p+1}2$, which is in accordance with table \ref{lightestnew}.
\end{itemize}

More generally, since the most general operator of depth $w=\tfrac12$ 
is characterized by $2\Diag$ parameters, subject 
to a single constraint \eqref{ns:cons12}, the space
of depth--$\tfrac12$ dressing functions is of dimension
$2\Diag-1$. The general solution can be simply obtained
by solving one of the $2\Diag$ parameters $\Coeff{u}_k$
and $\Coeff{v}_k$ in the general Ansatz 
\eqref{eq:Ansatz_w=1/2} in terms of the others
in \eqref{ns:cons12}. Another way to proceed
is to notice that \emph{only two} non-zero terms
in the Ansatz \eqref{eq:Ansatz_w=1/2} are necessary
in order to produce a solution. We can therefore
find a basis of this space of solutions made out
of simple linear combinations of two operators,
such as for instance,
\begin{equation}
    f_k = \left\{
    \begin{aligned}
        & \tfrac1k\,Q_k{}^{k+1}
        - \tfrac{\ell_k+h_{k+1}}{\ell_k+h_k}\,Q^k{}_k\,,
        & 1 \leq k \leq \Diag\,,\\
        & Q^{k+1}{}_{k+1}
        - \tfrac1k\,\tfrac{\ell_{k+1}+h_{k+1}}{\ell_k+h_{k+1}}\,
        Q_k{}^{k+1}\,,
        & \quad \Diag+1 \leq k \leq 2\Diag-1\,.
    \end{aligned}
    \right.
\end{equation}
In particular, the dressing functions presented above
for the leading Regge trajectory and the hook-type diagrams
are multiples of $f_1$.

\paragraph{\NS sector at depth $w=1$.}
To derive dressing functions of depth $w=1$,
we can repeat the above steps, starting with
spelling out the most general form
of such a dressing function,
\begin{equation}
    \begin{aligned}
        f & = \sum_{k\geq1} \tfrac1k\,
        \big(\Coeff{u}_k\,T^{k+1}{}_k
        + \Coeff{v}_k\,M^{k+1}{}_k\big) \\
        & \hspace{50pt} + \tfrac12\,\sum_{i,j\geq1}
        \Big(\Coeff{u}_{ij}\,Q^i{}_i\,Q^j{}_j
        + \tfrac1{ij}\,\Coeff{v}_{ij}\,
        Q_i{}^{i+1}\,Q_j{}^{j+1}
        + \tfrac1j\,\Coeff{t}_{ij}\,
        \big[Q^i{}_i, Q_j{}^{j+1}\big]\Big)\,,
    \end{aligned}
\end{equation}
where $\Coeff{u}_{ij}$ and $\Coeff{v}_{ij}$
are antisymmetric, and computing its anti/commutator
with $G_{\nicefrac12}$ and $G_{\nicefrac32}$.
The former reads
\begin{eqnarray}
    \big[G_{\nicefrac12}, f\big]
    & = & \sum_{k\geq1} \Big(\big[\Coeff{v}_k
    - \Coeff{u}_{k-1}
    +\tfrac12\,\big(\Coeff{t}_{k\,k-1}
    -\Coeff{t}_{kk}\big)\big]\,Q^k{}_k
    + \tfrac1k\,\big[\Coeff{u}_k-\Coeff{v}_k
    +\tfrac12(\Coeff{t}_{k+1\,k}
    -\Coeff{t}_{kk})\big]\,Q_k{}^{k+1}\Big) 
    \qquad\nonumber\\
    && - \sum_{k,l\geq1} \Big(Q^k{}_k\,
    \big[\Coeff{u}_{kl}\,(T^l{}_l+M^l{}_l)
    + \Coeff{t}_{kl}\,(T^l{}_l+M^{l+1}{}_{l+1})\big]\\
    && \hspace{125pt} + Q_k{}^{k+1}\,\tfrac1k\,
    \big[\Coeff{v}_{kl}\,(T^l{}_l+M^{l+1}{}_{l+1})
    -\Coeff{t}_{lk}\,(T^l{}_l+M^l{}_l)\big]\Big)
    \nonumber
\end{eqnarray}
and hence leads to the conditions
\begin{subequations}
    \begin{eqnarray}
        0 & = & \Coeff{v}_k - \Coeff{u}_{k-1}
        +\tfrac12\,\big(\Coeff{t}_{k\,k-1} - \Coeff{t}_{kk}\big)
        - \sum_{l=1}^{\Diag} \Big(\Coeff{u}_{kl}(\ell_l+h_l)
        + \Coeff{t}_{kl}(\ell_l+h_{l+1})\Big)\,,\\
        0 & = & \Coeff{u}_k - \Coeff{v}_k
        + \tfrac12\,(\Coeff{t}_{k+1\,k} - \Coeff{t}_{kk})
        - \sum_{l=1}^{\Diag} \Big(\Coeff{v}_{kl}(\ell_l+h_{l+1})
        - \Coeff{t}_{lk}\,(\ell_l+h_l)\Big)\,,
    \end{eqnarray}
\end{subequations}
when evaluated on a principally embedded state
with diagram $(\ell_1, \ell_2, \dots | h_1, h_2, \dots)$.
As for the depth $w=\tfrac12$ case, the anti/commutator
of $f$ with $G_{\nicefrac32}$ vanishes identically
on a principally embedded state and hence does not yield
any new condition on the dressing function. Although we will not discuss the general solutions
of these conditions here, let us consider a few examples.
\begin{itemize}
\item For diagrams of one row,
the solution is unique and given by
\begin{equation}
    f= (\ell-\tfrac12)\,M^2{}_1
    + \tfrac12\,\big[Q^1{}_1, Q_1{}^2\big]\,,
\end{equation}
up to a normalization coefficient. However,
it is easy to see that $f$ vanishes identically
upon acting on the one--row subclass of \eqref{princNS}.
The leading Regge cannot namely be uplifted to $w=1$
and this is in accordance with the fact that we see
no states of spin $s=0,1,2$ at $w=1$ in table 
\ref{lightestnew}.

\item For hook--shaped diagrams $(\ell|h)$ in the (modified) 
Frobenius notation, one also finds a unique solution,
\begin{equation}
    f = -(\ell-1)\,\Big((h-1)\,T^2{}_1
    - (\ell+\tfrac12)\,M^2{}_1
    - \tfrac12\,\big[Q^1{}_1,Q_1{}^2\big]\Big)\,,
\end{equation}
up to normalization. In the simple case of the smallest hook diagram
$\sYng{;;,;}$, this yields a vertex operator whose polynomial
is given by
\begin{equation}
    F = \varepsilon_{\mu\pmb|\nu\rho}(p)\,
    \big(i\partial X^\mu\,\partial\psi^{[\nu}\,\psi^{\rho]}
    - \tfrac12\,i\partial^2 X^\mu\,\psi^\nu \psi^\rho
    - \partial\psi^\mu\,i\partial X^{[\nu}\,\psi^{\rho]}\big)\,,
\end{equation}
which account for the hook--type physical state that appears
(with multiplicity $1$) at level $N=3$. The latter
is however absent after the $\mathsf{GSO}$ projection.

\item For $p$--forms of height $h=p$, we find that $f=0$:
like the leading Regge, $p$--forms cannot be uplifted
to $w=1$ and this is in accordance
with table \ref{lightestnew}.

\end{itemize}

\paragraph{\Ram sector at depth $w=1$.}
In the Ramond sector, a dressing function
of depth $w=1$ is of the form
\begin{equation}\label{eq:Ansatz_w=1_R}
    f = \sum_{k\geq1}\,
    \Big(T^{k+1}{}_k\, P^1_k
    + M^{k+1}{}_k\, P^2_k
    + Q^{k+1}{}_k\,P^3_k
    + Q_k{}^{k+1} P^4_k\Big) +M^1 P^5 + Q^1 P^6\,,
\end{equation}
where the polynomials $P$ are a priori
arbitrary polynomials of $Q^n{}_n$.
For the first trajectory, namely the hook of one row
and one column, the dressing function simplifies to%
\footnote{Notice that the dressing function $f$
should be applied on the lowest weight state of the $\Ram$ sector,
namely on $\ket{\Y}_{\nicefrac12}$. However, this is equivalent to 
having $f$ act on \textit{physical} principally embedded states,
as the latter take the form  $P\ket{\Y}_{\nicefrac12}
=\big( 1 \pm \tfrac{1}{\sqrt{\ell+h}} Q^1{}_1 \big)
\ket{\Y}_{\nicefrac12}$ for example for the first trajectory,
and so this option amounts to mere linear redefinitions
of the coefficients of $f$ acting on $\ket{\Y}_{\nicefrac12}$.
On the other hand, as previously highlighted,
the lowest weight states of the \NS sector \textit{are} physical,
so there is no such ambiguity in the \NS sector.}
\begin{equation}\label{Rw1}
    \begin{aligned}
        f &= T^{2}{}_1 \big( \Coeff{u}_1
        + \Coeff{u}_2 \,Q^1{}_1 \big)
        + M^{2}{}_1 \big( \Coeff{v}_1
        + \Coeff{v}_2 \,Q^1{}_1 \big)
        + Q^{2}{}_1 \big( \Coeff{s}_1
        + \Coeff{s}_2 \,Q^1{}_1 \big)
        + Q_1{}^{2}\, \big( \Coeff{t}_1
        + \Coeff{t}_2 \,Q^1{}_1 \big) \\
        &  \qquad + M^1 \big( \Coeff{a}_1
        + \Coeff{a}_2 \,Q^1{}_1 \big)
        + Q^1 \big( \Coeff{c}_1
        + \Coeff{c}_2 \,Q^1{}_1 \big) \,. 
    \end{aligned}
\end{equation}
The constraint \eqref{VirTR1}, $G_0 f \ket{\Y}_{\nicefrac12} =0$,
enforces then the conditions\footnote{More precisely, each of the terms in the Ansatz \eqref{Rw1} induces one condition, hence we have twelve conditions in total, half of which are consequences of the other half.}
\begin{equation} \label{VirTR1depth1}
    \begin{aligned}
        \pm\sqrt{\ell+h+1}\,\Coeff{u}_2
        & =\Coeff{u}_1+\Coeff{s}_2+2\Coeff{t}_2 \,,
        \qquad
        & \pm\sqrt{\ell+h+1}\,\Coeff{v}_2 
        & = \Coeff{t}_2+\Coeff{v}_1  +\Coeff{s}_2\,, \\
        \pm\sqrt{\ell+h+1}\,\Coeff{s}_2
        & = -\Coeff{s}_1 -\Coeff{u}_2 +2\Coeff{v}_2\,,
        \qquad 
        & \pm\sqrt{\ell+h+1}\,\Coeff{t}_2
        & = \Coeff{u}_2 -\Coeff{t}_1 -\Coeff{v}_2\,,\\ 
        \pm\sqrt{\ell+h+1}\,\Coeff{a}_2
        & = \Coeff{a}_1 -  \Coeff{c}_2 \,,
        \qquad
        & \pm\sqrt{\ell+h+1}\,\Coeff{c}_2
        & = - \Coeff{a}_2-\Coeff{c}_1\,,
    \end{aligned}
\end{equation}
while the constraint \eqref{VirTR2}
$G_1 f \ket{\Y}_{\nicefrac12} =0$ imposes that
\begin{align} \label{VirTR2depth1}
\begin{aligned}
     \big( \ell + \tfrac{d}{2}\big) \Coeff{c}_1+  \ell  \Coeff{t}_1 +h \Coeff{s}_1  + (\ell+h)( \Coeff{u}_2+\Coeff{a}_2) &=0\,, \\
    \Coeff{v}_1 +\big(\ell+ \tfrac{d}{2}+1\big) \Coeff{c}_2 + (\ell+1) \Coeff{t}_2  +(h-1) \Coeff{s}_2 &=0\,.
\end{aligned}
\end{align}
The solutions can be written as
\begin{align}
\begin{aligned}\label{generalRw1}
    \Coeff{s}_1&= (h+\ell) (\Coeff{u}_2-2 \Coeff{v}_2)\mp\sqrt{h+\ell+1} (\Coeff{u}_1-2 \Coeff{v}_1)\,,\quad \Coeff{s}_2=\mp\sqrt{h+\ell+1} (\Coeff{u}_2-2 \Coeff{v}_2)+\Coeff{u}_1-2 \Coeff{v}_1\,,\\
\Coeff{t}_1&= \pm\sqrt{h+\ell+1} (\Coeff{u}_1-\Coeff{v}_1)-(h+\ell) (\Coeff{u}_2-\Coeff{v}_2)\,,\quad
\Coeff{t}_2= \pm\sqrt{h+\ell+1} (\Coeff{u}_2-\Coeff{v}_2)-\Coeff{u}_1+\Coeff{v}_1\,,\\
\Coeff{a}_1&= \tfrac{1}{d-2h} \Big\{ -2(3h+\ell)\,\Coeff{u}_1 + 6(2h+\ell)\,\Coeff{v}_1 \pm 6 (h+\ell) \sqrt{h+\ell+1}(\Coeff{u}_2-\Coeff{v}_2)  \Big\}\,,\\
\Coeff{a}_2&= \tfrac{2}{(d-2h)(2+d+2\ell)}\Big\{\pm 2 \sqrt{h+\ell+1} \Big[- (d+h+\ell)\Coeff{u}_1+ (2 d+ 2 h+3 \ell) \Coeff{v}_1\Big] + 2(d h+2 d \ell \\
& \qquad +d+h^2+2 h \ell+h+3 \ell^2+3 \ell)\,\Coeff{u}_2 - (d h+4 d \ell+3 d+4 h^2+4 h \ell+6 \ell^2+6 \ell) \,\Coeff{v}_2\Big\} \,,\\
\Coeff{c}_1&= \tfrac{2}{(d-2h)(2+d+2\ell)}\Big\{ \pm \sqrt{h+\ell+1} \Big[- (-d h+d \ell+2 h^2-2 h \ell-6 h-2 \ell)\,\Coeff{u}_1 \\
& \qquad + (-2 d h+d \ell+4 h^2-2 h \ell-12 h-6 \ell)\, \Coeff{v}_1 \Big] + (h+\ell) \Big[ (-d h+d \ell-d+2 h^2\\
& \qquad -2 h \ell-4 h-6 \ell-6)\, \Coeff{u}_2  -(-2 d h+d \ell+4 h^2-2 h \ell-6 h-6 \ell-6) \,\Coeff{v}_2 \Big] \Big\}\,, \\
\Coeff{c}_2&= \tfrac{2}{2+d+2\ell} \Big\{\Big[(2-h+\ell)\,\Coeff{u}_1+(-4+2h-\ell)\,\Coeff{v}_1\Big] \\
& \qquad \pm \sqrt{h+\ell+1} \Big[(h-\ell-2)\Coeff{u}_2-(2h-\ell-3)\,\Coeff{v}_2 \Big] \Big\}\,,
\end{aligned}
\end{align}
where the $d$--dependence is due to the $d$--contribution to the anticommutator $\{Q^m,Q_n\}$ given in \eqref{alg:mod3}.

To consider an example,  let as look at the lightest member
of this trajectory, namely the vector--spinor
highlighted in \textcolor{Dandelion}{dandelion}
in table \ref{lightestnew}, for which $\ell=0\,,h=1$.
Its Ansatz simplifies further to
\begin{equation} \label{Rw1vs}
    \begin{aligned}
        f &= \Coeff{u}_2\, T^{2}{}_1 Q^1{}_1
        + \Coeff{v}_1\, M^{2}{}_1 + \Coeff{s}_1\, Q^{2}{}_1
        + \Coeff{t}_2\, Q_1{}^{2}Q^1{}_1
        + M^1 \big( \Coeff{a}_1 + \Coeff{a}_2\, Q^1{}_1 \big)
        + Q^1 \big( \Coeff{c}_1 + \Coeff{c}_2\, Q^1{}_1 \big)\,.
    \end{aligned}
\end{equation}
The first four conditions of \eqref{VirTR1depth1}
now have to be replaced by the system
\begin{equation} \label{VirTR1depth1sp}
    \begin{aligned}
        \pm\sqrt{2}\,\Coeff{u}_2
        & =2\Coeff{t}_2 \,,
        \qquad
        & \pm\sqrt{2}\,\Coeff{v}_1 
        & =\Coeff{s}_1\,,
    \end{aligned}
\end{equation}
as the choice of independent constraints in \eqref{VirTR1depth1} involves constraints that are not present for the vector--spinor, since four terms of \eqref{Rw1} now vanish trivially.\footnote{More precisely, the first four terms in \eqref{Rw1vs} induce four conditions, half of which are consequences of the other half, which can be written as \eqref{VirTR1depth1sp}.} In the remaining two conditions of \eqref{VirTR1depth1}, as well as the conditions \eqref{VirTR2depth1}, it suffices to set $\Coeff{u}_1=\Coeff{v}_2=\Coeff{s}_2=\Coeff{t}_1 =0 $. Solving the complete system yields the physical vector--spinor
\begin{align}\label{vectorsol}
\begin{aligned}
   & \varepsilon_{\mu}^{\Ram, A\pm} \,
    \Big[\pm\tfrac{1}{\sqrt{2}}\,b_{-2}^\mu +\tfrac12 \alpha_{-2}^\mu \pm\tfrac{3}{d-2}\gamma\cdot b_{-1} b^\mu_{-1}  \mp \tfrac{1}{d+2}\gamma\cdot \alpha_{-1} \alpha^\mu_{-1} \\
  & \qquad \qquad +\tfrac{\sqrt{2}}{(d^2-4)}\big[2 (d+1)\,\gamma\cdot b_{-1} \,\alpha^\mu_{-1} -(d+4) \,\gamma\cdot \alpha_{-1}  \,b^\mu_{-1}\big] \Big]  \ket{p; A;0}_\Ram\,,
\end{aligned}
\end{align}
up to an overall coefficient. Let us sketch how to find
the polynomial of the respective vertex operator
in the $(-\nicefrac12)$ picture after the \GSO projection.
First, for the terms of \eqref{vectorsol} which contain one
or no oscillator $b_{-1}$, the dictionary \eqref{dictionarybos}
and \eqref{dictionaryR1} yields\footnote{We use the symbol
$\gamma$ for all gamma matrices, irrespectively of whether
they carry Dirac or Weyl indices.}
\begin{align}
\begin{aligned}
    \alpha_{-2}^\mu\ket{\alpha;0}_\Ram \quad & \rightarrow \quad  \partial^2 X^\mu S_\alpha \,,\\
\gamma\cdot \alpha_{-1} \alpha^\mu_{-1} \ket{\dot{\alpha};0}_\Ram  \quad & \rightarrow \quad \partial X^\mu  (\partial \slashed{X})_{\beta \dot{\alpha}} S^{\dot \alpha}\,, \\
    \ov{\gamma} \cdot \alpha_{-1}\, b_{-1}^\mu \ket{\alpha;0}_\Ram \quad & \rightarrow \quad \partial X^\nu \, \ov{\gamma}_\nu^{\dot{\gamma} 
    \alpha} (\psi^\mu \slashed{\psi})_{\alpha \dot{\beta}}S^{\dot{\beta}}\,,\quad \partial X^\nu \, \ov{\gamma}_\nu^{\dot{\gamma} 
    \alpha} (\gamma^\mu \slashed{\psi} \slashed{\psi})_{\alpha \dot{\beta}}S^{\dot{\beta}}\,,\\
     \ov{\gamma} \cdot b_{-1}\, \alpha_{-1}^\mu \ket{\alpha;0}_\Ram \quad & \rightarrow \quad \partial X^\mu \,  (\slashed{\psi}\slashed{\psi})^{\dot{\gamma}}{}_{\dot{\beta}}\,S^{\dot{\beta}}\,,
\end{aligned}
\end{align}
so, up to the spin--field, terms of the schematic form $\partial^2 X^\mu$, $\partial X^\mu \partial X^\nu$ and $\partial X^\mu \psi^\nu \psi^\lambda$. In addition, using the dictionary \eqref{partialdictionaryR} for $r=1$, as well as the covariant form, in terms of $S^\mu_\alpha$ and $\partial S^{\dot{\beta}}$, of the OPE of $\psi(z)S_A(w)$ \cite{Feng:2012bb}, finding the terms arising from $b_{-2}^\mu \ket{p; \dot{\alpha};0}_\Ram $ and  $\gamma \cdot b_{-1} b_{-1}^\mu  \ket{p; \alpha;0}_\Ram $  amounts to  calculating
\begin{align} \label{intvec}
  \oint \frac{\md z}{2 \pi i}
    \frac{1}{z^{\nicefrac52}} \psi^\mu(z) S^{\dot{\alpha}}(0)\,,\quad  \oint \frac{\md z}{2 \pi i} \frac{1}{z^{\nicefrac32}}
    \psi^\mu(z) S_\alpha(0)
    \qquad  \textrm{and} \qquad
    \oint \frac{\md z}{2 \pi i} \frac{1}{z^{\nicefrac32}} 
    \psi^\mu(z) \partial S_\alpha(0)\,.
\end{align}
Up to subleading order, the OPEs $\psi^\mu(z) S^\nu_\alpha(0)$  
and $\psi^\mu(z) \partial S_\alpha(0)$ can be found
in \cite{Feng:2012bb}; subsubleading orders, however,
can also obviously contribute and they would have to be determined 
using the bosonization 
\eqref{bosonization} of the fermions $\psi$ and spin--fields $S$. 
Schematically, it is easy to see the evaluation
of the integrals \eqref{intvec} will produce terms
involving  $S^\mu_\alpha$, $\partial S^{\dot{\beta}}$
and their derivatives, which can be rewritten in the form
$\psi^\mu \psi^\nu \psi^\lambda \psi^\kappa$
and $\psi^\mu \partial \psi^\nu\,$,
suitably contracted to produce a vector--spinor.
Picture--changing is then straightforward
by means of \eqref{picture}.

For a state $\ket{\Y}_{\nicefrac12}$ with an arbitrary
number of boxes $\Diag\geq1$ on the diagonal
of the Young diagram $\Y$, the condition
\begin{equation}
    G_0\,f\,\ket{\Y}_{\nicefrac12} = 0\,,
\end{equation}
with $f$ the general $w=1$ Ansatz \eqref{eq:Ansatz_w=1_R},
yields
\begin{equation}
    \begin{aligned}
        \mu_{\pm} P^1_k & = -P^3_k - (k+1)\,P^4_k\,,\\
        \mu_{\pm} P^2_k & = -P^3_k - k\,P^4_k\,,\\
        \mu_{\pm} P^5 & = P^6\,,
    \end{aligned}
    \hspace{50pt}
    \begin{aligned}
        \mu_{\mp} P^3_k & = -k\,P^1_k + (k+1)\,P^2_k\,,\\
        \mu_{\mp} P^4_k & = P^1_k - P^2_k\,,\\
        \mu_{\mp} P^6 & = -P^5\,,
    \end{aligned}
\end{equation}
with $\mu_\pm := \pmb{m} \mp \sqrt{\mathsf{N}_\Y^{\mathsf{R}}+1}$.
In fact, only half of these equations are independent:
a simple way to see it is to notice that three equations
on the right column are obtained as consequences
of those on the right, upon using $\mu_+\,\mu_- P \, \ket{\Y}_{\nicefrac12}  = - P \, \ket{\Y}_{\nicefrac12} $ for $P$ any of the polynomials in the Ansatz \eqref{eq:Ansatz_w=1_R}. The second condition,
\begin{equation}
    G_1\,f\,\ket{\Y}_{\nicefrac12} = 0\,,
\end{equation}
amounts to
\begin{equation}
    \begin{aligned}
        & \sum_{k\geq1} \Big(Q^k{}_k\,\big[-(k-1)\,P^1_{k-1}
        + k\,P^2_k\big] + Q_k{}^k\,\big[P^1_k - P^2_{k-1}
        + \delta_{k,1}\,P^5\big]\\
        & \hspace{50pt} + T^k{}_k\,\big[P^3_{k-1}
        + k\,P^4_k + \delta_{k,1}\,P^6\big] + \tfrac{d}2\,P^6
        + M^k{}_k\,\big[P^3_k + (k-1)\,P^4_{k-1}\big] \Big)\,
        \ket{\Y}_{\nicefrac12} = 0\,,
    \end{aligned}
\end{equation}
where the dimension--dependent term $\tfrac{d}2\,P^6$
comes from the anticommutator between $Q_1$ and $Q^1$,
which receives a correction as mentioned previously.
Since this equation is a polynomial in $Q^k{}_k$,
we can use the dictionary \eqref{eq:Q-theta}
to rewrite it in terms of the anticommuting variables
$\theta_k$ introduced earlier, so as to simplify
computations. Since we mainly aim at outlining
how to produce physical subleading trajectories
in our formalism, we will not derive the general solution
of this equation. Instead, let us simply stress
that using the representation of the above equation
as a polynomial in $\theta_k$, one obtains a linear system
consisting of the conditions that the coefficient
of each monomial vanishes, and this system can be solved
as shown in the case of the vector--spinor discussed previously.

\section{Conclusions}
\label{sec:conclusion}

String spectra comprise infinitely many \textit{physical} states, 
the vast majority of which are massive tensors of mixed symmetry. 
Traditional methodologies of constructing physical states,
such as the light cone quantization or the DDF formulation,
come with limitations such as lack of manifest covariance
or the output not immediately given in terms of irreducible representations of the massive little group.
On the other hand, the partition function
\cite{Curtright:1986di, Hanany:2010da}, offers access
to the characters of the representations that appear,
in principle, at any level along with their multiplicities,
while the exact form of the respective states remains obscure.
In \cite{Markou:2023ffh}, it was observed that the Virasoro
constraints in bosonic string theory, which select physical
states from functions of bosonic oscillators,
are given in terms of linear combinations of lowering operators
of a bigger (infinite--dimensional) symplectic algebra.
The fact that the latter commutes with the spacetime
Lorentz algebra, with which it forms a Howe dual pair,
was further employed to develop a covariant and efficient technology
of excavating entire physical trajectories
deeper in the spectrum by dressing simpler trajectories
with suitable combinations of the raising operators
of the symplectic algebra in question.

In this work, we address the same question for superstring
theory, namely, how do excited superstring states look like
and how can we construct physical states deep
inside the spectrum in a systematic manner? More specifically,
we explore the realization of Howe duality in the case
of the superstring and develop a covariant technology
of excavating physical open superstring trajectories.
With respect to the bosonic string, a first difference
is that now the Fock space of states is constructed
by means of pairs of bosonic \textit{and} fermionic oscillators.
While the former are integer modes in both sectors
of the superstring, the latter are half--integer
and integer in the \NS and \Ram sectors respectively.
Now, the superstring Virasoro constraints 
are expressed in terms of lowering operators
of the orthosymplectic algebra, whose generators
are formed by pairs of bosonic and fermionic oscillators
and which commutes with the spacetime Lorentz algebra.
In particular, the algebra $\osp(\oLim|\spLim)$ is realized
in the \NS and \Ram sectors as a limit of the algebras
$\osp(2\oRank|2\spRank,\R)$ and $\osp(2\oRank+1|2\spRank,\R)$ 
respectively, when $\oRank,\spRank \longrightarrow \infty$;
the bullets stand for the infinity of the pairs of oscillators
that are necessary to describe the full spectrum,
while they correspond to specific integers,
enumerating the subsets of oscillators that are excited,
once a certain trajectory is chosen. 
After working out the form of the physical trajectories
whose member--states appear at the lowest possible mass level
that the respective Young diagrams can find themselves at,
we can dress the former with suitable combinations
of the raising operators of the orthosymplectic algebra.
With this dressing, we can construct entire physical trajectories
deeper in the spectrum in a completely covariant way
in the critical dimension of the superstring.%
\footnote{We do not explicitly set $d=10$, but it is implicit
that no compactification has been effectuated throughout
this work.}
We illustrate the algorithm for small values of the ``depth'',
which parametrizes how ``deep'' inside the spectrum
a certain diagram or set of diagrams find themselves,
as in \cite{Markou:2023ffh} for the bosonic string.

Interestingly, while in both the \NS sector
and the bosonic string the lowest weight states
of the orthosymplectic algebra are \emph{by construction}
physical, this is no longer true for the \Ram sector:
there, a linear combination of the lowest weight states
together with other states at the same level is necessary
to form physical states in the principal embedding.
Crucially, this implies a nontrivial multiplicity
for states of depth zero in the \Ram sector, unlike \NS states, 
which matches the prediction of the partition function 
\cite{Hanany:2010da}, at least at low levels. Intuitively,
we can motivate the difference in multiplicities through 
spacetime supersymmetry: to close the supersymmetry algebra
on massive supermultiplets, fermions of a given spin appear
in several disparate copies in the same multiplet,
as can be illustrated in the toy example
of the massive spin--$2$ multiplet of $\mathcal{N}=1$, $d=4$,
that takes the form $(2,\nicefrac32,\nicefrac32,1)$.
In addition, the technology is developed in the Fock space
of superstring states, which are then suitable functions
of the creation modes of bosonic and fermionic oscillators,
with no gauge--fixing, light cone or otherwise, necessary.
In the \NS sector, the expressions for physical trajectories
can easily be translated into their respective vertex operators
by means of the simple dictionary between oscillators
and the primary fields $\partial X$ and $\psi$
and their descendants, which we illustrate by specific examples,
including the two superghost pictures, $(-1)$ and $(0)$,
that are necessary and sufficient for the calculation 
of scattering amplitudes of \NS states.%
\footnote{We also produce a formula for computing
the $(0)$ picture of any \NS vertex operator,
once the latter is given in the $(-1)$ picture.}
In the \Ram sector, to the best of our knowledge,
the dictionary is not currently available in a closed form
for any fermionic creation mode, because of the fact
that the CFT of the field $\psi$ and the spin--field $S$
is interacting. Consequently, in order to find
the vertex operator creating a certain physical trajectory,
the dictionary for every mode that is excited
in order to construct the trajectory has to be computed first.
This implies a necessity for the knowledge of the corresponding 
subleading terms in the OPE of, for example, $\psi(z)S_A(w)$, 
which have to be computed first; we sketch the procedure
for the second--lightest vector--spinor in the main text.
Let us also note that the technology can be applied to construct
closed superstring trajectories in a straightforward manner:
one has to employ two copies of the technology,
one to the left-- and one to the right--moving sector
of the closed string and then apply the level--matching condition.

In view of the method for the construction of entire physical 
trajectories deep in the superstring spectrum presented in this work, several future directions can be discussed,
as for example its formulation for compactifications.
In addition, the decay of \textit{coherent}
or of \textit{superpositions of highly excited} 
bosonic string states has been under thorough investigation,
with evidence for chaotic traits in its dependence on the angle
between the emitted tachyons and initial states
having been presented in bosonic strings \cite{Bianchi:2019ywd,Gross:2021gsj,Rosenhaus:2021xhm,Firrotta:2022cku,Bianchi:2022mhs,Firrotta:2023wem,Bianchi:2023uby,Firrotta:2024qel,Bianchi:2024fsi}.\footnote{See \cite{Hindmarsh:2010if,Skliros:2011si,Skliros:2016fqs} and \cite{Aldi:2019osr} for the construction of bosonic and of superstring \textit{coherent} states respectively.}
The states built with our technology correspond
by construction to irreducible Lorentz representations 
and the question of whether their decay 
and, more generally, string scattering amplitudes can exhibit 
chaotic characteristics remains open. Moreover,
it would be interesting to explore whether trajectories
or sets of states deeper in the superstring spectrum
can reproduce, for example, amplitudes of Kerr black holes,
given that it was recently shown that leading Regge \NS states, 
in the classical limit, do not \cite{Cangemi:2022abk},
while there has been an aggregation of results indicating
that higher--spin \textit{fields} can model aspects
of black hole scattering \cite{Arkani-Hamed:2017jhn, Guevara:2018wpp,Guevara:2019fsj, Maybee:2019jus, Cangemi:2022bew, Cangemi:2023ysz, Cangemi:2023bpe}.
Lastly, it would be fascinating to use massive higher--spin
superstring states deep inside the spectrum to probe
black--hole microstates, with past investigations
having been focused on the massless level
\cite{Bianchi:2017sds, Bianchi:2018kzy}. After all,
the (super)massive states constructed with our technology
are excitations of (here spacetime--filling) D--branes and,
as such, their scattering amplitudes with closed string states
can, in certain configurations, be thought of as describing
the absorption or emission of Hawking radiation from black holes, 
while for suitable compactifications to four spacetime dimensions
they can be neglected \cite{Maldacena:1996ds,Maldacena:1996ky}. 
Ultimately, a deeper understanding of the string spectrum
may perhaps shed light on string theory per se.

\section*{Acknowledgments}

We are very grateful to Evgeny Skvortsov for several enlightening discussions of formative impact and comments on the draft. We have also greatly benefited from discussions with Nicolas Boulanger, Euihun Joung, Renann Lipinski Jusinskas, Dieter L\"ust, Pouria Mazloumi, Oliver Schlotterer and Stephan Stieberger. The work of T.B. was supported by the European Union’s Horizon 2020 research and innovation program under the Marie Sk\l{}odowska Curie grant agreement No 101034383. The work of C.M. was supported by the European Research Council (ERC) under the European Union’s Horizon 2020 research and innovation programme (grant agreement No 101002551).

\appendix

\section{Light cone construction of the lightest fermions}
\label{app:lcRS}

In this appendix, to illustrate how light cone oscillators 
combine into physical Wigner irreps on a level--by--level 
fashion in the superstring, we give the construction
of the part of the first three levels with $(-)^F=+1$ 
``generalized'' chirality in the $\mathsf{R}$ sector
in table \ref{little_dec}, with the ``generalized'' chirality
operator given by
\begin{align}
    (-1)^F= 16 b_0^2\dots b_0^9\,(-1)^{\sum_{n>0}b_{-n}^ib_n^i}\,,
\end{align}
see for example \cite{Blumenhagen:2013fgp}.
Let us further recall that, after the GSO projection,
the physical Wigner irreps can also be found by taking
the tensor product of the lightest massive supermultiplet,
namely the spin--$2$ multiplet at level $1$,
with another $\mathfrak{so}(9)$ representation,
see for example \cite{Hanany:2010da}. The simplest case
is to take the product
\begin{align}
   \bigg( \gyoung(;;)\, \oplus\, \gyoung(;,;,;)\,
   \oplus\, \gyoung(;)_{\nicefrac12} \bigg)
   \otimes \gyoung(;)\,,
\end{align}
which produces precisely the physical state content
at level $2$.

\begin{table}[ht!]
\centering 
\renewcommand{\arraystretch}{1.5}
  \begin{tabular}{ c || c | c | c  }
   $N$ & $\mathfrak{gl}(d-2)$ tensors & $\mathfrak{so}(d-2)$ irreps & Wigner  \\ \hline \hline
   \multirow{2}{*}{$0$}  & $|\alpha\rangle$  & \multirow{2}{*}{$\textbf{8}_s$  } & \multirow{2}{*}{$\textbf{8}_s$ } \\ 
   & $\textbf{8}_s$  &  \\ \hline
   \multirow{2}{*}{$1$} & $\alpha_{-1}^i|\alpha\rangle \qquad b_{-1}^i|\dot{\alpha}\rangle $ & \multirow{2}{*}{$ \textbf{8}_c \oplus  \textbf{8}_s \oplus \textbf{56}_c  \oplus \textbf{56}_s$} & \multirow{2}{*}{$\textbf{128}$}  \\ 
   & $\textbf{8}_v\otimes \textbf{8}_s  \qquad \textbf{8}_v\otimes \textbf{8}_c $ &  \\ \hline
   \multirow{4}{*}{$2$} & $\alpha^{i}_{-1}\alpha^{j}_{-1}|\alpha\rangle \quad \alpha^{i}_{-2}|\alpha\rangle \quad$ & \multirow{2}{*}{$ \big( \textcolor{NavyBlue}{\textbf{8}_s}   \oplus \textcolor{NavyBlue}{\textbf{56}}_s \oplus \textcolor{NavyBlue}{\textbf{224}}_{vs} \big) \oplus  \big(\textbf{8}_c \oplus \textbf{56}_c \big)$} & \multirow{2}{*}{$\textcolor{NavyBlue}{\textbf{576}} \oplus \textcolor{BrickRed}{\textbf{432}}$} \\ 
   & $\textbf{36}_v \otimes \textbf{8}_s \quad  \textbf{8}_v  \otimes \textbf{8}_s$ & \\ 
   & $ b^{i}_{-2}|\dot{\alpha}\rangle  \quad b^{i}_{-1}\alpha^{j}_{-1}|\dot{\alpha}\rangle \quad b^{i}_{-1}b^{j}_{-1}|\alpha\rangle $ & $  \oplus \big( \textbf{8}_s  \oplus \textbf{56}_s \big) \oplus \big(\textcolor{NavyBlue}{\textbf{8}_c} \oplus \textcolor{Dandelion}{\textbf{8}_c}  \oplus \textcolor{NavyBlue}{\textbf{56}_c} \oplus \textcolor{BrickRed}{\textbf{56}_c} $ & \multirow{2}{*}{$\oplus \textbf{128} \oplus \textcolor{Dandelion}{\textbf{16}}$} \\ 
   & $ \quad \textbf{8}_v  \otimes \textbf{8}_c \quad  \textbf{8}_v \otimes \textbf{8}_v \otimes \textbf{8}_s \quad \textbf{28}_v \otimes \textbf{8}_s$   & $ \oplus  \textcolor{BrickRed}{\textbf{160}_c}  \oplus \textcolor{NavyBlue}{\textbf{224}_{vc}} \big) \oplus \big( \textcolor{Dandelion}{\textbf{8}_s} \oplus \textcolor{BrickRed}{\textbf{56}_s} \oplus \textcolor{BrickRed}{\textbf{160}_s}  \big)$ & 
  \end{tabular}
\renewcommand{\arraystretch}{1}
\caption{Decomposition and recombination up to $N=2$, $(-)^F=1$ section of the $\mathsf{R}$ sector} \label{little_dec}
\end{table}

\section{Tensors and Young diagrams}
\label{app:Young}
In this paper, we have to deal with tensors of arbitrary ranks
and symmetries. To do so, it is convenient to denote
indices that are all symmetrized, respectively antisymmetrized,
by the same letter, with the number of indices
in round, respectively square, brackets. More explicitly,
\begin{equation}
    S_{\mu(m)} = \tfrac1{m!}\,\sum_{\sigma\in\mathcal{S}_m} 
    S_{\mu_{\sigma_1} \dots \mu_{\sigma_m}}\,,
    \qquad 
    A_{\nu[n]} = \tfrac1{n!}\,\sum_{\sigma\in\mathcal{S}_n}
    (-1)^\sigma\,A_{\nu_{\sigma_1} \dots \nu_{\sigma_n}}\,,
\end{equation}
where $(-1)^\sigma$ is the signature of the permutation $\sigma$.
We use a vertical bar to separate different groups
of symmetrized or antisymmetrized indices,
groups which do not have any symmetry properties
between one another: for instance, the tensor
$T_{\mu_1(m_1)|\mu_2(m_2)|\dots\pmb|\nu_1[n_1]|\nu_2[n_2]|\dots}$
has several groups of totally symmetric indices, denoted by
$\mu_i$ and containing $m_i$ indices, and several groups
of totally antisymmetric indices, denoted by $\nu_i$
and containing $n_i$ indices. There is, however, no symmetry
assumed between the indices with different names,
i.e. in different groups.

Such tensors are \emph{not} irreducible under the action
of the general linear algebra $\mathfrak{gl}_d$, 
as can be easily seen in the simplest case 
of a rank--$2$ tensor $T_{\mu|\nu}$:
both its symmetric and antisymmetric parts,
$T_{(\mu|\nu)}$ and $T_{[\mu|\nu]}$,
are preserved independently by the action
of $\mathfrak{gl}_d$, and one can check that
there exists no other invariant subspace.
For tensors of rank $r \geq 3$, the situation 
becomes slightly more subtle,
in that their decomposition under $\mathfrak{gl}_d$
irreps contains more general projections
than the totally symmetric or totally antisymmetric ones.
For instance, for a rank--$3$ tensor of the form
$T_{\mu(2)|\nu}$, i.e. symmetric in its first two indices,
one finds on top of its symmetric part,
\begin{equation}
    S_{\mu\nu\rho} := T_{(\mu\nu|\rho)}
    \equiv \tfrac13\big(T_{\mu\nu|\rho}
    + 2\,T_{\rho(\mu|\nu)}\big)\,,
\end{equation}
the projection 
\begin{equation}
    H_{\mu\nu,\rho} := \tfrac23\,\big(T_{\mu\nu|\rho}
    - T_{\rho(\mu|\nu)}\big)\,,
\end{equation}
which is again symmetric in the first two indices,
but now obeys the over--symmetrization condition
\begin{equation}\label{eq:over-sym}
    H_{(\mu\nu,\rho)} = 0\,.
\end{equation}
This makes it clear that this projection
is orthogonal to the symmetric one,
and one readily sees that the original tensor
is given by
$T_{\mu\nu|\rho} = S_{\mu\nu\rho} + H_{\mu\nu,\rho}$.
One can also check that the projection $H$
does not contain any invariant subspace
so that the previous decomposition does correspond
to that of the tensor $T$ into irreps of $\mathfrak{gl}_d$.
Similarly, one could start from a rank--$3$ tensor
of the form $\widetilde T_{\mu[2],\nu}$,
i.e. one whose first two indices are antisymmetric,
and find in addition to its antisymmetric part
\begin{equation}
    A_{\mu\nu\rho} := \widetilde T_{[\mu\nu|\rho]}
    \equiv \tfrac13\,\big(\widetilde T_{\mu\nu|\rho}
    + 2\,\widetilde T_{\rho[\mu|\nu]}\big)\,,
\end{equation}
the projection
\begin{equation}
    \widetilde H_{\mu\nu,\rho}
    := \tfrac13\,\big(\widetilde T_{\mu\nu,\rho}
    - \widetilde T_{\rho[\mu|\nu]}\big)\,,
\end{equation}
which now obeys an over--antisymmetrization condition,
\begin{equation}\label{eq:over-antisym}
    \widetilde H_{[\mu\nu,\rho]} = 0\,.
\end{equation}
The situation mirrors that of the previous example:
we see that $\widetilde H$ is orthogonal
to the antisymmetric projection, and such that
$\widetilde T_{\mu\nu|\rho} = A_{\mu\nu\rho}
+ \widetilde H_{\mu\nu,\rho}$
with $\widetilde H$ corresponding to another irrep
of $\mathfrak{gl}_d$. Summing the two tensors,
\begin{equation}
    \mathcal T_{\mu|\nu|\rho} = T_{\mu\nu|\rho}
    + \widetilde T_{\mu\nu|\rho}
    = S_{\mu\nu\rho} + H_{\mu\nu,\rho}
    + \widetilde H_{\mu\nu,\rho} + A_{\mu\nu\rho}\,,
\end{equation}
we obtain a generic tensor of rank--$3$,
i.e. one whose indices have no symmetry,
and its full decomposition into irreducible
representations of $\mathfrak{gl}_d$.
The two projections $H$ and $\widetilde H$
are examples of \emph{mixed--symmetry} tensors,
which are tensors with several groups of symmetric
or antisymmetric indices, which obey conditions
of over--anti/symmetrization similar to,
but generalizing, conditions \eqref{eq:over-sym}
and \eqref{eq:over-antisym}.

The symmetry pattern of such tensors is encoded by
\emph{Young diagrams}, which are graphical representations
of an ordered sequence of integers,
$n_1 \geq n_2 \geq \dots n_p \geq 0$, 
has $p$ rows (respectively columns) of $n_k$ boxes
stacked in decreasing order from top to bottom
(respectively from left to right). Young diagrams 
with a total of $n$ boxes classify the irreps
of the symmetric group $\mathcal{S}_n$.
For instance, if $n=3$ there exists three Young diagrams,
\begin{equation}
    \gyoung(;;;) \qquad\qquad \gyoung(;;,;) 
    \qquad\quad\text{and}\quad\qquad \gyoung(;,;,;)\,,
\end{equation}
corresponding to the three inequivalent irreps
of $\mathcal{S}_3$. Young diagrams also classify
finite--dimensional irreps of $\mathfrak{gl}_d$\,,
a result known as Schur--Weyl duality
(which can be considered a precursor of Howe duality,
see e.g. the classical textbook of Weyl \cite{Weyl1946},
or the more recent \cite{Goodman2009}). The correspondence
is obtained by Young projectors which,
given a rank--$n$ tensor and a Young diagram with $n$ boxes,
produce an irreducible $\mathfrak{gl}_d$ tensor by
\begin{enumerate}[label=$(\roman*)$]
\item First filling in the diagram with indices of the tensor
such that the position of the index both increases
when read from left to right and from top to bottom;
\item Second, symmetrizing over the indices in each row;
\item And third, antisymmetrizing the result over indices
in each column.
\end{enumerate}
Alternatively, one can swap the last two steps,
i.e. first antisymmetrize indices in each column,
and then symmetrize over the result in each row.
Tensors obtained with the first method
(symmetrization, then antisymmetrization)
are said to be in the \emph{symmetric basis},
whereas those obtained by the second method
in the \emph{antisymmetric basis}, but both
belong to the \emph{same} $\mathfrak{gl}_d$--irrep
--- as the name suggest, the two method merely
correspond to different basis for constructing
the same representation.

In the previous $n=3$ example, the totally symmetric tensor
corresponds to the one--row diagram,
\begin{equation}
    \Yboxdim{15pt}
    S_{\mu\nu\rho}
    \qquad\longleftrightarrow\qquad
    \gyoung(\mu\nu\rho)\,,
\end{equation}
the totally antisymmetric one to the one--column diagram,
\begin{equation}
    \Yboxdim{15pt}
    A_{\mu\nu\rho}
    \qquad\longleftrightarrow\qquad 
    \gyoung(\mu,\nu,\rho)\,,
\end{equation}
and the last two projections to the two possibilities
of filling in the ``hook'' diagram,
\begin{equation}
    \Yboxdim{15pt}
    H_{\mu\nu,\rho}
    \qquad\longleftrightarrow\qquad
    \gyoung(\mu;\nu,\rho)
    \hspace{50pt}\text{and}\hspace{50pt}
    \widetilde H_{\mu\nu,\rho}
    \qquad\longleftrightarrow\qquad
    \gyoung(\mu;\rho,\nu)\,.
\end{equation}
Note that $H_{\mu\nu,\rho}$ is written
in the symmetric basis, whereas $\widetilde H_{\mu\nu,\rho}$
is in the antisymmetric basis. Let us also stress
once more that both tensors correspond
to the same $\mathfrak{gl}_d$--irrep,
the one characterized by the Young diagram $\gyoung(;;,;)$.
For more details on mixed--symmetry tensors and Young diagrams,
see the pedagogical review \cite{Bekaert:2006py}.

Concretely, in the symmetric basis we will write 
\begin{equation}
    \phi_{\mu(s_1), \nu(s_2), \dots, \rho(s_p)}\,,
\end{equation}
for a tensor which possesses a group of $s_1$ indices
collectively denoted by $\mu$, a group of $s_2$ indices
collectively denoted by $\nu$, etc, all of which symmetric
with respect to permutations with other indices
in the same group. On top of that, these groups of symmetric
indices obey the conditions
\begin{equation}
    \phi_{\dots, \mu(s_i), \dots, \mu\nu(s_j-1), \dots} = 0\,,
    \qquad 1 \leq i < j \leq p\,,
\end{equation}
i.e. the symmetrization of all indices in the $i$th group
with any one index in the $j$th group, where $i<j$, 
identically vanishes. We usually say such a tensor
has the symmetries of a Young diagram whose $k$th row
has length $s_k$. In the antisymmetric basis, we will write
\begin{equation}
    F_{\mu[h_1],\nu[h_2],\dots,\rho[h_p]}\,,
\end{equation}
for a tensor which possesses a group of $h_1$ indices
collectively denoted by $\mu$, a group of $h_2$ indices
collectively denoted by $\nu$, etc, all of which antisymmetric
with respect to permutations with other indices
in the same group. On top of that, these groups
of antisymmetric indices obey the conditions
\begin{equation}
    F_{\dots, \mu[s_i], \dots, \mu\nu[s_j-1], \dots} = 0\,,
    \qquad 1 \leq i < j \leq p\,,
\end{equation}
i.e. the antisymmetrization of all indices in the $i$th group
with any one index in the $j$th group, where $i<j$, 
identically vanishes. We usually say such a tensor
has the symmetries of a Young diagram whose $k$th column
has height $h_k$.

\newpage
\bibliographystyle{utphys}
\bibliography{biblio}

\end{document}